\documentclass[aps,nofootinbib,showpacs,showkeys,preprint
] {revtex4}

\usepackage{epsf,epsfig,subfigure,axodraw,graphicx,amsmath,amssymb}
\usepackage{color}

\begin{document}
%\draft

 \begin{flushright}
{\tt KIAS-P07087
\\SNUTP 07/014}
\end{flushright}

\title{\Large\bf Kaluza-Klein masses in nonprime orbifolds:
\\${\bf Z}_{12-I}$ compactification and threshold correction
}

\author{ Jihn E. Kim$^{(a)}$\footnote{jekim@phyp.snu.ac.kr} and Bumseok Kyae$^{(b)}$\footnote{bkyae@kias.re.kr} }
\address{$^{(a)}$ Department of Physics and Astronomy and
Center for Theoretical Physics, Seoul National University, Seoul 151-747, Korea
\\
$^{(b)}$ School of Physics, Korea Institute for Advanced Study,
207-43 Cheongryangri-dong, Dongdaemun-gu, Seoul 130-722, Korea}

%\maketitle

\begin{abstract}
Analyzing the one-loop partition function, we discuss possible Kaluza-Klein (KK) states in the orbifold compactification of the heterotic string theory, toward the application to the threshold correction. The KK massive states associated with (relatively) large extra dimensions can arise only in non-prime orbifolds. The GSO projection condition by a shift vector $V^I$ is somewhat relaxed above the compactification scale $1/R$. We also present the other condition on Wilson line $W$, $P\cdot W={\rm integer}$.  With the knowledge of the partition function, we obtain the threshold corrections to gauge couplings, which include the Wilson line effects. We point out the differences in string and field theoretic orbifolds.
\end{abstract}

 \maketitle

\def\lsim{\lower.7ex\hbox{$\;\stackrel{\textstyle<}{\sim}\;$}}
\def\lsl{ l \hspace{-0.45 em}/}
\def\ksl{ k \hspace{-0.45 em}/}
\def\qsl{ q \hspace{-0.45 em}/}
\def\psl{ p \hspace{-0.45 em}/}
\def\ppsl{ p' \hspace{-0.70 em}/}
\def\dsl{ \partial \hspace{-0.5 em}/}
\def\Dsl{ D \hspace{-0.55 em}/}
\def\N{$\cal N$}
\def\tphi{\tilde\phi}

\def\Qem{{$Q_{\rm em}$}}
  \def\SMSSM{${\cal S}$MSSM\ }
\def\CPT{${\cal CPT}$\ }

 \def\N{{$\cal N$}}
 \def\Z{{\bf Z}}
 \def\MG{{$M_{\rm GUT}$}}

\def\EE{E$_8\times$E$_8^\prime$}

 \def\N{{$\cal N$}}
 \def\Z{{\bf Z}}
 \def\MG{{$M_{\rm GUT}$}}

\def\EE{E$_8\times$E$_8^\prime$}
\def\Eo{E$_8$}
\def\Eh{E$_8'$}
\def\MGUT{$M_{\rm GUT}$}
\def\ie{{\it i.e.}\ }

\def\fourb{\overline{\bf 4}}
\def\four{{\bf 4}}
\def\two{{\bf 2}}
\def\fiveb{\overline{\bf 5}}
\def\five{{\bf 5}}
\def\tenb{\overline{\bf 10}}
\def\ten{{\bf 10}}
\def\one{{\bf 1}}
\def\six{{\bf 6}}
\def\threeb{\overline{\bf 3}}
\def\three{{\bf 3}}

%%%%%%%%%%%%%%%%%%%%%%%%%%%%%%%%%%%%%%%%%%%%%%%%%%%%%%%%%%%
%%%%%%%%%%%%%%%%%%%%%%%%%%%%%%%%%%%%%%%%%%%%%%%%%%%%%%%%%%%%%
\section{introduction}

The original Kaluza-Klein (KK) method for obtaining a gauge
symmetry from the compactification process introduces a mass scale of the inverse compactification radius, $m/R$ for a charge $m$ particle with the compactification radius $R$.  Thus the attempt to obtain QED from the KK process is phenomenologically unsuccessful \cite{KKlein} if the compactification radius is of order the Planck scale. This requires the compactification process with chiral fermions so that there would be massless particles whose masses are independent of the compactification radius. The smallness of the masses of the standard model (SM) fields is
interpreted basically due to the appearance of chiral
representations in the compactification process.

While the original KK compactification on torus does not admit such chiral fermions \cite{ShelterII}, orbifolding of torus makes it possible to obtain chiral fermions. Orbifold has been introduced first in the heterotic string theory \cite{DHVW}, where the compactification process toward chiral fermions has been very successful \cite{IKNQ}.

The boundary conditions used in string orbifolds are also
applicable to higher dimensional field theory where useful
phenomenological consequences such as the doublet-triplet
splitting have been obtained \cite{kawamura}. However, orbifold field theory cannot be considered a fundamental theory and it has been hoped that string orbifolds would provide some successful field theoretic orbifold models. To obtain field theoretic orbifold models with a large radius $R$ from string theory, one needs two effective radii for compactification, which has led to the recent interest in nonprime orbifolds \cite{nonprime}.

\begin{figure}[t]
\resizebox{0.5\columnwidth}{!}
{\includegraphics{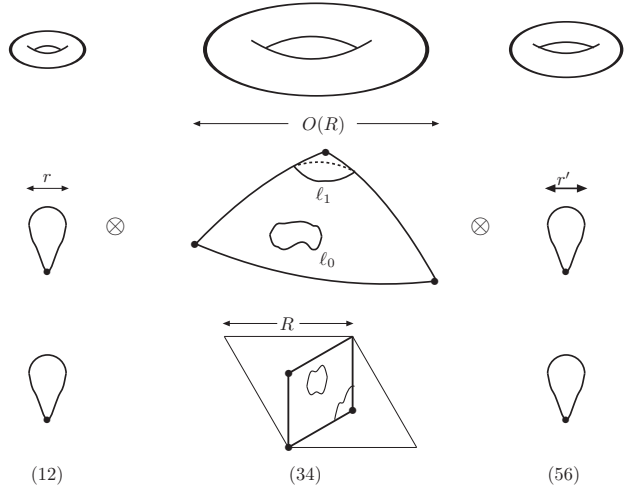}}
\caption{The 6d internal space of $T_{1,2,4,7}$ sectors of the $\Z_{12-I}$ orbifold: two teardrop topologies and one triangular ravioli topology. In the (34)-torus, untwisted string $\ell_0$ and twisted string $\ell_1$ are also shown. The area of the lower
(12)-orbifold [(34)-orbifold, (56)-orbifold] is
$\frac{1}{12}\left[\frac13,\frac{1}{12}\right]$ of the upper
torus. The fundamental region in the (34) torus is the
parallelogram (the second of the bottom figures) bounded by thick lines with three fixed points marked as bullets. This
parallelogram is the spread out version of the triangular
ravioli.}\label{fig:Z12I}
\end{figure}

Orbifolds are tori moded out by discrete actions $\Z_N$.
Therefore, the fundamental region of an orbifold is $1/N$ of the area of the original torus. In the orbifold topology, there are fixed points which are located at the boundary of the fundamental region. A typical cartoon for for a $\Z_{12-I}$ orbifold is shown in Fig. \ref{fig:Z12I}.
%%%%%%%%%%%%%%%%%%%%%%%%%%%%%%%%%%%%%%%%%%%%%%%%%%%%%%%%%%%%%%
%%%%%%%%%%%%%%%%%%%%%%%%%%%%%%%%%%%%%%%%%%%%%%%%%%%%%%%%%%%%%%
For string orbifolds the orbifold boundary conditions are used for the $(\sigma,\tau)$ directions of the string world-sheet coordinates, and for field theoretic orbifolds the boundary conditions are used for the extra coordinates of the wave functions. In an orbifold, strings can propagate in the extra space (untwisted string) or located at the fixed points (twist string). For a string orbifold, they are shown in Fig. \ref{fig:Z12I}.

In Fig. \ref{fig:Z12I}, we consider that one radius of two tori are large $R\gg r\sim r'$. The KK states in four spacetime dimensions (4D) have masses quantized in units of $1/R$. It is useful to obtain the explicit KK masses for the purpose of obtaining the running equations for gauge couplings. In field theoretic calculation, there is no hint how the threshold correction appears in the evolution equation \cite{Dienes}. In string theory, however, the threshold correction can be reliably calculated due to the exact knowledge on the spectrum in principle. Indeed, the calculation is possible with the string partition function \cite{Kaplunovsky,Dixon:1990pc,Antoniadis}. As
for the gauge model in six spacetime dimensions (6D) compactified from 10D, a simple method is to compactify just four internal spaces (4d$_{\rm i}$) to obtain a 6D model. Equivalently, it can be obtained from the 4D model by taking the limit $R\to\infty$. However, toward the gauge coupling running, we need the {\it exact} $R$ dependence of the massive 4D KK particles. So one of our objectives in this paper is to obtain the exact $R$ dependence of the KK modes from string compactification.

By employing the nonprime orbifold compactification of the
heterotic string theory , we obtain 5D or 6D orbifold SUSY GUTs. The KK excited states and the relevant (relatively) large extra dimensions should be necessarily discussed in the above framework. The radii of the extra dimensions ($R$s) should be moduli in the string orbifold compactification such that one can take the large limit of the radii. In the infinitely large limit of the extra 2d$_{\rm i}$ in a 4D model, the theory would be expected to become
an effective 6D string theory with ${\cal N}=2$ or 4 SUSY (in terms of 4D SUSY). A 4D ${\cal N}=1$ orbifold model realizing a 6D ${\cal N}=2$ theory in the limit of $R\rightarrow\infty$ should possess a 6D orbifold model.

In the orbifold compactification, the presence of the ${\cal N}=2$ SUSY sector implies the presence of a sub-lattice invariant under a given twist action. An invariant sub-lattice is just a torus.
Hence, there is no twisting for strings on the invariant
sub-lattice. Background moduli such as radius (or metric) can be encoded only in the zero modes' momenta of untwisted bosonic strings~\cite{Narain:1986am}.  Since twisted strings are stuck to orbifold fixed points, they cannot accommodate moduli. In order to introduce a modulus such as arbitrarily large radius and to discuss the relevant KK excitations, therefore, we need to employ orbifold compactification providing invariant sub-lattices~\cite{Antoniadis}.

In nonprime orbifold compactifications, some higher twist sectors turn out to behave like the invariant sub-lattice preserving ${\cal N}=2$ SUSY.  KK excited states can arise from such higher twist sectors in nonprime orbifold compactifications \cite{Kaplunovsky,Dixon:1990pc,Antoniadis}. If orbifolding is associated with gauge symmetry breaking, the gauge symmetry could be enhanced by including KK massive states above the compactification scale. Hence, it is possible to construct a higher dimensional SUSY GUT with the help of KK states. Along this line, 6D SUSY GUTs based on string theory can be realized in nonprime orbifolds. As an explicit example for a concrete presentation, we choose $\Z_{12-I}$ ~\cite{Z12I, IWKim, Rparity,
SMZ12I, GMSBun, GMSBst, Katsuki}. We will discuss its partition function in details, from which we will read information on KK modes. Once we obtain the partition function, we can discuss the renormalization group (RG) evolutions of the gauge couplings in the context of string theory.

The knowledge on the SUSY breaking scale is very important toward the TeV scale phenomenology of the minimal supersymmetric standard model (MSSM). Gravity mediation needs a hidden sector confining
around $10^{13}$ GeV, while the gauge mediated SUSY breaking (GMSB) needs a hidden sector confining at a scale less than $10^{12}$ GeV \cite{GMSBun}. For the 4D calculation of the unification of gauge couplings to make sense, we assume that the 4D GUT scale is below the compactification scale $\frac1R$, leading to the following hierarchy
\begin{equation}
 M_{\rm GUT} \le\textstyle\frac1R\le\frac1r<M_s, M_P
\end{equation}
where we have not specified the hierarchy between the string scale $M_s$ and the Planck scale $M_P$.

We will employ the $\Z_{12-I}$ compactification model
\cite{SMZ12I} for a concrete discussion if needed. The first
impression for the $\Z_{12-I}$ might be that it is too
complicated. But it is not. In any orbifold compactification, the final number of fields are of order 100--300. The seemingly simple $\Z_3$ orbifold has a complexity in its Wilson line choices because 27 fixed points of $\Z_3$ are distinguished only by Wilson lines. The seemingly complicated $\Z_{12-I}$ has only 3 fixed points and hence the $\Z_{12-I}$ Wilson line is very simple. In addition,  the $\Z_{12-I}$ orbifold has an invariant torus so that there results a meaningful $R$ dependence of the KK masses. Actually the $\Z_{12-I}$ orbifold requires only one modulus. In a
sense, for an example toward the threshold correction, $\Z_{12-I}$ models are the simplest. Some important physical observations such as the axion coupling constants and $R$-parity embedding with three family MSSM have been explicitly studied in $\Z_{12-I}$ \cite{IWKim,Rparity}.

The organization of the paper is the following. In.
Sec. \ref{sec:LargeR}, we discuss the limit of a large
compactification radius. In Sec. \ref{sec:6DGUT}, we compactify 4d$_{\rm i}$ internal space to obtain a 6D model. In the limit of $R\to\infty$ from a 4D model, we expect this 6D spectrum. In Secs. \ref{sec:Partition}, \ref{sec:Modular}, and \ref{sec:KKTower}, we take the partition function approach. In Sec. \ref{sec:Partition}, we pay attention to the S and T transformation properties of the partition function. In Sec. \ref{sec:Modular}, we discuss the
modular invariance of the partition function in detail. In Sec. \ref{sec:KKTower}, we derive the masses and the GSO projection valid for the KK states from the partition function. In Sec. \ref{sec:couplings}, we obtain the gauge coupling evolution and
the threshold correction. Sec. \ref{sec:Conclusion} is a
conclusion. In Appendix, we present some calculations for a
coefficient in the beta function.

%%%%%%%%%%%%%%%%%%%%%%%%%%%%%%%%%%%%%%%%%%%%%%%%%%%%%%%%%%%%%%
%%%%%%%%%%%%%%%%%%%%%%%%%%%%%%%%%%%%%%%%%%%%%%%%%%%%%%%%%%%%%%%
\section{Large compactification radii}\label{sec:LargeR}

Partition functions of nonprime orbifolds are the key toward introducing KK masses. As a simple example, we take the $\Z_{12-I}$ orbifold model \cite{SMZ12I} which seems to have interesting properties \cite{Z12I}.  Our discussion here could be generalized to other nonprime orbifolds. Before presenting partition functions, let us get a physical intuition for the internal space of nonprime orbifolds.

\begin{figure}[t]
\resizebox{0.5\columnwidth}{!}
{\includegraphics{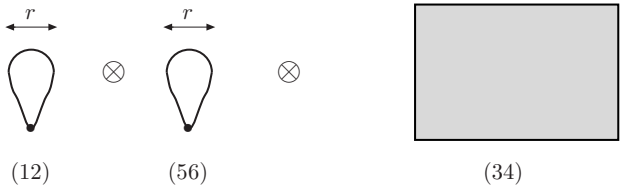}}
\caption{The 4d internal space.}\label{fig:4dint}
\end{figure}

The $\Z_{12}$ orbifold is not an irreducible one, and we consider it as a direct product type orbifold $\Z_M\times\Z_N$ in 4D with $M=4$ and $N=3$
\cite{DHVW,ChoiKimBk}. Thus, the $\Z_{12-I}$ shift vector we will introduce is
\begin{align}
{\rm \Z_{12-I}\ shift}:\quad \phi&=\textstyle(\frac{5}{12}
~\frac{4}{12}~\frac{1}{12})\label{shif}
\end{align}
which is the sum of $\Z_4$ and $\Z_3$ shifts
\begin{align}
&\Z_4: \textstyle(\frac{-1}{4}~0~\frac{-1}{4})\\
&\Z_3: \textstyle(\frac23~\frac{1}{3}~\frac{1}{3})
\end{align}
In the limit $R\to\infty$, we may consider a 4D $\Z_4\times\Z_3$ orbifold with
\begin{align}
&\Z_4: \textstyle(\frac{-1}{4}~\frac{-1}{4})\label{Z4shift}\\
&\Z_3: \textstyle(\frac23~\frac{1}{3})
\end{align}
where the modding is for the first and the third tori. The
$\Z_4\times\Z_3$ orbifold has only one fixed point at the origin, which is a kind of teardrop topology. In \cite{GMSBun}, a relatively small gauge coupling for the hidden sector compared to that of the observable sector needed in some gauge mediated supersymmetry breaking is expected to arise due to the KK mode contribution, which can be studied in the 6D model. The second torus in Eq. (\ref{shif}) is a $\Z_3$ orbifold shift and has three
fixed points. But when we consider the 6D theory, the second torus is treated uncompactified as depicted in Fig. \ref{fig:4dint}.
%%%%%%%%%%%%%%%%%%%%%%%%%%%%%%%%%%%%%%%%%%%%%%%%%%%%%%%%%%%%%%
%%%%%%%%%%%%%%%%%%%%%%%%%%%%%%%%%%%%%%%%%%%%%%%%%%%%%%%%%%%%%%
The compactification radius $r$ of (12) and (56) tori are assumed to be close to the string scale $R_s\equiv r$ and the compactification radius $R$ of the (34) torus is large with a hierarchy of radii $R\gg r$. With the 6d$_{\rm i}$ compactified, we obtain a four dimensional (4D)
model\footnote{We use d$_{\rm i}$ for the internal space dimension and D for the uncompactified dimension.} which must give the phenomenologically successful SM and a confining hidden sector gauge group.

We are familiar with the 4D gauge group from orbifold
compactification by considering massless conditions satisfying $P\cdot V=0$ and $P\cdot W=0$ mod Z in the untwisted sector. Of course, this gauge group can be obtained also by considering the common {\it intersection} of gauge groups obtained at each fixed point. What will be an appropriate approximate gauge group when the radius of the second torus tends to infinity? The Wilson line in the second torus distinguishes three fixed points there. If we
do not want to distinguish the three fixed points, then we simply neglect the Wilson line, and consider only $P\cdot V=0$. Even though we neglect the Wilson line $W$, still there must be three fixed points in the second torus with the same spectrum at each fixed point. When the radius of the second torus tends to infinity, the gauge bosons moving in the torus, e.g. $l_0$ of Fig. \ref{fig:Z12I}, are still the gauge boson. In addition, there can be additional gauge bosons. A string of the type $l_1$ in  Fig. \ref{fig:Z12I} cannot be a gauge boson. But, if the fixed points are not present in some twisted sector, there can be additional gauge bosons. In this case, we must consider {\it the union} of
gauge bosons of the $l_0$ type gauge bosons of the untwisted
sector and $l_1$ type gauge bosons in twisted sectors without fixed points. The $l_1$ type gauge boson is in the fixed point in the first and second tori.  For the $l_1$ type gauge bosons of $T_3$ sector to be truly 6D gauge bosons, it should not belong to the fixed points of the first and the third tori.

The case when fixed points are removed appears in the tori where twisting is zero mod integer. For $\Z_{12-I}$, this happens in $T_{3k}\ (k=1,2,3)$ twisted sectors with shift vector $3kV$. The gauge group of the $T_6$ sector therefore contains the gauge group of the $T_3$ sector. Thus, for the 6D gauge group it is appropriate to take the common intersection of the gauge groups of $T_{3k}$ sectors of 4D, which is simply the gauge group of $T_3$. Then, the shift vector $3V$ with $\Z_{12-I}$ shift $V$ gives a
$\Z_4$ orbifold model in 6D since the second torus is the fixed torus. The teardrop fixed points of (12) and (56) tori are the fixed points of the $\Z_4$ orbifold, and the $l_1$ type fields become the untwisted sector fields of the $\Z_4$ orbifold and become 6D gauge bosons. This kind of analysis is the strategy we take when the
second torus is enlarged.

A similar method can be used in other nonprime orbifolds,
$\Z_4,\Z_{6-I},\Z_{6-II},\Z_{8-I},\Z_{8-II}$, and $\Z_{12-II}$. Even though $\Z_{12-I}$ looks complicated, in fact it is the simplest in discussing fixed torus since the fixed torus is obtained just by removing $\Z_3$. In stating $\lq\lq$simplest'', we consider the number of Wilson line conditions and the number of fixed points being simplest. It is known that $\Z_3$ is simpler than $\Z_2$ in string compactification.

%%%%%%%%%%%%%%%%%%%%%%%%%%%%%%%%%%%%%%%%%%%%%%%%%%%%%%%%%%%%%%
%%%%%%%%%%%%%%%%%%%%%%%%%%%%%%%%%%%%%%%%%%%%%%%%%%%%%%%%%%%%%%
\section{6D Grand Unification}\label{sec:6DGUT}

In this section, we obtain the 6D massless spectrum by considering orbifold compactification of 4d internal space, following the method given in \cite{ChoiKimBk}. Let us take the following forms of the shift vector and Wilson lines~\cite{SMZ12I}
\begin{equation}\label{ShiftSM}
\begin{array}{l}
V=\textstyle\left( \frac14~ \frac14~ \frac14~ \frac{1}{4}~
\frac14~ \frac{5}{12}~\frac{5}{12}~  \frac{1}{12}~ \right)\left(\frac{1}{4}~
\frac{3}{4}~ 0~ 0~0~0~0~0 \right) ,\\
W_3=W_4\equiv W=\textstyle\left(
\frac23~\frac23~\frac23~\frac{-2}{3}~\frac{-2}{3}~\frac23~
0~\frac23 \right)\left( 0~ \frac23~\frac{2}{3}~0~0~0~0~0 \right),
\end{array}
\end{equation}
and $W_1=W_2=W_5=W_6=0$.

%%%%%%%%%%%%%%%%%%%%%%%%%%%%%%%%%%%%%%%%%%%%%%%%%%%%%%%%%%%%%
\subsection{$\Z_4$ in two tori}\label{subsec:6Dmodel}

Thus, toward a 6D model we study the $\Z_4$ twist given in Eq. (\ref{Z4shift})
$$
\phi=\textstyle(\frac{1}{4}~\frac{1}{4}).
$$
We split the shift Eq.~(\ref{ShiftSM}) in terms of $\Z_4$ and
$\Z_3$ shifts $V=V_4+V_3$ where
\begin{align}\label{VZ4}
V_4 &= \textstyle\left( \frac{1}{4}~\frac{1}{4}~
\frac{1}{4}~\frac{1}{4} ~\frac{1}{4}~  \frac{-1}{4}~
\frac{-1}{4}~\frac34~ \right) \left(\frac{1}{4}~\frac{3}{4}~ 0~
0~0~0~0~0 \right) ,
\nonumber\\
V_3 &=\textstyle\left( 0~0~0~0~0~ \frac{2}{3}~\frac{2}{3}~
\frac{-2}{3}~ \right)\left(0~0~ 0~ 0~0~0~0~0 \right).
\end{align}
Note $3V\sim -V_4$. For a 6D theory, we consider only the $\Z_4$
shifts.

\vskip 0.3cm
%%%%%%%%%%%%%%%%%%%%%%%%%%%%%%%%%%%%%%%%%%%%%%%%%%%%%%%%%%%%%
\centerline{\it Untwisted sector}
 The massless gauge fields in the
untwisted sector satisfy the condition $P\cdot V_4=0$ mod
integer with $P^2=2$. These are
\begin{align}
{\bf 63}&:\ P=\left\{
\begin{array}{ll}
 \textstyle\left( \underline{-1~1~0~0~0~0~0~}0 \right)\left(0^8 \right) &:42\\
\textstyle\left( \underline{\pm1~0~0~0~0~0~0~}\pm1 \right)
\left(0^8 \right) &:14\\
 (0^8)(0^8)\ {\rm KK\ modes}&:{\rm rank\ 7}
\end{array}\right.\\
{\bf 3}&:\ P=\left\{
\begin{array}{ll}
 \textstyle\left(0^8 \right)\left( \underline{-1~-1};~0^6 \right) &:2\\
(0^8)(0^8)\ {\rm KK\ modes}&:{\rm rank\ 1}
\end{array}\right.\\
{\bf 66}&:\ P=\left\{
\begin{array}{ll}
 \textstyle\left(0^8 \right)\left(0~0~; \underline{\pm1~\pm1~0~0~0~0}
 \right) &:60\\
 (0^8)(0^8)\ {\rm KK\ modes}&:{\rm rank\ 6}
\end{array}\right.
\end{align}
Thus, the 6D gauge group is
\begin{equation}
{\rm SU(8)\times U(1)\times SU(2)'\times SO(12)'\times U(1)'}.
\end{equation}

%%%%%%%%%%%%%%%%%%%%%%%%%%%%%%%%%%%%%%%%%%%%%%%%%%%%%%%%%%%%%%%
%
\begin{table}
\begin{tabular}{c|ccc}
\hline %\noalign{\smallskip}
$P\cdot V_4$ (mod Z) &States ($P$) & 4D chirality
& ~$(SU(8); SU(2)', SO(12)')$~ \\
\hline $\frac14$  &$\left(\underline{++---};\underline{++-}\right)
(0~0;0^6)'$ & L, R &\\
$\frac14$  & $\left(\underline{+++--};+++\right) (0~0;0^6)'$  & L, R &\\
$\frac14$  &$\left(\underline{+----};\underline{+--}\right)
(0~0;0^6)'$ & L, R & $({\bf 56};\one,\one)$\\
$\frac14$  &$\left(-----;---\right) (0~0;0^6)'$ & L, R &
\\
 \hline $\frac14$
&$\left(\underline{+----};+++\right) (0~0;0^6)'$ & L, R &\\
$\frac14$ & $\left(-----;\underline{-++}\right) (0~0;0^6)'$ & L, R& $({\bf 8};\one,\one)$\\
\hline
$\frac{-1}{4}$
&$\left(\underline{--+++};\underline{--+}\right) (0~0;0^6)'$ & R, L &\\
$\frac{-1}{4}$ &$\left(\underline{---++};---\right) (0~0;0^6)'$  &R, L & CTP conjugate\\
$\frac{-1}{4}$ &$\left(\underline{-++++};\underline{-++}\right)
(0~0;0^6)'$ & R, L & $(\overline{\bf 56};\one,\one)$\\
$\frac{-1}{4}$  &$\left(+++++;+++\right) (0~0;0^6)'$ & R, L &\\
\hline
$\frac{-1}{4}$&$\left(\underline{-++++};---\right)
(0~0;0^6)'$ & R, L & CTP conjugate\\
$\frac{-1}{4}$ & $\left(-----;\underline{+--}\right) (0~0;0^6)'$ &R, L  & $(\overline{\bf 8};\one,\one)$\\
\hline
$\frac14$ & $(0^8)(+1~0;\underline{\pm 1,0^5})'$ & L, R &\\
$\frac14$ & $(0^8)(0,-1;\underline{\pm 1,0^5})'$ & L, R& $(\one;\two',{\bf 12'})$\\
\hline
$\frac14$ & $(0^8)(-+;\underline{+-----})'$ & L, R &\\
$\frac14$ & $(0^8)(-+;\underline{+++---})'$ & L, R &
$(\one;\one',{\bf 32'})$\\
$\frac14$ & $(0^8)(-+;\underline{+++++-})'$ & L, R &
\\
\hline
$\frac{-1}{4}$ & $(0^8)(-1~0;\underline{\pm 1,0^5})'$ & R, L & CTP conjugate\\
$\frac{-1}{4}$ & $(0^8)(0,+1;\underline{\pm 1,0^5})'$ & R, L &$(\one;\two',{\bf 12'})$ \\
\hline
$\frac{-1}{4}$ & $(0^8)(+-;\underline{-+++++})'$ & R, L &
 CTP conjugate\\
$\frac{-1}{4}$ & $(0^8)(+-;\underline{---+++})'$ & R, L &
$(\one;\one',{\bf 32'})$\\
$\frac{-1}{4}$ & $(0^8)(+-;\underline{-----+})'$ & R, L &
\\ \hline
\end{tabular}
\caption{6D untwisted matter states. } \label{tab:6Duntwst}
\end{table}
Matter representations satisfy $P\cdot V={k}/4\ (k=  {\rm odd\ integer})$ with $P^2=2$, which are shown in Table \ref{tab:6Duntwst}. The sector for $P\cdot V={(4-k)}/4$ contains the CTP conjugates of $P\cdot V={k}/4$. In Table
\ref{tab:6Duntwst}, we also list 4D chiralities from the SO(8) spinor $s$, i.e. L for $s\cdot\phi$ ($=\pm\frac14$) and R for $s\cdot \phi$ ($=\mp\frac14$). One L and one R of a 4D spinor make up one 6D spinor. Thus, $({\bf 56};\one,\one)_L+({\bf 56};\one,\one)_R$ compose a 6D spinor, and $({\bf 8};\one,\one)_L+({\bf 8};\one,\one)_R$ compose a 6D spinor. Their CTP conjugates are $(\overline{\bf 56};\one,\one)_R+(\overline{\bf
56};\one,\one)_L$ and  $(\overline{\bf 8};
\one,\one)_R+(\overline{\bf 8};\one,\one)_L$.

\vskip 0.3cm
%%%%%%%%%%%%%%%%%%%%%%%%%%%%%%%%%%%%%%%%%%%%%%%%%%%%%%%%%%%%%
\centerline{\it Twisted sectors} For the $\theta^k$ twists the left and right movers have the following vacuum energy
contributions
\begin{align}
2\tilde c_k=\left\{\begin{array}{c} \frac{13}{8},\ k=1\\
 \frac{3}{2},\ k=2
 \end{array}
 \right.\label{ctildeL4}
\end{align}
from which we can calculate $2c=2\tilde c-1$,
\begin{align}
2 c_k=\left\{\begin{array}{c} \frac{5}{8},\ k=1\\
 \frac{1}{2},\ k=2\ .
 \end{array}
 \right.\label{cZ4R}
\end{align}

For a 6D theory, we consider the $T_1,T_2,T_3$ twisted sectors, i.e. $k=1,2,3$ with the shift (\ref{VZ4}). Since the spectrum of $T_3$ is the CTP conjugates of that of $T_1$, it is enough to calculate the massless spectrum of $T_1$ and $T_2$ sectors. The masslessness condition in the twisted sectors is
\begin{equation}
(P+k V_4)^2=2\tilde c-2\tilde N_L
\end{equation}

%%TTTTTTTTTTTTTTTTTTTTTTTTTTTTTTTTTTTTTTTTTTTTTTTTTTTTTTTTTTTTTTTTTTTTTTTT
\begin{table}
\begin{center}
\begin{tabular}{|c|cc|}
\hline %& & & & & & $k$ & $=$ & & & & & \\
 $\Z_2:k\diagdown l$~ & ~~${0}$~ & ~${1}$~
\\
\hline $\quad\ \ {0}$ & ~${1}$  & ${1}$
\\
$\quad\ \ {1}$ & ~${16}$  & ${16}$\\
\hline
\end{tabular}\hskip 0.4cm
\begin{tabular}{|c|ccc|}
\hline %& & & & & & $k$ & $=$ & & & & & \\
 $\Z_3:k\diagdown l$~ & ~~${0}$~ & ~${1}$~ & ~${2}$~\\
\hline $\quad\ \ {0}$ & ~${1}$  & ${1}$  & ${1}$\\
$\quad\ \ {1}$ & ~${9}$  & ${9}$& ${9}$\\
\hline
\end{tabular}\hskip 0.4cm
\begin{tabular}{|c|cccc|}
\hline %& & & & & & $k$ & $=$ & & & & & \\
 $\Z_4:k\diagdown l$~ & ~~${0}$~ & ~${1}$~ &~${2}$~  &~${3}$~\\
\hline $\quad\ \ {0}$ & ~${1}$ & ${1}$ & ${1}$ & ${1}$\\
$\quad\ \ {1}$ & ~${4}$ & ${4}$ & ${4}$ & ${4}$\\
$\quad\ \ {2}$ & ~${16}$ & ${4}$ & ${16}$  & ${4}$\\
\hline
\end{tabular}\vskip 0.2cm
\begin{tabular}{|c|cccccc|}
\hline %& & & & & & $k$ & $=$ & & & & & \\
 $\Z_6:k\diagdown l$~ & ~~${0}$~ & ~${1}$~ &
~${2}$~ & ~${3}$~ & ~${4}$~ & ~${5}$~\\
\hline $\quad\ \ {0}$ & ~${1}$ & ${1}$ & ${1}$ & ${1}$ & ${1}$ & ${1}$\\
$\quad\ \ {1}$ & ~${1}$ & ${1}$ & ${1}$ & ${1}$ & ${1}$ & ${1}$\\
$\quad\ \ {2}$ & ~${9}$ & ${3}$ & ${9}$ & ${3}$& ${9}$ & ${3}$\\
$\quad\ \ {3}$ & ~${16}$ & ${1}$ & ${1}$ & ${16}$& ${1}$ & ${1}$\\
\hline
\end{tabular}

\end{center}
\caption{The degeneracy factor $\tilde\chi(\theta^k,\theta^l)$ of
${\bf Z}_{N}$ orbifolds in $\rm d_i=4$ {\it compact} space dimensions.
}\label{tb:degnKK}
\end{table}
%%TTTTTTTTTTTTTTTTTTTTTTTTTTTTTTTTTTTTTTTTTTTTTTTTTTTTTTTTTTTTTTTTTTTTTTTTTTTT
%

 With 4d$_{\rm i}$ internal space, there are only four kinds of orbifolds, with orders 2, 3, 4, and 6. The order 6 orbifold is too simple so that there is only one kind of $\Z_6$ in 4d$_{\rm i}$ compared to two kinds, $\Z_{6-I}$ and $\Z_{6-II}$, in 6d$_{\rm i}$.

%%%%%%%%%%%%%%%%%%%%%%%%%%%%%%%%%%%%%%%%%%%%%%%%%%%%%%%%%%%%%%%
%
\begin{table}
\begin{tabular}{c|c|c|c|c}
\hline
States ($P+kV_4$) & ~Sector~ & ~${\cal P}_k$~ & ~4D $\chi$~
& ~Representations~\\
\hline
$\left(\underline{\frac{-3}{4}~\frac{1}{4}~\frac{1}{4}
\frac{1}{4}~\frac{1}{4}}~; ~\frac{-1}{4}~\frac{-1}{4}~\frac{-1}{4}\right)
(\{\frac{1}{4}~\frac{3}{4}\};0^6)'$ & $T_1$ & 4 & L, R &
\\
$\left(\frac{1}{4}~\frac{1}{4}~\frac{1}{4}~\frac{1}{4}
~\frac{1}{4}~;\underline{\frac{3}{4}~
\frac{-1}{4}~\frac{-1}{4}}\right)
(\{\frac{1}{4}~\frac{3}{4}\};0^6)'$ &  $T_1$ & 4 & L, R & $4\times
(\overline{\bf 8};\two',\one)$
\\ [0.3em]
\hline $\left(\frac{-1}{4}~\frac{-1}{4}~\frac{-1}{4}~
\frac{-1}{4}~\frac{-1}{4}~;~
\frac{1}{4}~\frac{1}{4}~\frac{1}{4}\right)
(\frac{1}{4}~\frac{-1}{4};\underline{\pm 1~0~0~0~0~0})'$ & $T_1$ &
4 & L, R & $4\times (\one;\one,{\bf 12}')$
\\
\hline $\left(\frac{-1}{4}~\frac{-1}{4}~\frac{-1}{4}~
\frac{-1}{4}~\frac{-1}{4}~;~
\frac{1}{4}~\frac{1}{4}~\frac{1}{4}\right)
(\{\frac{1}{4}~\frac{3}{4}\};0^6)'$ & $T_1(
N_i^L=1_1,1_{\bar{2}})$ & 4 & L, R  & $8\times (\one;\two',\one)$\\
\hline
$\left(\underline{1~0~0~0~0};~0~0~0\right)(+-;0^6)'$ & $T_2$ & 6 & L, R &\\
$\left(0~0~0~0~0~0~;\underline{-1~0~0}\right) (+-;0^6)'$ &  $T_2$& 6  & L, R & $6\times ({\bf 8};\one,\one)$\\
\hline
$\left(\underline{1~0~0~0~0};~0~0~0\right)(-+;0^6)'$ & $T_2$ & 10& L, R &\\
$\left(0~0~0~0~0~0~;\underline{-1~0~0}\right) (-+;0^6)'$ &  $T_2$& 10 & L, R & $10\times ({\bf 8};\one,\one)$\\
\hline
$\left(\underline{-1~0~0~0~0};~0~0~0\right)(+-;0^6)'$ & $T_2$ & 10& R, L & CTP conjugate\\
$\left(0~0~0~0~0~0~;\underline{1~0~0}\right) (+-;0^6)'$ &  $T_2$ &10 & R, L & $10\times (\overline{\bf 8};\one,\one)$\\
\hline $\left(\underline{-1~0~0~0~0};~0~0~0\right)(-+;0^6)'$ &$T_2$ & 6& R, L & CTP conjugate\\
$\left(0~0~0~0~0~0~;\underline{1~0~0}\right) (-+;0^6)'$ &  $T_2$ &6 & R, L & $6\times (\overline{\bf 8};\one,\one)$\\
 \hline
\end{tabular}
\caption{6D twisted sector massless matter states under SU(8)$\times$SU(2)$'\times$SO(12)$'$. The
$\{\frac{1}{4}~\frac{3}{4}\}$ means a doublet of
$(\frac{1}{4}~\frac{3}{4})$ and $(\frac{-3}{4}~\frac{-1}{4})$. The CTP conjugates of the $T_1$ sector fields appear in the $T_3$ sector.}
\label{tab:6Dtwisted}
\end{table}
The 6D massless fields from the twisted sector are shown
in Table \ref{tab:6Dtwisted}.

%%%%%%%%%%%%%%%%%%%%%%%%%%%%%%%%%%%%%%%%%%%%%%%%%%%%%%%%%%%%%%%
\subsection{6D anomaly-free combinations}\label{subsec:6Danomaly}

In this subsection, we show that the massless fields obtained in \ref{subsec:6Dmodel} form an anomaly free set.

\vskip 0.3cm
\noindent{(i)  SO($2N$) with adjoint plus $m$ spinors and $n$ vectors}:

The anomaly cancellation condition for SO($2N$) is
\begin{equation}
2N-8+2^{N-5}m-n=0.
\end{equation}
Namely, for some SO($2N$) groups  we have
\begin{align}
\begin{array}{l}
{\rm SO(10)}:\ 2+m_{\rm 16}-n_{10}=0\\
{\rm SO(12)}:\ 4+2m_{\rm 32}-n_{12}=0\\
{\rm SO(14)}:\ 6+4m_{\rm 64}-n_{14}=0\\
{\rm SO(16)}:\ 8+8m_{\rm 128}-n_{16}=0.
\end{array}
\end{align}
For the model discussed in the preceding subsection, the SO(12)$'$ anomaly is absent since $m_{32}=1$ and $n_{12}=6$.
\vskip 0.3cm

\noindent{(ii) SU($N$) with an adjoint}:

The anomaly cancellation condition with a few simple representations of SU($N$) is
 \begin{equation}
2N-n_f-(N-8)n_{a_2}-\frac{1}{2}(N^2-17N+54)n_{a_3}
-(N+8)n_{s_2}=0
\end{equation}
where $n_f$ is the fundamentals, $n_{s_2}(n_{a_2})$ is the two-index (anti)symmetric tensors, $n_{a_3}$ is the three-index antisymmetric tensors. We did not write contributions from more complex representations. Symmetric tensors do not appear in the Kac-Moody level 1 algebra. Thus, the relevant SU(5), SU(6), SU(7) and SU(8) formulae for string compactification derived from the Kac-Moody level 1 are
\begin{align}
\begin{array}{l}
{\rm SU(5)}:\ 10-n_5+3n_{10}=0\\
{\rm SU(6)}:\ 12-n_6+2n_{15}+6n_{20}=0\\
{\rm SU(7)}:\ 14-n_7+n_{21}+8n_{35}=0\\
{\rm SU(8)}:\ 16-n_8\quad\quad +9n_{56}=0.
\end{array}
\end{align}
For the model of the preceding subsection, the SU(8) anomaly is absent since $n_{8,\bar 8}=25$ and $n_{56}=1$.

%%%%%%%%%%%%%%%%%%%%%%%%%%%%%%%%%%%%%%%%%%%%%%%%%%%%%%%%%%%%%%
%%%%%%%%%%%%%%%%%%%%%%%%%%%%%%%%%%%%%%%%%%%%%%%%%%%%%%%%%%%%%%
\section{${\bf Z}_{12-I}$ Orbifold with Background fields}
\label{sec:Partition}
%%%%%%%%%%%%%%%%%%%%%%%%%%%%%%%%%%%%%%%%%%%%%%%%%%%%%%%%%%%%%%%
%%%%%%%%%%%%%%%%%%%%%%%%%%%%%%%%%%%%%%%%%%%%%%%%%%%%%%%%%%%%%%%

Starting from this section, we reproduce the results obtained in Sec. \ref{sec:6DGUT} by studying the partition functions and furthermore derive the $R$ dependence of KK masses. In Sec. \ref{sec:6DGUT}, we could not obtain the $R$ dependence of the KK spectrum since we took the $R\to\infty$ limit.
 A field theoretic method with the spectrum of Sec. \ref{sec:6DGUT} may enable us to guess the $R$ dependence, which may not be reliable. Following this section, we now derive a much more reliable calculation of the $R$ dependence of the KK spectrum.

The orbifold compactification is a kind of torus compactification by modding out with a discrete action, namely a 6d torus $T^6$ is modded out by a twist $\theta$. The twist $\theta$ is an automorphism of the lattice $\Lambda$. The 6d torus $T^6$ can be studied easily if it is assumed to be factorized as $T^2\times T^2\times T^2$, which is followed here. Then the twist $\theta$ can be given as eigenvalues in the three complex planes
\begin{equation}
(e^{2\pi i\phi_1},e^{2\pi i\phi_2},e^{2\pi i\phi_3}).\label{twistphi}
\end{equation}
For the \N=1 4D SUSY, we require \cite{SMZ12I,Z12I,ChoiKimBk}
\begin{equation}
-\phi_1+\phi_2+\phi_3={\rm integer}.
\end{equation}
Strings appear as untwisted or twisted strings. The twisted string in the $i$th torus is closed with the orbifold identification
\begin{equation}
X(e^{2\pi i\phi_i} z,e^{-2\pi i\phi_i}\bar z)=\theta^kX(z,\bar z)+v, \quad v\in\Lambda
\end{equation}

In this paper, we are particularly interested in the nonprime orbifold $\Z_{12-I}$ which has some interesting features
\cite{Z12I,SMZ12I,GMSBun,GMSBst}. This method can be easily
extended to the other nonprime orbifold compactifications.

The $\Z_{12-I}$ orbifold is defined with the twist vector defined in (\ref{twistphi}) as
\begin{eqnarray}
 \phi=\left(\frac{5}{12},\frac{4}{12},\frac{1}{12}\right) ~,
\end{eqnarray}
acting on the complexified three dimensional $X^i$. Similarly, $\theta^*$ acts on $X^{i*}$. This orbifold can be the ${\rm SO(8)\times SU(3)}$ or ${\rm F_4\times SU(3)}$ lattices, where the SU(3) lattice corresponds to the second complex plane~\cite{Katsuki}. We will assume that the size $R$ of the second torus is much larger than the sizes $r$ of the first and the second tori as shown in Fig. \ref{fig:Z12I}. So, in the limit $R/r\to\infty$ we will have an effective 6D spacetime.

In nonprime orbifolds $\Z_N$, we can consider multiple twisted sectors $\theta^k$ with the sector $k=\frac{N}{2}$ included. The $k$th twisted sector is denoted as $T_k$. The untwisted sector $U$ corresponds to the $k=0$ case. For the $\Z_{12-I}$, there are 12 sectors including the untwisted sector.

In the $T_3$, $T_6$, and $T_9$ sectors of $\Z_{12-I}$, where the boundary conditions are given by $3\phi$, $6\phi$, and $9\phi$, respectively, the second (complexified) sub-lattice remains untwisted.  Namely, the second sub-lattice is just an ordinary torus in the $T_3$, $T_6$, and $T_9$ sectors. The KK massive states can arise from the sectors associated with invariant sub-lattices, where \N=2 4D SUSY is
preserved~\cite{Kaplunovsky,Dixon:1990pc,Antoniadis}. This means that the spectrum is vectorlike in 4D. Moreover
such sectors are decoupled from the other sector under modular transformations as will be shown. Thus, our discussion  will be mainly focused on the $U$, $T_3$, $T_6$, and $T_9$ sectors.

Let us consider the one-loop partition function in orbifiold. Much important physical information such as the mass-shell condition, GSO projection~\cite{GSO}, and so on, can be extracted from the one-loop partition function. The one-loop amplitude by closed strings has the topology of a torus.  A world-sheet torus can be parametrized by $\sigma_1+\tau\sigma_2$
($0\leq\sigma_1,~\sigma_2\leq 2\pi$), where $\tau$ ($\equiv
\tau_1+i\tau_2$) is the modular parameter. On the  world-sheet torus, we have two boundary conditions, $[g^k,~g^l]$ (or simply $[k,l]$), where $g^k$ ($g^l$) implies the action of the order $k$ ($l$) twisted string boundary conditions in the $\sigma_1$ ($\sigma_2$) direction. In the $[k,l]$ sectors with $k~and~l= 0,3,6,9$, $X_{L,R}^i$ in the longitudinal directions of the second complex plane are untwisted, and so KK massive states can arise
from there.  On the other hand, in the $[k,l]$ sectors with
$k~or~l\neq 0,3,6,9$, where only ${\cal N}=1$ SUSY is preserved, there do not appear KK excited states.

The partition function in ${\bf Z}_{12-I}$ orbifold
compactification is given by \cite{Ibanez:1987pj,ChoiKimBk}
\begin{eqnarray} \label{partition}
\int\frac{d\tau^2}{\tau_2^2} \frac{1}{12}\sum_{k,l;\zeta,\zeta'}
\lvert Z^X_{[k,l]}\rvert^2\cdot
Z^G_{[k,l;\zeta,\zeta']}\cdot\overline{Z}^\psi_{[k,l;\zeta,\zeta']}
~,
\end{eqnarray}
where $\lvert Z^X_{[k,l;\zeta,\zeta']}\rvert^2$,
$Z^G_{[k,l;\zeta,\zeta']}$ and
$\overline{Z}^\psi_{[k,l;\zeta,\zeta']}$ indicate the
contributions to the full partition function by the world-sheet bosons $X_{L,R}^\mu$ ($\mu=1,2,\cdots,6;0,7,8,9$), $X_L^I$
($I=10,11,\cdots,25$), and world-sheet fermion $\psi_R^\mu$
in the light cone gauge. The $Z^X_{[k,l]}$ takes the following form:
\begin{align} \label{Z^X}
&Z^X_{[k,l]}(\tau)=
\frac{\tilde{\chi}^{1/2}(\theta^k,\theta^l)}{\left(
2\pi\tau_2^{1/2}\left[\eta(\tau)\right]^2\right)}
\times\prod_{i=1}^3\frac{\eta(\tau)}{\vartheta\left[
\begin{array}{c}
1/2+k\phi_i \\
1/2+l\phi_i
\end{array} \right] },
\quad {\rm for}~k\neq
0 ~,
\\ \label{untwistZ^X}
&Z^X_{[k=0,l]}(\tau)=
\frac{\tilde{\chi}^{1/2}(\theta^0,\theta^l)}{\left(
2\pi\tau_2^{1/2}\left[\eta(\tau)\right]^2\right)}
\times\left\{\chi^{1/2}_{l}\prod_{i=1}^3\frac{\eta(\tau)
}{\vartheta\left[
\begin{array}{c}
1/2 \\
1/2+l\phi_i \end{array} \right] }\right\}, \quad
{\rm for}~k=0~,
\end{align}
where $\eta(\tau)$ and $\vartheta[\cdots]$ denotes the well-known Dedekind (``eta'') and Jacobi (``theta'') functions. Their definitions are found e.g. in
Refs.~\cite{Ibanez:1987pj,ChoiKimBk}.
$\tilde{\chi}(\theta^k,\theta^l)$ is the degeneracy factor, which is displayed in Table~\ref{tb:degn}.
%
%%TTTTTTTTTTTTTTTTTTTTTTTTTTTTTTTTTTTTTTTTTTTTTTTTTTTTTTTTTTTTTTTTTTTTTTTT
\begin{table}
\begin{center}
\begin{tabular}{c|cccccccccccc}
\hline %& & & & & & $k$ & $=$ & & & & & \\
 $k\diagdown l$ & ${\bf 0}$ & $1$ & $2$ & ${\bf 3}$ & $4$ & $5$ &
${\bf 6}$ & $7$ & $8$ & ${\bf 9}$ & $10$ & $11$
\\
\hline ${\bf 0}$ & ${\bf 1}$ & $1$ & $1$ & ${\bf 1}$ & $1$ & $1$ &
${\bf 1}$ & $1$ & $1$ & ${\bf 1}$ & $1$ & $1$
\\
$1$ & $3$ & $3$ & $3$ & $3$ & $3$ & $3$ & $3$ & $3$ & $3$ & $3$ &
$3$ & $3$
\\
$2$ & $3$ & $3$ & $3$ & $3$ & $3$ & $3$ & $3$ & $3$ & $3$ & $3$ &
$3$ & $3$
\\
${\bf 3}$ & ${\bf 4}$ & $1$ & $1$ & ${\bf 4}$ & $1$ & $1$ & ${\bf 4}$ & $1$ & $1$ & ${\bf 4}$ & $1$ & $1$
\\
$4$ & $27$ & $3$ & $3$ & $3$ & $27$ & $3$ & $3$ & $3$ & $27$ & $3$ & $3$ & $3$
\\
$5$ & $3$ & $3$ & $3$ & $3$ & $3$ & $3$ & $3$ & $3$ & $3$ & $3$ & $3$ & $3$
\\
${\bf 6}$ & ${\bf 16}$ & $1$ & $1$ & ${\bf 4}$ & $1$ & $1$ & ${\bf 16}$ & $1$ & $1$ & ${\bf 4}$ & $1$ & $1$
\\
\hline
\end{tabular}
\end{center}
\caption{Degeneracy factor $\tilde\chi(\theta^k,\theta^l)$ in the ${\bf Z}_{12-I}$ orbifold.}\label{tb:degn}
\end{table}
%%TTTTTTTTTTTTTTTTTTTTTTTTTTTTTTTTTTTTTTTTTTTTTTTTTTTTTTTTTTTTTTTTTTTTTTTTTTTT
%
$\chi^{1/2}_l$ in Eq.~(\ref{untwistZ^X}) appears only in the
untwisted sector. By modular invariance, it should be taken as
$\chi_{l}^{1/2}=\tilde{\chi}(\theta^l,\theta^0)$. Thus, the part
of
$\{\chi^{1/2}_l\prod_{i=1}^3\frac{\eta(\tau)}{\theta[\cdots]}\}$
in Eq.~(\ref{untwistZ^X}) is expanded as
$q^{-1/4}\prod_{i=1}^{3}\prod_{n=1}^{\infty}(1-q^ne^{2\pi il\phi_i
})^{-1}(1-q^ne^{-2\pi il\phi_i})^{-1}$, where $q\equiv e^{2\pi
i\tau}$. In Eqs.~(\ref{Z^X}) and (\ref{untwistZ^X}),
``$1/(2\pi\tau_2^{1/2}\left[\eta(\tau)\right]^2)$'' corresponds to
the contribution of the $X_{L,R}^{\mu}$ with the non-compact
spacetime components ($\mu=0,9$). The remaining part comes from the twisted bosonic strings $X_{L,R}^{1}, X_{L,R}^{2},\cdots, X_{L,R}^{6}$.  Note that $k\phi_i$ and $l\phi_i$ are associated with the {\it twisted} boundary conditions for them in the $(\sigma_1,\sigma_2)$ directions:
\begin{eqnarray}
\Bigg\{
\begin{array}{c}
X_{L,R}^i(\sigma_1=2\pi)=(\theta^k X_{L,R})^i(\sigma=0)+z^ae_a^i
\\
X_{L,R}^i(\sigma_2=2\pi)=(\theta^l X_{L,R})^i(\tau=0)+z^{\prime
a}e_a^i
\end{array} ~.
\end{eqnarray}
As mentioned in Introduction, the twisted strings can not
accommodate moduli such as radii of the extra dimensions.

Under ``S-transformation'' $\tau\rightarrow -1/\tau$ and ``T-transformation'' $\tau\rightarrow\tau+1$,
$\lvert Z^X_{[k,l]}\rvert^2$ is transforming as
\begin{eqnarray}
{\rm S}~&:&~ \lvert Z^X_{[k,l]}\rvert^2\longrightarrow \lvert Z^X_{[l,-k]}\rvert^2 , \label{XS}
\\
{\rm T}~&:&~ \lvert Z^X_{[k,l]}\rvert^2\longrightarrow \lvert Z^X_{[k,l+k]}\rvert^2 ,\label{XT}
\end{eqnarray}
unless both $k\phi_i$ and $l\phi_i$ are integers. If both
$k\phi_i$ and $l\phi_i$ are ${\rm integers}$, however, the theta function in Eq.~(\ref{Z^X}) becomes singular. This problem happens generically in non-prime orbifolds. In fact, the case of $k\phi_i,~l\phi_i={\rm integer}$ corresponds to torus compactification. For the sectors of $k\phi_i,~l\phi_i={\rm integer}$, thus, it would be natural to reflect the results discussed in torus compactification.
In the second torus of ${\bf Z}_{12-I}$, $k\phi_2$ and $l\phi_2$ are integers for $k,l=0,3,6,9$.  Hence, in Eq.~(\ref{Z^X}),
\begin{eqnarray}
\frac{\eta(\tau)}{\vartheta\left[
\begin{array}{c}
1/2+k\phi_2 \\
1/2+l\phi_2 \end{array} \right]} \quad\quad\quad {\rm
for}~~k,l=0,3,6,9
\end{eqnarray}
needs to be replaced by the result from the torus
compactification.

The string partition function in the presence of background gauge fields $A_\mu$ and $B_{\mu\nu}$ in $D$-dimensional space compactified on a {\it torus} was studied in \cite{Ginsparg}. We will employ its result with some modifications. We will see that the sectors $\{k,l=0,3,6,9\}$ are not mixed with the other sectors by the modular transformations, and hence they are distinguished
from the other sectors.

In a $D$-dimensional internal space compactified on a torus
$\Lambda_D$, in our case $D=2$, which is generated with basis vectors $\vec{e}_i$ ($i=1,2,\cdots,D$), the metric on the space is constructed as $g_{ab}=\vec{e}_a\cdot\vec{e}_b$. Its dual lattice $\Lambda^*_D$ is generated with the dual basis vectors $\vec{e}^{*a}$ satisfying $\vec{e}^{*a}\cdot\vec{e}_b=\delta^a_b$.
Similarly, the metric of $\Lambda_D^*$ is given by
$g^{ab}=\vec{e}^{*a}\cdot\vec{e}^{*b}$, which is the inverse
metric of $g_{ab}$, i.e. $g^{ab}=g^{-1}_{ab}$. In particular, in the two dimensional SU(3) lattice $\Lambda_2$, one can take (dual) basis vectors and (inverse) metric as
\begin{align} \label{metric}
&\Bigg\{
\begin{array}{c}
\vec{e}_1=\left(\sqrt{2}, 0\right) \quad\quad~
\\
\vec{e}_2=\left(-\sqrt{\frac{1}{2}},\sqrt{\frac{3}{2}}\right)
\end{array} , \quad g_{ab} = \left(
\begin{array}{cc}
2 & -1 \\
-1 & ~2 \end{array} \right),\\
 &\Bigg\{
\begin{array}{c}
\vec{e}^{*1}=\left(\frac{1}{\sqrt{2}},
\frac{1}{\sqrt{6}}\right)\\
\vec{e}^{*2}=\left(0, \sqrt{\frac{2}{3}}\right)
\end{array} ,\quad
g^{ab} = \frac{1}{3}\left(
\begin{array}{cc}
2 & 1 \\
1 & 2 \end{array} \right) . ~
\end{align}

The partition function by $X_L^i$ and $X_R^i$ in the lattice is proportional to~\cite{Ginsparg}
\begin{eqnarray} \label{6dpart}
\left[2\pi\tau_2^{-1/2}|\eta(\tau)|^{-2}\right]^D
\sum_{{\bf\omega},\omega'\in\Lambda_D}{\rm
exp}\left[-2\pi\tau_2{\omega}^2-\frac{2\pi}{\tau_2}
(\omega'+\tau_1\omega)^2+4\pi i\omega'B\omega\right] ~.
\end{eqnarray}
Here a background gauge field $B=B_{ab}\vec{e}^{*a}\wedge
\vec{e}^{*b}$ in the sublattice $(x^3,x^4)$ is introduced.  But $B$ is not essential in the following discussion.  $\omega$ ($=z^{a}\vec{e}_a$, $z^{a}={\rm integer}$) and $\omega'$ ($=z^{\prime a}\vec{e}_a$, $z^{\prime a}={\rm integer}$) denote the windings of $X_{L,R}^{3,4}$ in the $(\sigma_1,\sigma_2)$ directions:
\begin{eqnarray} \label{winding}
\Bigg\{
\begin{array}{c}
X_{L,R}^i(\sigma_1=2\pi)=X_{L,R}^i(\sigma_1=0)+z^ae_a^i
\\
X_{L,R}^i(\sigma_2=2\pi)=X_{L,R}^i(\sigma_2=0)+z^{\prime a}e_a^i
\end{array} ~,
\end{eqnarray}
where $i=3,4$, and $a$ also runs $a=3,4$. The pre-factor
$\left[\tau_2^{-1/2}|\eta(\tau)|^{-2}\right]^D$ in
Eq.~(\ref{6dpart}) is indeed the modular invariant one-loop
partition function by $X_L^i$ and $X_R^i$ in the {\it non-compact} $D$-dimensional space. It contains all the contributions by oscillator appearing in the mode expansion of $X_{L,R}^i$. Its modifications by torus compactification appear in the exponent which contain the $\omega$ and $\omega'$ dependence, i.e. the compactification size dependence.

The exponent of Eq.~(\ref{6dpart}) is simplified as
\begin{eqnarray} \label{simple6d}
-\frac{2\pi}{\tau_2}~\bigg|\omega'+\tau\omega\bigg|^2 +4\pi
i\omega'B\omega ~.
\end{eqnarray}
Thus, one can see that in the first term, $\tau\rightarrow
-1/\tau$ (``S-transformation'') exchanges $\omega$ and $\omega'$ as $\omega\rightarrow \omega'$ and $\omega'\rightarrow -\omega$, and $\tau\rightarrow\tau+1$ (``T-transformation'') just replaces $\omega'$ by $\omega'\rightarrow\omega'+\omega$. The second term
``$4\pi i\omega'B\omega$'' in Eq.~(\ref{simple6d}) remains
invariant. Thus, under S and T transformations of $\tau$, a sector of $[\omega,\omega']$ should be transformed to other sectors as
\begin{align}
 &{\rm S} ~:~ \left(
\begin{array}{c}
\omega \\
\omega' \end{array}\right) \longrightarrow
\left( \begin{array}{c}
\omega' \\
-\omega \end{array} \right)\label{S}
\\
 &{\rm T} ~:~ \left(
\begin{array}{c}
\omega \\
\omega' \end{array} \right)\longrightarrow\left(
\begin{array}{c}
\omega \\
\omega'+\omega \end{array} \right) ~.\label{T}
\end{align}

The components in $X_L^I$ and $\overline{Z}^\psi$ in the partition function in Eq.~(\ref{partition}),
i.e. $Z^G_{[k,l;\zeta,\zeta']}\cdot
\overline{Z}^\psi_{[k,l;\zeta,\zeta']}$, takes the following form
\begin{eqnarray} \label{orbifoldpartition}
&& \left\{
\begin{array}{c}
kV+\zeta^aW_a \\
lV+\zeta^{'a}W_a \end{array} \right\}\equiv
\prod_{I=1}^{16}\frac{\vartheta \left[
\begin{array}{c}
kV^I+\zeta^aW_a^I \\
lV^I+\zeta^{'a}W_a^I  \end{array} \right] ~ e^{-\pi
i(kV+\zeta^aW_a)\cdot(lV+\zeta^{'a}W_a)} }{\eta(\tau)} \times\prod_{i=0}^{3}\frac{\bar{\vartheta} \left[
\begin{array}{c}
k\phi_i \\
l\phi_i \end{array} \right] ~ e^{\pi
ikl\phi^2}}{\bar{\eta}(\bar{\tau})} \nonumber
\\ [1em]
&&
\quad\quad = \sum_{P\in \Lambda_{16}}
\frac{q^{[P+kV+\zeta^aW_a)]^2/2}e^{2\pi
i[P+\frac{k}{2}V +\frac{\zeta^a}{2}W_a]
\cdot[lV+\zeta^{'b}W_b]}} {[\eta(\tau)]^{16}}
\sum_{s\in\Lambda_8}\frac{\bar{q}^{(s+k\phi)^2/2}e^{-2\pi
i[(s+\frac{k}{2}\phi)\cdot l\phi]}}{[\bar{\eta}(\bar{\tau})]^4} ~,
\end{eqnarray}
where $q\equiv e^{2\pi i\tau}$. $\phi_0$ ($=0$) acts on the
components ($\mu=0,9$) of the non-compact 4D directions of
$\psi^\mu_R$ in the light cone gauge. Here, for simplicity we neglect the spin structure.  $P$ and $s$ indicate the \EE\ and the SO(8) weight vectors, respectively. $kV+\zeta^aW_a$ and $lV+\zeta^{\prime a}W_a$ are associated with the twist boundary conditions of $X_L^I$ in the $(\sigma_1,\sigma_2)$ directions on the world-sheet torus,
\begin{eqnarray} \label{wilson}
\Bigg\{
\begin{array}{c}
X^I_L(\sigma_1=2\pi)=X^I_L(\sigma_1=0)+kV^I+\zeta^a W^I_a
\\
X^I_L(\sigma_2=2\pi)=X^I_L(\sigma_2=0)+lV^I+\zeta^{\prime a} W^I_a
\end{array} ~,
\end{eqnarray}
where $V^I$ and $W^I_a$ ($I=10, 11,\cdots, 15$, $a=1, 2,\cdots, 6$) stand for the shift vector and Wilson line, respectively. They shift \EE\ lattice vectors. $\zeta^a$ and $\zeta^{\prime a}$ in Eq.~(\ref{orbifoldpartition}) are proper integers. The homomorphism $\vec{e}_a\rightarrow W^I_a$ in Eqs.~(\ref{winding}) and (\ref{wilson}) leads to the identification of $z^{(\prime)a}$ in Eq.~(\ref{6dpart}) and $\zeta^{(\prime)a}$ in Eq.~(\ref{orbifoldpartition}). Under S and T transformations, the $\{kV+\zeta^aW_a,lV+\zeta^{\prime a}W_a\}$ sector transforms
itself in the same manner as given with the components of
$X_{L,R}^\mu$  in Eqs.~(\ref{XS}), (\ref{XT}), (\ref{S}), and (\ref{T}),
\begin{eqnarray}
 {\rm S} ~:~ \left\{
\begin{array}{c}
kV+\zeta^aW_a \\
lV+\zeta^{\prime a}W_a \end{array} \right\}&\longrightarrow&
\left\{
\begin{array}{c}
lV+\zeta^{\prime a}W_a \\
-kV-\zeta^{a}W_a \end{array} \right\}
\\
{\rm T} ~:~ \left\{
\begin{array}{c}
kV+\zeta^aW_a \\
lV+\zeta^{\prime a}W_a \end{array} \right\}&\longrightarrow&
\left\{
\begin{array}{c}
kV+\zeta^aW_a \\
(k+l)V+(\zeta^{\prime a}+\zeta^{a})W_a \end{array} \right\} .
\end{eqnarray}
Since the full partition function of Eq.~(\ref{partition})
contains all possible sectors for $k,l,\zeta,\zeta'$, it is
modular invariant.

%%%%%%%%%%%%%%%%%%%%%%%%%%%%%%%%%%%%%%%%%%%%%%%%%%%%%%%%%%%%%%%
%%%%%%%%%%%%%%%%%%%%%%%%%%%%%%%%%%%%%%%%%%%%%%%%%%%%%%%%%%%%%%%
\section{Modular Invariance in ${\bf Z}_{12-I}$}\label{sec:Modular}

In ${\bf Z}_{12-I}$, two identical order three Wilson lines can be introduced,
\begin{eqnarray}
W^I_3=W^I_4\equiv W^I~,
\end{eqnarray}
and $W_1=W_2=W_5=W_6=0$. For consistency, $12\times V^I$ and
$3\times W^I_a\ (a=3,4)$ should be \EE\ weight vectors,
and $V^I$ and $W^I_a$ should satisfy the following modular
invariance conditions~\cite{Ibanez:1987pj,Forste,ChoiKimBk}:
\begin{align}
&12(V^2-\phi^2) = {\rm even ~integer} ~,
\label{v^2} \\
&12 V\cdot W = {\rm even ~integer} ~,
\label{vw} \\
&12 W^2 = {\rm even ~integer} ~. \label{w^2}
\end{align}
In fact, the general modular invariance condition to Eq. (\ref{vw}) is $12 V\cdot W = {\rm integer}$. However, the case with $12 V\cdot W = {\rm odd ~integer}$ can be always converted to the case with $12 V\cdot W = {\rm even ~integer} $ by a proper lattice shifting of $W$: $W\to W'+\omega_{\rm E_8\times E_8'}$ where $\omega_{\rm E_8\times E_8'} (\in \Lambda_{\rm E_8\times E_8'})$ denotes a weight vector of ${\rm E_8\times E_8'}$, leaving field spectra intact.
 With Eq.~(\ref{w^2}) and $(3W^I_a)^2=({\rm E_8\times E_8' ~weight~vectors})^2={\rm even~integer}$, we get
\begin{eqnarray}
 W^2=\frac{2}{3}\times {\rm integer} \label{w^2-2}
\end{eqnarray}
which is the same as the $\Z_3$ condition.

One can easily see that under the S transformation $\tau\rightarrow -1/\tau$, the $[k,l]$ sectors\footnote{Sometimes we will call the $[kV+\zeta
V,lV+\zeta^\prime W]$ sector briefly as $[k,l]$ sector.} with $k,l=0,3,6,9$ are transformed as
\begin{align}
\one= \left[
\begin{array}{c}
k=0
\\[-0.5em] ^{(\zeta)}
\\[-0.5em] l=0
\\[-0.5em] ^{(\zeta')}
\\[-0.3em]
\end{array} \right],~
 \one=\left[
\begin{array}{c}
k=6
\\[-0.5em] ^{(\zeta)}
\\[-0.5em]  l=6
\\[-0.5em] ^{(\zeta')}
\\[-0.3em]
\end{array} \right]
\end{align}
%%%%%%%%%%%%%%%%%%%%%%%%%%%%%%%%%%%%%%%%%%%%%%%%%%%%%%
%%%%%%%%%%%%%%%%%%%%%%%%%%%%%%%%%%%%%%%%%%%%%%%%%%%%%%
\begin{align}
\two=\left[
\begin{array}{c}
k=0
\\[-0.5em] ^{(\zeta)}
\\[-0.5em]  l=6
\\[-0.5em] ^{(\zeta')}
\\[-0.3em]
\end{array} \right] \longleftrightarrow \left[
\begin{array}{c}
k=6
\\[-0.5em] ^{(\zeta')}
\\[-0.5em] l=0
\\[-0.5em] ^{(-\zeta)}
\\[-0.3em]
\end{array} \right]
\label{S-untwist}
\end{align}
%%%%%%%%%%%%%%%%%%%%%%%%%%%%%%%%%%%%%%%%%%%%%%%%%%%%
%%%%%%%%%%%%%%%%%%%%%%%%%%%%%%%%%%%%%%%%%%%%%%%%%%%%
\begin{align}
{\bf 4}=\left[\begin{array}{c} k=0
\\[-0.5em] ^{(\zeta)}
\\[-0.5em]  l=3
\\[-0.5em] ^{(\zeta')}
\\[-0.3em]
\end{array} \right]
\rightarrow \left[
\begin{array}{c}
k=3
\\[-0.5em] ^{(\zeta')}
\\[-0.5em] l=0
\\[-0.5em] ^{(-\zeta)}
\\[-0.3em]
\end{array} \right] \rightarrow \left[
\begin{array}{c}
k=0
\\[-0.5em] ^{(-\zeta)}
\\[-0.5em] l=9 \\[-0.5em] ^{(-\zeta')}
\\[-0.3em]
\end{array} \right] \rightarrow \left[
\begin{array}{c}
k=9
\\[-0.5em] ^{(-\zeta')}
\\[-0.5em]  l=0
\\[-0.5em] ^{(\zeta)}
\\[-0.3em]
\end{array} \right] \rightarrow \left[
\begin{array}{c}
k=0
\\[-0.5em] ^{(\zeta)}
\\[-0.5em]  l=3
\\[-0.5em] ^{(\zeta')}
\\[-0.3em]
\end{array} \right]  \label{S-twist1}
\end{align}
%%%%%%%%%%%%%%%%%%%%%%%%%%%%%%%%%%%%%%%%%%%%%%%%%%%%%%%%%
%%%%%%%%%%%%%%%%%%%%%%%%%%%%%%%%%%%%%%%%%%%%%%%%%%%%%%%%%
\begin{align}
{\bf 4}=\left[\begin{array}{c} k=3
\\
l=3
\end{array} \right]
\rightarrow \left[
\begin{array}{c}
k=3
\\
l=9
\end{array} \right] \rightarrow \left[
\begin{array}{c}
k=9
\\
l=9
\end{array} \right] \rightarrow \left[
\begin{array}{c}
k=9
\\
l=3
\end{array} \right] \rightarrow \left[
\begin{array}{c}
k=3
\\
l=3
\end{array} \right]
\label{S-twist2}
\end{align}
%%%%%%%%%%%%%%%%%%%%%%%%%%%%%%%%%%%%%%%%%%%%%%%%%%%%%%%%
%%%%%%%%%%%%%%%%%%%%%%%%%%%%%%%%%%%%%%%%%%%%%%%%%%%%%%%%
\begin{align}
{\bf 4}=\left[\begin{array}{c} k=3
\\
l=6
\end{array} \right] \rightarrow \left[
\begin{array}{c}
k=6
\\
l=9
\end{array} \right] \rightarrow \left[
\begin{array}{c}
k=9
\\
l=6
\end{array} \right] \rightarrow \left[
\begin{array}{c}
k=6
\\
l=3
\end{array} \right] \rightarrow \left[
\begin{array}{c}
k=3
\\
l=6
\end{array} \right] \label{S-twist3}
\end{align}
and under the T transformation $\tau\rightarrow\tau+1$,
\begin{align}
\one=\left[\begin{array}{c} k=0
\\
l=0
\end{array} \right],~
\one=\left[
\begin{array}{c}
k=0
\\
l=3
\end{array} \right],~
\one=\left[
\begin{array}{c}
k=0
\\
l=6
\end{array} \right],~
\one=\left[
\begin{array}{c}
k=0
\\
l=9
\end{array} \right]
\label{T-U}
\end{align}
%%%%%%%%%%%%%%%%%%%%%%%%%%%%%%%%%%%%%%%%%%%%%%%%%%%%%%%%%%%%%
%%%%%%%%%%%%%%%%%%%%%%%%%%%%%%%%%%%%%%%%%%%%%%%%%%%%%%%%%%%%%
\begin{align}
{\bf 4}=\left[\begin{array}{c} k=3
\\[-0.5em] ^{(\zeta)}
\\[-0.5em]  l=0
\\[-0.5em] ^{(\zeta')}
\\[-0.3em]
\end{array} \right]
\rightarrow \left[
\begin{array}{c}
k=3
\\[-0.5em] ^{(\zeta)}
\\[-0.5em]  l=3
\\[-0.5em] ^{(\zeta'+\zeta)}
\\[-0.3em]
\end{array} \right] \rightarrow \left[
\begin{array}{c}
k=3
\\[-0.5em] ^{(\zeta)}
\\[-0.5em]  l=6
\\[-0.5em] ^{(\zeta'+2\zeta)}
\\[-0.3em]
\end{array} \right] \rightarrow \left[
\begin{array}{c}
k=3
\\[-0.5em] ^{(\zeta)}
\\[-0.5em]  l=9
\\[-0.5em] ^{(\zeta'+3\zeta)}
\\[-0.3em]
\end{array} \right] \rightarrow \left[
\begin{array}{c}
k=3
\\[-0.5em] ^{(\zeta)}
\\[-0.5em]  l=0
\\[-0.5em] ^{(\zeta'+4\zeta)}
\\[-0.3em]
\end{array} \right]
\label{T-T3}
\end{align}
%%%%%%%%%%%%%%%%%%%%%%%%%%%%%%%%%%%%%%%%%%%%%%%%%%%%%%%%%%%%%
%%%%%%%%%%%%%%%%%%%%%%%%%%%%%%%%%%%%%%%%%%%%%%%%%%%%%%%%%%%%%
\begin{align}
{\bf 2}=\left[\begin{array}{c} k=6
\\[-0.5em] ^{(\zeta)}
\\[-0.5em]  l=0
\\[-0.5em] ^{(\zeta')}
\\[-0.3em]
\end{array}
\right] \longleftrightarrow \left[
\begin{array}{c}
k=6
\\[-0.5em] ^{(\zeta)}
\\[-0.5em]  l=6
\\[-0.5em] ^{(\zeta'+\zeta)}
\\[-0.3em]
\end{array} \right] ~,\quad\quad
\two=\left[\begin{array}{c} k=6
\\[-0.5em] ^{(\zeta)}
\\[-0.5em]  l=3
\\[-0.5em] ^{(\zeta')}
\\[-0.3em]
\end{array} \right]
\longleftrightarrow \left[
\begin{array}{c}
k=6
\\[-0.5em] ^{(\zeta)}
\\[-0.5em]  l=9
\\[-0.5em] ^{(\zeta'+\zeta)}
\\[-0.3em]
\end{array} \right]
\label{T-T6}
\end{align}
%%%%%%%%%%%%%%%%%%%%%%%%%%%%%%%%%%%%%%%%%%%%%%%%%%%%%%%%%%%%%
%%%%%%%%%%%%%%%%%%%%%%%%%%%%%%%%%%%%%%%%%%%%%%%%%%%%%%%%%%%%%
\begin{align}
{\bf 4}=\left[\begin{array}{c} k=9
\\[-0.5em] ^{(\zeta)}
\\
[-0.5em] l=0
\\[-0.5em] ^{(\zeta')}
\\[-0.3em]
\end{array} \right]
\rightarrow \left[
\begin{array}{c}
k=9
\\[-0.5em] ^{(\zeta)}
\\
[-0.5em] l=9
\\[-0.5em] ^{(\zeta'+\zeta)}
\\[-0.3em]
\end{array} \right] \rightarrow \left[
\begin{array}{c}
k=9
\\[-0.5em] ^{(\zeta)}
\\
[-0.5em] l=6
\\[-0.5em] ^{(\zeta'+2\zeta)}
\\[-0.3em]
\end{array} \right] \rightarrow \left[
\begin{array}{c}
k=9
\\[-0.5em] ^{(\zeta)}
\\
[-0.5em] l=3
\\[-0.5em] ^{(\zeta'+3\zeta)}
\\[-0.3em]
\end{array} \right] \rightarrow \left[
\begin{array}{c}
k=9
\\[-0.5em] ^{(\zeta)}
\\
[-0.5em] l=0
\\[-0.5em] ^{(\zeta'+4\zeta)}
\\[-0.3em]
\end{array} \right] ~,
\label{T-T9}
\end{align}
where we wrote the  transformation behavior of $\zeta$ and $\zeta'$ explicitly in Eqs.~(\ref{S-untwist}), (\ref{S-twist1}), (\ref{T-T3}), (\ref{T-T6}), and (\ref{T-T9}) for future discussions. For $\tau\rightarrow -1/\tau$ and $\tau\rightarrow\tau+1$, therefore, the $[k,l]$ with sectors $k,l=0,3,6,9$  are interchanged with each other only inside $\{[k,l]$ sectors ; $k,~ {\rm and~}l=0,3,6,9\}$, decoupled from the other sectors with $k,~ {\rm or~}l\neq 0,3,6,9$.

Let us proceed to discuss some modular invariance conditions on $\zeta$ and $\zeta'$, which are integers. As seen from
Eqs.~(\ref{T-T3}) and (\ref{T-T9}), the $[k=3,l=0]$ and
$[k=9,l=0]$ sectors return to themselves after 4 times
T-transformations. Hence, the $\zeta^\prime$ in the $T_3$ and $T_9$ sectors ($k=3,9$) should be of the form
\begin{eqnarray}
\zeta^{\prime a}_4={\rm 4\times integer}+\delta^a_4 ,
\end{eqnarray}
where $\delta^a_4=0,1,2,3$ ($a=3,4$).\footnote{$T_3$ and $T_9$
sectors are related through the CTP conjugation.} Similarly, from
Eq.~(\ref{T-T6}),
\begin{eqnarray}
\zeta^{\prime a}_{2}={\rm 2\times integer}+\delta^a_2 ,
\end{eqnarray}
where $\delta^a_2=0,1$ ($a=3,4$) in the $T_6$ sector. As seen in Eqs.~(\ref{S-twist1}) and (\ref{S-untwist}), the $[k=3,l=0]$ and $[k=6,l=0]$ sectors are related with $[k=0,l=3]$ and $[k=0,l=6]$ sectors, respectively, via S-transformations. Therefore, considering both S and T transformations of the
partition function in Eqs.~(\ref{S}) and (\ref{T}), one can find proper $\zeta^a$ and $\zeta^{\prime a}$ for the other $[kV+\zeta W, lV+\zeta^\prime W]$ sectors. For the case of
$(\vec{\delta}_4,\vec{\delta}_2)=(0,0)$, they are displayed in Fig. 3.
\begin{figure}[t]
\resizebox{1\columnwidth}{!}
{\includegraphics{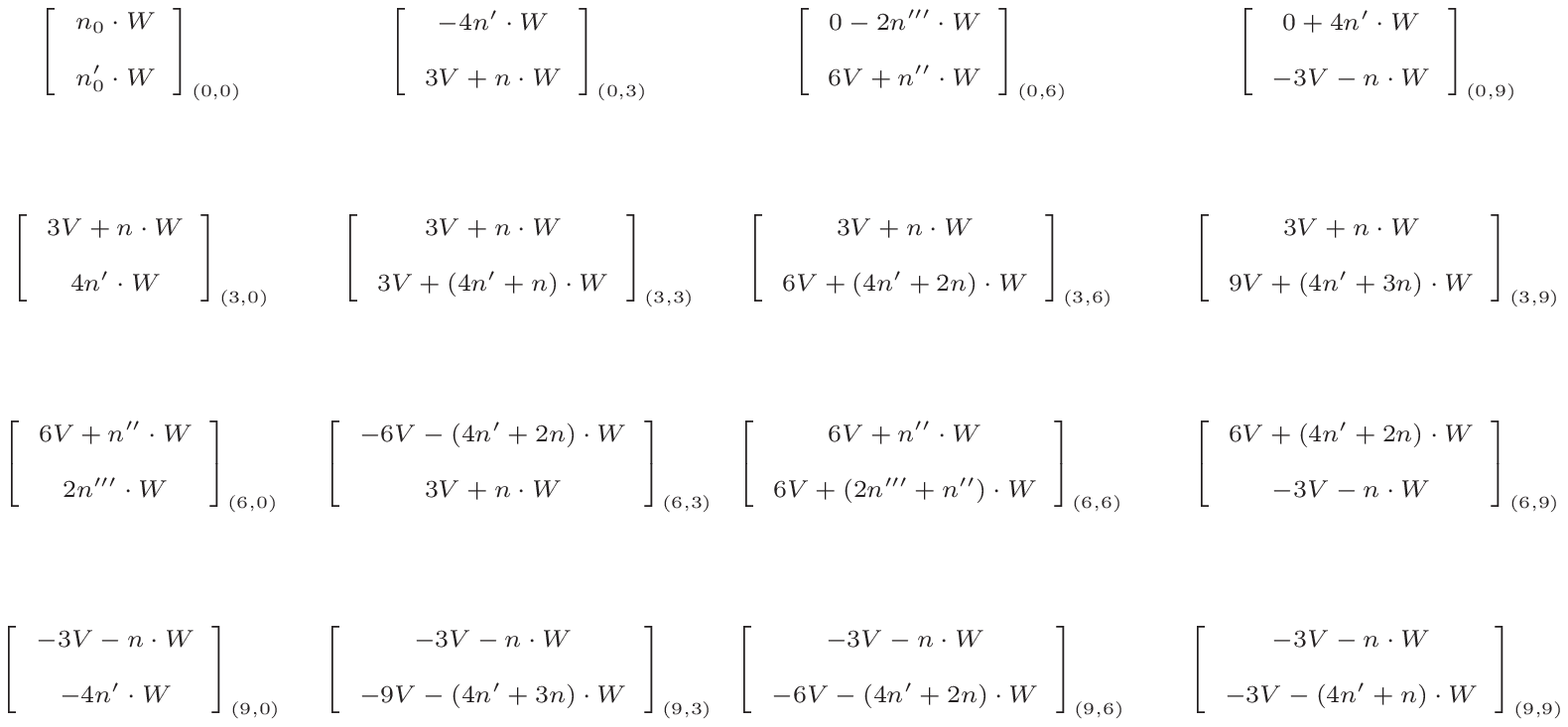}}
\caption{Boundary conditions for $X_L^I$
in the $[k,l]$ sectors ($k,l=0,3,6,9$) for
$(\delta_4,\delta_2)=(0,0)$.}
\label{Boundary}
\end{figure}
%%%%%%%%%%%%%%%%%%%%%%%%%%%%%%%%%%%%%%%%%%%%%%%%%%%%%%%%%%%%%%%%%%%%%%%%%%%%%%
%%%%%%%%%%%%%%%%%%%%%%%%%%%%%%%%%%%%%%%%%%%%%%%%%%%%%%%%%%%%%%%%%%
\vspace{0.4cm}
It is obvious that the set in Fig. 3 is consistent with Eqs.~(\ref{S-twist1}), (\ref{S-twist3}) and (\ref{T-U}), (\ref{T-T3}) (\ref{T-T9}), and also $[k=6,l=0]\leftrightarrow [k=6,l=6]$ in Eq.~(\ref{T-T6}).  Note that $12V$ and $12\phi$ are ${\rm E_8\times E_8'}$ and SO(8) weight vectors and so $12P\cdot V$ and $12s\cdot \phi$ are in general integers. With Eqs.~(\ref{v^2})--(\ref{w^2}) we obtain
\begin{eqnarray}
&&\left(P+\frac{3}{2}V+\frac{n}{2}W\right)12V
-\left(s+\frac{3}{2}\phi\right)12\phi ={\rm integer,}\quad {\rm and}\label{Mshell1}
\\ [1.0em]
&&\big[P+3V+(n+2n')W\big]12V-\big(s+3\phi\big)12\phi={\rm
integer} .\label{Mshell2}
\end{eqnarray}
Using these and Eq.~(\ref{orbifoldpartition}), one can show
$[k=3,l=9]\rightarrow [k=3,l=0]$ and $[k=6,l=6]\rightarrow
[k=6,l=0]$ under the T-transformation.

The S-transformation of the $[k=6,l=0]$ sector gives $[2n''' W, -6V-n'' W]$, which can be shown to be identified with $[-2n''' W, 6V+n'' W]$ by redefining $P\rightarrow -P$ in
Eq.~(\ref{orbifoldpartition}). So Eq.~(\ref{S-untwist}) is easily shown.

One can see also that in the above the $[k=3,l=3]$ sector is
transformed into the $[k=3,l=9]$: The S-transformation of the $[k=3,l=3]$ gives
\begin{eqnarray}
\sum_{n,n',P}\left[
\begin{array}{c}
3V+(4n'+n)W
\\
-3V-nW
\end{array} \right]  =
\sum_{x,y,P}\left[
\begin{array}{c}
3V+xW
\\
-3V+(4y+3x)W
\end{array} \right] , ~~
\label{S-33}
\end{eqnarray}
where we redefine $x\equiv 4n'+n$ and $y\equiv -3n'-n$, which preserves $|{\rm Det.}A|$ where $A=\left(\begin{array}{cc} 4&1\\ -3&-1 \end{array}\right)$. It should be identified with the $[k=3,l=9]$ sector.
The definition of Eq.~(\ref{S-33}), which can be obtained from Eq.~(\ref{orbifoldpartition}), derives a required condition in the exponent for identification of Eq.~(\ref{S-33}) and the $[k=3,l=9]$, $\textstyle(P+\frac{3}{2}V+\frac{x}{2}W)\cdot(12V)
-(s+\frac{3}{2}\phi)\cdot(12\phi) ={\rm integer}$. It is satisfied as shown above.  Through the similar procedure, one can show that $[k=9,l=9]\rightarrow [k=9,l=3]$ under S.

The S-transformation of the $[k=3,l=9]$ gives
\begin{eqnarray}
\sum_{n,n',P}\left[
\begin{array}{c}
9V+(4n'+3n)W
\\
-3V-nW
\end{array} \right]=
\sum_{x,y,P}\left[
\begin{array}{c}
9V-xW
\\
-3V-(4y+x)W
\end{array} \right] , ~
\label{S-39}
\end{eqnarray}
where we redefine $x\equiv -4n'-3n$ and $y\equiv n'+n$.  It can be identified with the $[k=9,l=9]$ sector, since
\begin{eqnarray}
\textstyle 6V[-3V-(x+4y)W]-6\phi(-3\phi)={\rm integer}
\end{eqnarray}
after redefining $P$ and $s$ as $P\rightarrow P-12V$ and
$s\rightarrow s-12\phi$ in Eq.~(\ref{orbifoldpartition}).
Similarly, the $[k=9,l=3]$ is transformed to the $[k=3,l=3]$. Together with Eq.~(\ref{S-33}), therefore, we have proven
Eq.~(\ref{S-twist2}).

The $[k=6,l=6]$ sector should be invariant under S. The S
transformation of the $[k=6,l=6]$ yields
\begin{eqnarray}
\sum_{n'',n''',P}\left[
\begin{array}{c}
6V+(2n'''+n'')W
\\
-6V-n''W
\end{array} \right]=
\sum_{x,y,P}\left[
\begin{array}{c}
6V+xW
\\
-6V+(2y+x)W
\end{array} \right] ,
\end{eqnarray}
where $x\equiv 2n'''+n''$ and $y\equiv -n'''-n''$. Since
$\textstyle(P+3V+\frac{x}{2}W)\cdot(12V) -(s+3\phi)\cdot(12\phi) ={\rm integer}$, the $[k=6,l=6]$ sector is S-invariant.

By redefinition of $P\rightarrow -P$, one can show the $[k=6,l=3]$ is identical with the $[k=6,l=9]$. Hence, the $[k=6,l=3]$ and $[k=6,l=9]$ should be T-invariant. The $[k=6,l=3]$ can be given in another form:
\begin{eqnarray} \label{63'}
\sum_{x,n'}\left[
\begin{array}{c}
-6V-2xW
\\
3V-(2n'-x)W
\end{array} \right]  ,
\end{eqnarray}
where $x\equiv 2n'+n$.  The T-transformation of it yields
\begin{eqnarray} \label{T-63'}
\sum_{x,n'}\left[
\begin{array}{c}
-6V-2xW
\\
-3V-(2n'+x)W
\end{array} \right]=
\sum_{x,y}\left[
\begin{array}{c}
-6V-2xW
\\
-3V-(2y-x)W
\end{array} \right]  ,
\end{eqnarray}
where $y\equiv n'+x$. We will show later that it is really
coincident with Eq.~(\ref{63'}).

Similar to the procedure for Fig. 3, one can get the sets of the boundary conditions for $\vec{\delta}_4\neq 0$ and $\vec{\delta}_2\neq 0$, namely, $4\vec{n}'\rightarrow
4\vec{n}'+\vec{\delta}_4$ or $2\vec{n}'''\rightarrow
2\vec{n}'''+\vec{\delta}_2$, where $\delta^a_4=1,2,3$ and
$\delta^a_2=1$ for $a=3,4$. Therefore, we have in total $4^2\times 2^2=64$ sets.

%%%%%%%%%%%%%%%%%%%%%%%%%%%%%%%%%%%%%%%%%%%%%%%%%%%%%%%%%%%%%%%
%%%%%%%%%%%%%%%%%%%%%%%%%%%%%%%%%%%%%%%%%%%%%%%%%%%%%%%%%%%%%%%
\section{Kaluza-Klein Tower}\label{sec:KKTower}

As mentioned before, the homomorphism $\vec{e}_a\rightarrow W^I_a$ between Eqs.~(\ref{winding}) and (\ref{wilson}) leads to the identification of $z^{(\prime)a}$ in Eq.~(\ref{6dpart}) and $\zeta^{(\prime)a}$ in Eq.~(\ref{orbifoldpartition}). By including the $\zeta^{\prime a}$ dependent piece appearing in Eq.~(\ref{orbifoldpartition}),  Eq.~(\ref{6dpart}) becomes
\begin{eqnarray} \label{2dpart}
\frac{4\pi^2}{\tau_2[\eta(\tau)\bar{\eta}(\bar{\tau})]^{2}}
\sum_{\zeta,\zeta^\prime}{\rm
exp}\left[-2\pi\tau_2g_{ab}\zeta^a\zeta^b
-\frac{2\pi}{\tau_2}g_{ab}(\zeta^{\prime a}+\tau_1
\zeta^a)(\zeta^{\prime b}+\tau_1\zeta^{b}) +2\pi i\zeta^{\prime
b}\nu_b\right] ,
\end{eqnarray}
where $\nu_b$ is given by
\begin{eqnarray} \label{nu}
\nu_b=2B_{ba}\zeta^a+\left[P^I+\frac{k}{2}V^I+
\frac{\zeta^a}{2}W^I_a\right]W^I_b ~.
\end{eqnarray}
As seen in Fig. 3 and Eq.~(\ref{63'}), $\vec{\zeta}^\prime$
($=\zeta^{\prime a}\vec{e}_a$) is given, in our case by
$\vec{\zeta}^\prime=\lambda\vec{n}^\prime+\sigma\vec{\zeta}$,
where $\lambda$s and $\sigma$s in the various sectors are listed
in Table~\ref{tb:zetaprime}.
\begin{table}
\begin{center}
\begin{tabular}{c||c|c|c|c|c}
\hline & $~~~[k=0,l~]~~~$ & $~~~[k=3,l~]~~~$ & $~[k=6,l=0,6]~$ &
$~[k=6,l=\pm 3]~$ & $~~~[k=9,l~]~~$
\\ \hline
~$\lambda$~~ & $1$ & $4$ & $2$ & $-2$ & $-4$
\\
~$\sigma$~~ & $0$ & $l/3$ & $l/6$ & $-1/2$ & $4-l/3$
\\
\hline
\end{tabular}
\end{center}
\caption{ ~$\vec{\zeta}^\prime$ ($
=\lambda\vec{n}^\prime+\sigma\vec{\zeta}$) in the various
sectors.} \label{tb:zetaprime}
\end{table}

Let us replace $\lambda\vec{n}^\prime$ (winding independent of $\vec{\zeta}$) by its dual vector (corresponding to the Kaluza-Klein momentum), $\vec{\mu}$ ($\equiv m_a\vec{e}^{*a}$), using the Poisson resummation formula~\cite{GSWbk}. For $\zeta^{\prime
a}=\lambda n^{\prime a}+\sigma \zeta^a$, the Poisson resummation formula states
\begin{align}
& F(\vec{x})=\sum_{n'}{\rm exp}\left[-\pi(\lambda
\vec{n}'+\vec{x})\cdot A\cdot (\lambda \vec{n}'+\vec{x})+2\pi i\lambda\vec{n}'\cdot\vec{\nu} \right]
\\
&\ \ =\left({\rm det}[\lambda^2A]\right)^{-1/2}\sum_{m}{\rm
exp}\left[-\pi\left(
\frac{1}{\lambda}\vec{\mu}-\vec{\nu}\right)\cdot
A^{-1}\cdot\left(\frac{1}{\lambda}\vec{\mu}-\vec{\nu}\right) +2\pi i\left(\frac{1}{\lambda}\vec{\mu}-\vec{\nu}\right)
\cdot\vec{x}\right] , \nonumber
\end{align}
where $\vec{x}=(\tau_1+\sigma)\vec{\zeta}$ and
$A_{ab}=(2/\tau_2)g_{ab}$. The full partition function $\lvert Z^X\rvert^2Z^G\overline{Z}^\psi$ in the
$(\vec{\delta}_4,\vec{\delta}_2)=(0,0)$ and $l=0,3,6,9$ parts of the $U$, $T_3$, $T_6$, and $T_9$ sectors is given by
\begin{eqnarray}
\frac{1}{\lambda^2}\Bigg[\sum_{\vec{\mu}\in
\Lambda^*_2,\vec{\zeta}\in\Lambda_2}
\frac{q^{(\vec{L}+\vec{\zeta})^2/2}}{\left[\eta(\tau) \right]^2}
~\frac{\bar{q}^{(\vec{L}-\vec{\zeta})^2/2}}
{\left[\bar{\eta}(\bar{\tau})\right]^2}\Bigg]
\Bigg[\sum_{P\in\Lambda_{16},s\in\Lambda_8}
\frac{q^{(L^I)^2/2}}{\left[\eta(\tau)\right]^{16}}
~\frac{\bar{q}^{(\tilde{L})^2/2}}{\left[\bar{\eta}
(\bar{\tau})\right]^4} \times e^{2\pi il\Theta_k}\Bigg]
\big|\widehat{Z}^X_{[k,l]}\big|^2 . \label{full}
\end{eqnarray}
Here we neglect again the spin structure for simplicity.
$\vec{L}$, $L^I$, and $\Theta_k$ are given by
\begin{eqnarray}
&&\vec{L}=\sum_{a,b=6,7}\left[
\frac{m_a\vec{e}^{*a}}{2\lambda}-B_{ab}\zeta^{[b}\vec{e}^{*a]}
-\left(P^I+\frac{kV^I}{2}+\frac{\zeta^bW^I_b}{2}\right)
\frac{W_a^I}{2}~\vec{e}^{*a}
\right] ~, \label{L^i}
\\
&&L^I=P^I+kV^I+\zeta^aW^I_a ~,\quad\quad \tilde{L}=s+k\phi ~,
\label{Rmom}
\\ \label{theta}
&&l\times\Theta_k=l\times\left[\bigg(P^I+\frac{k}{2}V^I
+\frac{\zeta^a}{2}W^I_a\bigg)V^I-\bigg(s+\frac{k}{2}\phi\bigg)
\phi\right]+\frac{\sigma}{\lambda}m_a\zeta^a ~.
\end{eqnarray}
The large radius ($R$) dependence of the spectrum can be read from Eqs. (\ref{full}) and (\ref{L^i}), through the definition of the unit length $\vec{e}_{a}\to (R/\sqrt{\alpha'})\vec{e}_{a}$ and $\vec{e}^{*a} \to (R^{-1}\sqrt{\alpha'})\vec{e}^{*a}$. In Eq. (\ref{full}), $\widehat{Z}^X_{[k,l]}$ is defined as
\begin{eqnarray}
\widehat{Z}^X_{[k,l]}(\tau)=
Z^X_{[k,l]}(\tau)\times\frac{\vartheta\left[
\begin{array}{c}
1/2+k\phi_2 \\
1/2+l\phi_2 \end{array} \right]}{\eta(\tau)}
\end{eqnarray}
where $Z^X_{[k,l]}(\tau)$ was given in Eqs.~(\ref{Z^X}) and
(\ref{untwistZ^X}). It does not contain KK states associated with internal four dimensional radii of $(x^1,x^2;x^5,x^6)$.

Eq.~(\ref{full}) is just the result by
$(\vec{\delta}_4;\vec{\delta}_2)
=(\delta_4^3,\delta_4^4;\delta_2^3,\delta_2^4)=(0,0;0,0)$. We should also take into account the parts by $\delta^a_4=1,2,3$ and $\delta^a_2=1$ ($a=3,4$), which redefine $4\vec{n}'\rightarrow 4\vec{n}'+\vec{\delta}_4$ and
$2\vec{n}'''\rightarrow2\vec{n}'''+\vec{\delta}_2$ in Fig.
3.

Indeed, the presence of background gauge fields $B_{ab}$ and $W^I_a$ affects only the momenta of the zero modes of {\it untwisted} bosonic strings~\cite{Narain:1986am}.
So modifications are possible only for the $U$, $T_3$, $T_6$, and $T_9$ sectors. The factors including $L^I$ and
$\vec{L}\pm\vec{\zeta}$ in Eq. (\ref{full}) modify such zero modes' momenta with the compactification radius ($R$) dependence.

%%%%%%%%%%%%%%%%%%%%%%%%%%%%%%%%%%%%%%%%%%%%%%%%%%%%%%%%%%%%%%
\subsection{Mass-shell Condition}

The powers of $q, \eta$ and $\bar{q}, \bar\eta$ in Eq.~(\ref{full}) read the
mass-shell formulae,
\begin{eqnarray} \label{Lmass}
\frac{\alpha'M_L^2}{4}&=&\frac{(\vec{L}+\vec{\zeta}~)^2}{2}
+\frac{(L^I)^2}{2}+\sum_{i=j,\bar{j}}N^L_{i}\tilde{\phi}_i
-1+\frac{c_k}{2}~,
\\ \label{Rmass}
\frac{\alpha'M_R^2}{4}&=&\frac{(\vec{L}-\vec{\zeta}~)^2}{2}
+\frac{(\tilde{L})^2}{2}+\sum_{i=j,\bar{j}}N^R_i\tilde{\phi}_i
-\frac{1}{2}+\frac{c_k}{2}~,
\end{eqnarray}
where $i$ runs over $\{1,2,3,\bar{1},\bar{2},\bar{3}\}$, and
$\tilde{\phi}_j\equiv k\phi_j$ mod Z such that
$0<\tilde{\phi}_j\leq 1$ and $\tilde{\phi}_{\bar{j}}\equiv
-k\phi_j$ mod Z such that $0<\tilde{\phi}_{\bar{j}}\leq 1$. If $k\phi_j$ is an integer, $\tilde{\phi}_j=1$.  $c_k$ in
Eqs.~(\ref{Lmass}) and (\ref{Rmass}) denotes the world sheet
vacuum energy $\sum_{i=1}^{3}|\widetilde{k\phi_i}|(1-|\widetilde{k\phi_i}|)$,
where $\widetilde{k\phi_i}=k\phi$ mod Z such that
$0<|\widetilde{k\phi_i}|\leq 1$. Since
$\phi=(\frac{5}{12},\frac{4}{12},\frac{1}{12})$ in ${\bf
Z}_{12-I}$, and KK states appear only in the $U$, $T_3$, $T_6$, and $T_9$ sectors, the relevant $c_k$ are listed as follows:
\begin{align}
2-c_k=\left\{\begin{array}{l} ~2\quad k=0\\
 \frac{13}{8}\quad k=3,9
 \\
 ~\frac{3}{2}\quad k=6 ~.
 \end{array}
 \right.\label{}
\end{align}

The overall size of the plane $(x^3,x^4)$ is a modulus. One can define the size $R$ in the unit of $\sqrt{\alpha'}$ e.g. such that ${\rm det}g_{ab}=1$ by scaling $\vec{e}_a\rightarrow (R/\sqrt{\alpha'})\times\vec{e}_a$ and $\vec{e}^{*a}\rightarrow (\sqrt{\alpha'}/R)\times \vec{e}^{*a}$, under which $g_{ab}$ and
$g^{ab}$ ($=g_{ab}^{-1}$) are rescaled as $g_{ab}\rightarrow
(R^2/\alpha')g_{ab}$ and $g^{ab}\rightarrow (\alpha'/R^2)g^{ab}$.

For non-critical radius $R$, the level matching condition
$M_L^2=M_R^2$ ($\equiv M_l^2+M_h^2$), composed of $R$-dependent
and $R$-independent pieces, reduces to
\begin{eqnarray} \label{M1}
\frac{\alpha'M_l^2}{4}\equiv\frac{(\vec{L}+\vec{\zeta}~)^2}{2}
&=&\frac{(\vec{L}-\vec{\zeta}~)^2}{2} ~,
\\ \label{M2}
\frac{\alpha'M_h^2}{4}\equiv\frac{(L^I)^2}{2}
+\sum_{i=j,\bar{j}}N^L_{i}\tilde{\phi}_i -1+\frac{c_k}{2}
&=&\frac{(\tilde{L})^2}{2} +\sum_{i=j,\bar{j}}N^R_i\tilde{\phi}_i
-\frac{1}{2}+\frac{c_k}{2}~.
\end{eqnarray}
Eq.~(\ref{M1}) implies $2\vec{L}\cdot\vec{\zeta}=0$, or
\begin{eqnarray} \label{level-matching}
\lambda\left(P^I+\frac{kV^I}{2}+\frac{\zeta^bW^I_b}{2}
\right)W_a^I\zeta^a=m_a\zeta^a={\rm integer} \quad\quad {\rm
for}~~\vec{\zeta}\neq 0 ~,
\end{eqnarray}
since $B_{ab}\zeta^a\zeta^b=0$. We should note that the level matching condition Eq.~(\ref{level-matching}) must be satisfied only for the states with $\vec{\zeta}\neq 0$. For a nonzero $\zeta$, the mass becomes large, $R/\sqrt{\alpha'}$ which is above the string scale. So, we will set $\vec{\zeta}=0$ below. Eq.~(\ref{M1}) is trivially satisfied if $\vec{\zeta}=0$.
After some algebra, one can see that Eq.~(\ref{M2}) becomes
\begin{align} \label{level-matching2}
\frac12(P^2+s^2-1)&+k\bigg[\Theta_k \pm
\sum_{i=j,\bar{j}}\left(N_i^L-N_i^R\right)\phi_i\bigg]
+\left(P+\frac{k}{2}V+\frac{\zeta^a}{2}W_a\right)\zeta^bW_b
 \nonumber\\
 &-\frac{k\sigma}{l\lambda}m_a\zeta^a= {\rm integer} ,
\end{align}
where $+$ ($-$) in $\pm$ corresponds to $\tilde{\phi}_j$
($\tilde{\phi}_{\bar{j}}$), and $\Theta_k$ is given in
Eq.~(\ref{theta}). Note that  $\frac{1}{2}\left(P^2-s^2-1\right)$
is generically an integer. If $\vec{\zeta}=0$, therefore,
Eq.~(\ref{level-matching2}) becomes
\begin{eqnarray} \label{tildetheta}
k\bigg[\Theta_k \pm
\sum_{i=j,\bar{j}}\left(N_i^L-N_i^R\right)
\phi_i\bigg]_{\vec{\zeta}=0} \equiv k\widetilde{\Theta}^\pm_k\big|_{\vec{\zeta}=0}={\rm integer}
\end{eqnarray}
where
\begin{eqnarray} \label{FullTheta}
\widetilde{\Theta}^\pm_k=\Theta_k +\sum_i(N^L_i-N_i^R)\hat{\phi}_{i}
~,
\end{eqnarray}
with $\hat{\phi}_j=\phi_j$ for $\Theta^+$ and $\hat{\phi}_{\bar{j}}=-\phi_j$ for $\Theta^-$,
and $\Theta_k$ was given in Eq.~(\ref{theta}).

For $(\vec{\delta}_4;\vec{\delta}_2)\neq 0$, $\vec{\zeta}$ cannot be zero in the untwisted sectors. Thus, the states in the untwisted sector with $(\vec{\delta}_4,\vec{\delta}_2)\neq (0;0)$
should be super-massive. On the other hand, the masses of the states in the $T_3$ and $T_9$ sectors are not affected by
$(\vec{\delta}_4;\vec{\delta}_2)\neq (0;0)$, but $\Theta_k$ in Eq.~(\ref{theta}) is modified to
\begin{eqnarray}
l\cdot\Theta_k\longrightarrow l \cdot\Theta_k
+\frac{m_a\delta_4^a}{\lambda} .
\end{eqnarray}
Therefore, $m^a$ should be $\lambda\times {\rm integer}$ for
physical states.

$m_a$ turns out to be $\lambda\times P^IW^I=\lambda\times {\rm integer}$ for massless states. We have, therefore, 15 more massless states in $T_3$ and $T_9$, and 3 more states (in total 4 states with $\delta_4^a=0,2$ for $a=3,4$) in the $[k=6,l=3]$ and $[k=6,l=9]$ sectors. The states from $\delta_4^a=1,3$ in the $[k=6,l=3]$ and $[k=6,l=9]$ sectors are super-massive. The presence of such massless states exactly cancel the overall denominator $\lambda^2$ ($16$ for $T_{3,6}$ sectors, and $4$ for $[k=6,l=3]$ and $[k=6,l=9]$) in Eq.~(\ref{full}). Similarly, we have 3 more states in the $[k=6,l=0]$ with $\vec{\delta}_2\neq 0$ and $[k=6,l=6]$ sectors double, and so their presence also cancels
the denominator $\lambda^2$ ($=4$).

%%%%%%%%%%%%%%%%%%%%%%%%%%%%%%%%%%%%%%%%%%%%%%%%%%%%%%%%%%%%%
\subsection{GSO Projection}

The generalized GSO projector \cite{GSO} is read from the coefficient of $q^{\alpha'M_L^2/4}\bar{q}^{\alpha'M_R^2/4}$ in the partition function. Including the contributions coming from the eta and theta functions in Eq.~(\ref{full}), the GSO projector in the $k=0,3,6,9$ sectors is given by
\begin{eqnarray} \label{GSO}
{\cal
P}_k=\frac{1}{N}\sum_{l}\tilde{\chi}(\theta^k,\theta^l)e^{2\pi i l\widetilde{\Theta}_k} ~,
\end{eqnarray}
where $N=12$ for the massless states. However, $N=4$ for the KK massive states, because ${\cal N}=2$ KK massive states appear only in the $U$, $T_3$, $T_6$, and $T_9$ sectors. Hence, $l$ in Eq.~(\ref{GSO}) runs $0,3,6,9$ for KK massive states, whereas $l=0,1,2,\cdots, 11$ for massless states. Note that $\tilde{\phi}_i$ in Eqs.~(\ref{Lmass}) and (\ref{Rmass}) is given by $\tilde{\phi}_i=k\hat{\phi}_i$ mod $Z$. The degeneracy factor $\tilde{\chi}(\theta^k,\theta^l)$ is determined only from the $\widehat{Z}^X_{[k,l]}$ part in the $U$, $T_3$, $T_6$, and $T_9$ sectors. It is summarized in Table~\ref{tb:degnKK}.
%
%
%%TTTTTTTTTTTTTTTTTTTTTTTTTTTTTTTTTTTTTTTTTTTTTTTTTTTTTTTTTTTTTTTTTTTTTTTT
\begin{table}
\begin{center}
\begin{tabular}{c|cccc}
\hline %& & & & & & $k$ & $=$ & & & & & \\
 $k\diagdown l$~ & ~~${0}$~ & ~${3}$~ &
~${6}$~ & ~${9}$~
\\
\hline ${0}$ & ~${1}$ & ${1}$ & ${1}$ & ${1}$
\\
${3}$ & ~${4}$ & ${4}$ & ${4}$ & ${4}$
\\
${6}$ & ~${16}$ & ${4}$ & ${16}$ & ${4}$
\\
${9}$ & ~${4}$ & ${4}$ & ${4}$ & ${4}$
\\
\hline
\end{tabular}
\end{center}
\caption{Degeneracy factor $\tilde\chi(\theta^k,\theta^l)$ for the KK massive states in the ${\bf Z}_{12-I}$
orbifold.}\label{tb:degnKK}
\end{table}
%%TTTTTTTTTTTTTTTTTTTTTTTTTTTTTTTTTTTTTTTTTTTTTTTTTTTTTTTTTTTTTTTTTTTTTTTTTTTT
%
%
The $T_9$ sector takes the same values of
$\tilde{\chi}(\theta^k,\theta^l)$ as those in $T_3$.
Since the same numbers are horizontally aligned at $l=0,3,6,9$ of the $U$ and $T_3$ sectors, the physical KK massive states surviving the projection should take
\begin{eqnarray} \label{physicalKK}
3\times\widetilde{\Theta}_{k=0,3,9}={\rm integer} ,
\end{eqnarray}
which from Eq.~(\ref{level-matching2}) implies
\begin{eqnarray}
\left(P+\frac{k}{2}V+\frac{\zeta^a}{2}W_a\right)
\zeta^bW_b={\rm integer} \quad {\rm for~physical~KK~states~in~}U,~T_3,~T_9 .
\end{eqnarray}
From Eqs.~(\ref{tildetheta}) and (\ref{physicalKK}) we note that if $\vec{\zeta}=0$, all KK {\it massive} states in $T_3$ are physical states.

Now we are ready to prove the T-invariance of Eq.~(\ref{63'}). In Eqs.~(\ref{63'}) and (\ref{T-63'}), $\lambda=-2$, $\sigma=-\frac{1}{2}$, and $\vec{\zeta}=2x^a\vec{e}_a$. They have the same masses. From Eqs.~(\ref{level-matching}) and (\ref{level-matching2}), thus, we get in the $[k=6,l=3]$ sector
\begin{eqnarray} \label{mass-shell63'}
6\times\bigg[\Theta_{k=6} \pm
\sum_{i=j,\bar{j}}\left(N_i^L-N_i^R\right)\phi_i\bigg]
=6\times\widetilde{\Theta}_{k=6}={\rm integer} .
\end{eqnarray}
Eq.~(\ref{63'}) and Eq.~(\ref{T-63'}) are different only in $l$: $l$ in Eq.~(\ref{63'}) is 3, whereas $-3$ in Eq.~(\ref{T-63'}). The difference appears in $l\times\Theta_k$ in Eq.~(\ref{full}) or more generally $l\times\widetilde{\Theta}_k$ from Eq.~(\ref{FullTheta}). For Eq.~(\ref{63'}) to be coincident with Eq.~(\ref{T-63'}), $3\widetilde{\Theta}_{k=6}$ should be the same as  $-3\widetilde{\Theta}_{k=6}$ upto an integer. They are indeed identical due to the level matching condition of
Eq.~(\ref{mass-shell63'}).  Thus, we have completed the proving of modular invariance for the sets in Fig. 3.

%%%%%%%%%%%%%%%%%%%%%%%%%%%%%%%%%%%%%%%%%%%%%%%%%%%%%%%%%%%%%%%
\subsection{Massless modes}

In fact, the mass-squared of the right mover, which is the
supersymmetrized string, is non-negative. Accordingly, for the massless right mover, $\vec{L}=\vec{\zeta}$ and
\begin{eqnarray} \label{mR}
\frac{(\tilde{L})^2}{2} +\sum_{i=j,\bar{j}}N^R_i\tilde{\phi}_i
-\frac{1}{2}+\frac{c_k}{2}=0
\end{eqnarray}
should be required in Eq.~(\ref{Rmass}). For non-critical radius $R$, however, Eqs.~(\ref{M1}) and (\ref{Lmass}) state that only the left mover with $\vec{L}=\vec{\zeta}=0$ and
\begin{eqnarray} \label{mL}
\frac{(L^I)^2}{2}+\sum_{i=j,\bar{j}}N^L_{i}\tilde{\phi}_i
-1+\frac{c_k}{2}=0
\end{eqnarray}
can be massless. Moreover, $\vec{L}=\vec{\zeta}=0$ gives
\begin{eqnarray} \label{Lzero}
\left[P^I+\frac{k}{2}V^I\right]W^I_b \times\lambda= m_b ={\rm integer}~.
\end{eqnarray}
Note that from Table~\ref{tb:degn}, $(k,\lambda)=(0,1)$, $(3,4)$, $(6,\pm 2)$, and $(9,4)$ for the $U$, $T_3$, $T_6$, and $T_9$ sectors, respectively, in the ${\bf Z}_{12-I}$ case. Since $3W_a^I$ is an ${\rm E_8\times E_8}'$ weight vector, $P\cdot W_a=\frac{1}{3}\times {\rm integer}$ in general. With Eq.~(\ref{vw}), therefore, Eq.~(\ref{Lzero}) can be simplified into\footnote{ In Refs.~\cite{pw,SMZ12I,Z12I}, ``$(P+kV)\cdot
W={\rm integer}$'' is suggested instead of Eq.~(\ref{massless2}). This statement has to be clraified as follows. The shifts $V\rightarrow V+v$ and $W\rightarrow W+w$, where $v$ and $w$ are \EE weight vectors, result in different spectra, which is inconsistent. On the other hand, the Eq.~(\ref{massless2}) as well as the other conditions
for massless states, i.e. Eqs. (\ref{mL}) and (\ref{masslessGSO})
are invariant under such shifts of $V$ and $W$.  One can prove those using the general fact that $\vec{r}\cdot \vec{r}'$ is an ${\rm even ~integer}$ if $\vec{r}=\vec{r}'$, and is just an ${\rm integer}$ if $\vec{r}\neq\vec{r}'$, where $\vec{r}$ and $\vec{r}'$ are \EE\ weight vectors. However, the results in Refs.~\cite{pw,SMZ12I,Z12I} are fortunately correct: In Ref.~\cite{pw} only the $V\cdot W=0$ cases are examined. Since $(P+kV)\cdot W$ becomes $P\cdot W$ in these cases, the results are correct. The Ref.~\cite{SMZ12I} is the case of $V\cdot W=1$. So
the spectra with $(P+kV)\cdot W={\rm integer}$ are also the same as the results by $P\cdot W={\rm integer}$. In Ref.~\cite{Z12I}, $V\cdot W=\frac{-1}{6}$ and it is argued that there is no massless state from $T_3$. Thus, it is naively expected that Eq.~(\ref{massless2}) change the massless spectrum in the $T_3$ sector while it does not affect the spectra of the $U$ and $T_6$ sectors. However, it turns out that even with $P\cdot W={\rm integer}$, still there is no massless matter in $T_3$.}
\begin{eqnarray} \label{massless2}
 P^IW_b^I={\rm integer}
\quad\quad\quad {\rm for~~the}~~U,~T_3,~T_6,~T_9~~{\rm sectors}.
\end{eqnarray}
Since $\vec{\zeta}=0$, the $\widetilde{\Theta}_k$ in the GSO
projector of Eq.~(\ref{GSO}) becomes just
\begin{eqnarray} \label{masslessGSO}
\widetilde{\Theta}_k=\bigg(P^I+\frac{k}{2}V^I\bigg)V^I
-\bigg(s+\frac{k}{2}\phi\bigg)\phi
+\sum_i(N^L_i-N_i^R)\hat{\phi}_{i} .
\end{eqnarray}

As seen in the degeneracy factors in Table~\ref{tb:degn}, the same numbers (1s) are horizontally aligned for the $U$ sector ($k=0$). Thus, the GSO projection for massless states ($P^2=2$) in the $U$ sector
\begin{eqnarray}
\widetilde{\Theta}_{k=0}=P^IV^I -s\cdot\phi ={\rm integer} .
\end{eqnarray}
In particular, since $s\cdot\phi=0$ for the gauge sectors, the GSO projection condition becomes
\begin{eqnarray}\label{masslessgauge}
P^IV^I = {\rm integer} \quad\quad {\rm for ~the ~(massless) ~gauge ~sector}.
\end{eqnarray}

%%%%%%%%%%%%%%%%%%%%%%%%%%%%%%%%%%%%%%%%%%%%%%%%%%%%%%%%%%%%%%
\subsection{KK massive modes}\label{subsec:massive}

The KK masses from $M_h^2$ in Eq.~(\ref{M2}) would be of order $1/\alpha'$. $\vec{\zeta}$ [$=\zeta^a(R/\sqrt{\alpha'})\vec{e}_a$]
makes contributions of order $(\alpha'/R)^2$ to $M_l^2$ in
Eq.~(\ref{M1}). For the case $R\gg\sqrt{\alpha'}$, thus,
relatively light KK states can be excited if $M_h^2=0$ and
$\vec{\zeta}=\zeta^a(R/\sqrt{\alpha'})\vec{e}_a=0$. As a result, Eqs.~(\ref{mR}) and (\ref{mL}) should be still satisfied as in the massless case. The relatively light KK mass-squareds are given by
\begin{eqnarray} \label{kkmass}
M_l^2 &=&\frac{2}{\alpha'}(\vec{L})^2\big|_{\vec{\zeta}=0}
=\sum_{m_a,m_b}\frac{g^{ab}}{2R^2\lambda^2} \left[m_a-\lambda
P^IW^I_a\right] \left[m_b-\lambda P^IW^I_b\right]
\nonumber \\
&=&\sum_{\overline{m}_a,\overline{m}_b}\frac{g^{ab}}{2R^2}
\left[\overline{m}_a- P^IW^I_a\right] \left[\overline{m}_b-
P^IW^I_b\right] ,
\end{eqnarray}
which is of order $1/R^2$ ($\ll 1/\alpha'\ll (R/\alpha')^2$)
because of the appearance of $g^{ab}$. Here we shifted $m_a$ (and also $m_b$) in Eq.~(\ref{L^i}) as $m_{a}\rightarrow
m_{a}-\frac{k\lambda}{2}V^IW^I_{a}={\rm integer}$ with
Eq.~(\ref{vw}). As mentioned before, $\lambda=1,4,2,4$ for the $U$, $T_3$, $T_6$, $T_9$ sectors, respectively, and for physical states $m_a$ should be $\lambda\times \overline{m}_a$, where $\overline{m}_a$s are integers. Since $\vec{\zeta}=0$, the $\widetilde{\Theta}_k$ in the GSO projector of Eq.~(\ref{GSO}) is given again by Eq.~(\ref{masslessGSO}).

From Eq.~(\ref{physicalKK}), therefore, physical KK massive states with $\vec{\zeta}=0$ satisfy
\begin{eqnarray}
\widetilde{\Theta}_{k}=\bigg(P^I+\frac{k}{2}V^I\bigg)V^I
-\bigg(s+\frac{k}{2}\phi\bigg)\phi
+\sum_i(N^L_i-N_i^R)\hat{\phi}_{i} =\frac{1}{3} \times {\rm
integer}~~ {\rm for}~k=0,3,9 . ~~
\end{eqnarray}
Particularly, the (KK massive) gauge sector with $s\cdot\phi=0$ of the $U$ sector ($k=0$) obeys
\begin{eqnarray} \label{massivegauge}
P^IV^I=\frac{1}{3} \times {\rm integer} .
\end{eqnarray}
Comparing Eq.~(\ref{massivegauge}) with Eq.~(\ref{masslessgauge}), one can see that the gauge symmetry can be enhanced by including KK states above the compactification scale $1/R$, because the condition Eq.~(\ref{massivegauge}) is less constrained than
$P^IV^I={\rm integer}$. The gauge group obtained in this way must coincide with the gauge group by orbifolding 4d internal space by $\Z_{12-I}$. It is reminiscent of the ${\bf Z}_{3}$ orbifold GUT in six dimensional effective {\it field} theory. In six dimensional orbifold field theory, all gauge fields satisfying $P\cdot V=\pm\frac13$ mod integer become massive by the boundary condition, while gauge fields with $P\cdot V={\rm integer}$ permit massless modes. In the ${\bf Z}_{12-I}$ orbifolded {\it string} theory, the second sub-lattice, from which KK massive states arise, is indeed the ${\bf Z}_3$ orbifold. However, the $V^I$ in
Eq.~(\ref{massivegauge}) is the shift vector of order 12 rather than 3 as in the ${\bf Z}_3$ orbifold. Hence, one could define a 6D effective shift vector as $V^I_{6d}\equiv 4\times V^I$ (and also a 6D effective twist vector as $\phi_{6d}\equiv 4\times \phi$), which is of order 3 and still consistent with Eqs.~(\ref{masslessgauge}) and (\ref{massivegauge}).

%%%%%%%%%%%%%%%%%%%%%%%%%%%%%%%%%%%%%%%%%%%%%%%%%%%%%%%%%%%%
\subsubsection{{\cal N}=2 gauge multiplet from the $U$ sector}

In 6D ${\cal N}=2$ (in terms of 4D SUSY) gauge theory, the gauge bosons $A_3$ and $A_4$ and their spin-1/2 superpartners (gauginos) are the ${\cal N}=2$ SUSY partners of a 4D ${\cal N}=1$ gauge multiplet. They compose an ${\cal N}=1$ chiral multiplet with the adjoint representation. If ${\cal N}=2$ SUSY is broken to ${\cal N }=1$ via a compactification mechanism, they all achieve masses of order $1/R$. Since their polarizations are in the $(x^3,x^4)$ directions, $s\cdot \phi$ is given by $\pm\frac13$ in the ${\bf
Z}_{12-I}$ orbifold compactification.

Suppose that a gauge group ${\bf G}$ and ${\cal N}=2$ SUSY in an effective 6D theory is broken to a smaller gauge group ${\bf H}$ and ${\cal N}=1$ SUSY below the compactification scale through the ${\bf Z }_{12-I}$ orbifold compactification. Then, the original 6D ${\cal N}=2$ gauge multiplet is split into
\begin{align}
&\left[\begin{array}{c} {\rm \tiny Gauge\ group\ {\bf H}}
\\ [-0.5em] {\tiny\rm {\cal N}=1 ~{\rm gauge\ multiplet} }
\\ \scriptsize
\begin{array}{c}  P\cdot V=0 ~{\rm mod~Z}
\\[-0.3em]
and
\\[-0.3em]
 P\cdot W=0 ~{\rm mod~Z} ;
\\[-0.3em]
 s\cdot\phi= 0
\end{array}
%\right)
\end{array} \right]_{\bf 0}+\left[\begin{array}{c}
{\rm \tiny Gauge\ group\ {\bf H}}
\\ [-0.5em] {\tiny\rm {\cal N}=1 ~{\rm chiral\ multiplet} }
\\ \scriptsize
%\left(
\begin{array}{c}  P\cdot V=0 ~{\rm mod~Z}
\\[-0.3em]
and
\\ [-0.3em]
P\cdot W=0 ~{\rm mod~Z} ;
\\ [-0.3em]
s\cdot\phi=\pm\frac13
\end{array}
%\right)
\end{array} \right]_{\bf KK}
\\[0.5em]
&+\left[\begin{array}{c} {\rm \tiny Coset\ {\bf G/H}}
\\ [-0.5em] {\tiny\rm {\cal N}=1 ~{\rm gauge\ multiplet} }
\\ \scriptsize
\begin{array}{c}  P\cdot V=\pm\frac13 ~{\rm mod~Z}
\\[-0.3em]
or
\\[-0.3em]
 P\cdot W=\pm\frac13 ~{\rm mod~Z} ;
\\[-0.3em]
 s\cdot\phi=0
\end{array}
\end{array} \right]_{\bf KK}+
\left[\begin{array}{c} {\rm \tiny Coset\ {\bf G/H}}
\\ [-0.5em] {\tiny\rm {\cal N}=1 ~{\rm chiral\ multiplet} }
\\ \scriptsize
\begin{array}{c}  P\cdot V=\pm\frac13 ~{\rm mod~Z}
\\[-0.3em]
or
\\ [-0.3em]
P\cdot W=\pm\frac13 ~{\rm mod~Z} ;
\\ [-0.3em]
s\cdot\phi=\pm\frac13
\end{array}
\end{array} \right]_{\bf 0}  ,
\end{align}
where the subscripts ``${\bf 0}$'' (``${\bf KK}$'') means
that the corresponding sectors can(not) contain massless states. Note that the massless chiral matter with ${\bf G/H}$ fulfilling $P\cdot V=\frac13$ mod Z and $s\cdot\phi=\frac13$ participates in a 6D ${\cal N}=2$ gauge sector above the compactification scale. In fact, the models suggested in Refs.~\cite{SMZ12I} and \cite{Z12I}, the electroweak Higgs doublets belong to this kind of matter. Thus, in such models, the gauge bosons and Higgs are unified, forming an ${\cal N}=2$ gauge multiplet together with other KK massive states above the compactification scale.

%%%%%%%%%%%%%%%%%%%%%%%%%%%%%%%%%%%%%%%%%%%%%%%%%%%%%%%%%%%%%%%%%
\subsubsection{{\cal N}=2 hypermultiplets in the $U$
sector}

From Eq.~(\ref{physicalKK}), the ${\cal N}=2$
hypermultiplet states in the untwisted sector satisfy
\begin{eqnarray} \label{hyperU}
3\times \left[P^IV^I-s\cdot\phi\right]={\rm integer} ,
\end{eqnarray}
or $[P^IV^I_{6d}-s\cdot \phi_{6d}]=\frac13\times {\rm integer}$. While matter states satisfying $s\cdot\phi =\pm\frac{4}{12}$ (and so $P\cdot V=\pm\frac{4}{12}$ mod $\frac{1}{3}$) have become a part of the ${\cal N}=2$ gauge sector, the other matter states with $s\cdot\phi=\pm\frac{1}{12},\pm\frac{5}{12}$ from the
untwisted sector compose ${\cal N}=2$ hypermultiplets. From
Eq.~(\ref{hyperU}), they obey
\begin{eqnarray} \label{KKUmatter}
P\cdot V= \bigg\{
\begin{array}{l}
\frac{-5}{12}~~~{\rm
mod}\quad\frac{1}{3}:\quad\frac{-5}{12},\quad\frac{-1}{12},
\quad\frac{3}{12},\cdots
~\\
\frac{1}{12}^{}\,_{}\quad {\rm mod}\quad \frac{1}{3} :\quad ~
\frac{5}{12},\quad\frac{1}{12},\quad\frac{-3}{12},\cdots ~,
\end{array}
\end{eqnarray}
or $3\times P\cdot V=\frac14$ mod Z.  They always form vector-like representations, because $P^I$ and $-P^I$ satisfy the same mass-shell condition, and if $P\cdot V$ is an allowed value, $-P\cdot V$ should be so from Eq.~(\ref{KKUmatter}). Thus, the KK matter from the untwisted sector always composes $N=2$ hypermultiplets above the compactification scale.

\subsubsection{$T_3$, $T_6$ sectors}

As seen in Table \ref{tb:degnKK}, the degeneracy factors of
$l=0,3,6,9$ in the $T_3$ sector are the same, and so the
projection condition should be Eq.~(\ref{physicalKK}). The SO(8) weight satisfying Eq.~(\ref{M2}) in $T_3$ gives $3\times (s_{\pm}+3\phi)\phi= \frac{1}{8}$ mod integer, where $\pm$ indicates the chirality. Note that $-\frac{3}{2}\phi^2$ in $\tilde{\Theta}_3$ is just a common term for the left and right chiral states. It means that $3\times \tilde{\Theta}_3$ in Eq.~(\ref{physicalKK}) gives the same value for the left and right handed chiral states. Hence, the physical KK states in $T_3$ are always vector-like and they can form ${\cal N}=2$ hypermultiplets.

In the $T_6$ sector, $6\times (s_{\pm}+6\phi)\phi= \mp 1$ mod integer. Because of the same reason, KK states in $T_6$ are always vector-like. Unlike in the $U$ and $T_3$ sectors, the degeneracy factors in $T_6$ are not the same as seen in Table \ref{tb:degnKK}. Thus one should consider the full GSO projection condition in Eq.~(\ref{GSO}) to determine the number of the physical KK states.

%%%%%%%%%%%%%%%%%%%%%%%%%%%%%%%%%%%%%%%%%%%%%%%%%%%%%%%%%%
\subsection{Decompactification limit} \label{subsec:Themodel}

In the decompactification limit $R\rightarrow\infty$, non-compact space-time dimension becomes six.  In this limit, winding is impossible $\vec{\zeta}=0$, and $\vec{L}$ and $M_l^2$ in Eqs.~(\ref{M1}) and (\ref{kkmass}) vanish. Hence, the KK massive
states in the $U$, $T_3$, $T_6$, $T_9$ sectors become massless. They reside in the 6 dimensional bulk, whereas the states in $T_1$, $T_2$, $T_4$, and $T_5$ sectors are still localized on 4 dimensional fixed points.

A ${\bf Z}_{12-I}$ shift and twist vectors can be always decomposed to those of ${\bf Z}_4$ and ${\bf Z}_3$:
\begin{eqnarray}
V=V_4+V_3 ~,\quad \phi =\phi_4+\phi_3 ~,
\end{eqnarray}
where $4V_4$ and $3V_3$ are \EE weights, and $4\phi_4$ and
$3\phi_3$ are SO(6) weights. For instance,
$\phi=(\frac{5}{12},\frac{4}{12},\frac{1}{12})
=(\frac{-1}{4},0,\frac{-1}{4})+
(\frac{8}{12},\frac{4}{12},\frac{4}{12})$, where
$(\frac{-1}{4},0,\frac{-1}{4})$ and
$(\frac{8}{12},\frac{4}{12},\frac{4}{12})$ are the ${\bf Z}_4$ and
${\bf Z}_{3}$ twist vectors, respectively.

Suppose that an SO(8) weight $s$ and an \EE\ weight $P$ satisfy the mass-shell conditions Eqs.~(\ref{mR}) and (\ref{mL}) in the ${\bf Z}_{12-I}$ orbifold, which are required even for KK massive states:
\begin{eqnarray}
&&\frac{(s+k\phi)^2}{2} +\sum_{i=j,\bar{j}}N^R_i\tilde{\phi}_i
-\frac{1}{2}+\frac{c_k}{2}=0 ,
\\
&&\frac{(P+kV)^2}{2}+\sum_{i=j,\bar{j}}N^L_{i}\tilde{\phi}_i
-1+\frac{c_k}{2}=0 ,
\end{eqnarray}
where $k=0,3,6,9$.  Note that $c_{k=0,3,6,9}$ in ${\bf Z}_{12-I}$ takes the same values as $c_{k=0,3,2,1}$ in ${\bf Z}_4$. Since $k\phi_3$, $kV_3$ are SO(6) and \EE\ weight vectors for $k=0,3,6,9$, the shifted weights, $s^\prime\equiv s+k\phi_3$ and $P^\prime\equiv P+kV_3$, are weights satisfying the mass-shell conditions in the $T_k$ sector of ${\bf Z}_4$: compare $2\tilde c$ of $\Z_{12-I}$ with Eqs. (\ref{ctildeL4}) and (\ref{cZ4R}) of $\Z_4$.

For \EE\ weight vectors $\vec{r}$ and $\vec{r}'$, in general, $\vec{r}\cdot \vec{r}'$ is an ${\rm even ~integer}$ if
$\vec{r}=\vec{r}'$, and is just an integer if
$\vec{r}\neq\vec{r}'$.  Using this, one can show that the phase in the GSO projector in Eq.~(\ref{GSO}),
\begin{align} \label{12-4}
3\tilde{\Theta}_3&=\left(P+\frac{3}{2}V\right)\cdot 3V
-\left(s+\frac{3}{2}\phi\right)\cdot
3\phi+3\sum_{i}(N_i^L-N_i^R)\hat{\phi}_i
\\
&=-\left(P^\prime +\frac{3}{2}V_4\right)\cdot V_4
+\left(s^\prime+\frac{3}{2}\phi_4\right)\cdot
\phi_4-\sum_{i}(N_i^L-N_i^R)\hat{\phi}_{4i} +{\rm integer} ,
\nonumber
\end{align}
where $P^\prime=P+3V_3$ and $s^\prime=s+3\phi_3$. Note $3\phi\sim -\phi_4$. Eq.~(\ref{12-4}) implies that the $[k=3,l=3]$ sector in the 4 dimensional ${\bf Z}_{12-I}$ corresponds to $[k=3,l=3]$ in the 6 dimensional ${\bf Z}_4$. Similarly, one can show that the $U$, $T_6$, and $T_9$ sectors in the 4 dimensional ${\bf Z}_{12-I}$ respectively correspond to $U$, $T_2$, and $T_1$ in the 6 dimensional ${\bf Z}_4$. Since the degeneracy factor $\tilde{\chi}(\theta^k,\theta^l)$ in Eq.~(\ref{GSO}) comes from the $|\widehat{Z}_X|^2$ in Eq.~(\ref{full}), which is from the compact extra dimensional part, $\tilde{\chi}(\theta^k,\theta^l)$ for $k,l=0,3,6,9$ in ${\bf Z}_{12-I}$ should be the same as those for $k,l=0,3,2,1$ in ${\bf Z}_4$. As argued already, the denominator $N$ in Eq.~(\ref{GSO}) is 4 for KK massive states. Therefore, we can conclude that the KK spectra in ${\bf Z}_{12-I}$ should coincide with massless spectra from the 6 dimensional ${\bf
Z}_4$ orbifold compactification.

Thus, we end up using the same shift vectors given in Eqs.
(\ref{ShiftSM}),
\begin{equation}
\begin{array}{l}
V=\textstyle\left( \frac14~ \frac14~ \frac14~ \frac{1}{4}~
\frac14~ \frac{5}{12}~\frac{5}{12}~  \frac{1}{12}~ \right)\left(
\frac{1}{4}~
\frac{3}{4}~ 0~ 0~0~0~0~0 \right) , \\
W_3=W_4\equiv W=\textstyle\left(
\frac23~\frac23~\frac23~\frac{-2}{3}~\frac{-2}{3}~\frac23~
0~\frac23 \right)\left( 0~ \frac23~\frac{2}{3}~0~0~0~0~0 \right) ,
\end{array}\label{Z12ISM}
\end{equation}
and $W_1=W_2=W_5=W_6=0$, which  satisfy Eqs.~(\ref{v^2})-(\ref{w^2}). The massless states in the gauge sector are given by the ${\rm E_8\times E_8'}$ root vectors ($P^2=2$) satisfying $P\cdot V={\rm integer}$ and $P\cdot W={\rm integer}$ with $s\cdot\phi=0$, which are listed as follows
\begin{eqnarray}
\textstyle (\underline{1~-1~0}~;~0~0~;~0^3)(0^8)'
~,~~(0~0~0~;~\underline{1~-1}~;~0^3)(0^8)'~,~~
(0^8)(0~0~;~\underline{\pm 1~\pm 1~0~0~0})' ~,
\end{eqnarray}
where the underlined entries allow all possible commutations. Thus, the resulting gauge group is
\begin{equation}
{\rm [\{SU(3)_c\times SU(2)_L\times U(1)_Y\}\times U(1)^4]\times
[SO(10)\times U(1)^3]^\prime }~.
\end{equation}

One definition of the hypercharge ${\rm U(1)_Y}$ is the standard one but the model has (vector-like) exotics \cite{SMZ12I},
\begin{eqnarray}
\textstyle
{\rm Model\ E}:\ Y=\sqrt{\frac35}~(\frac13~\frac13~\frac13~\frac{-1}{2}~
\frac{-1}{2};0^3)(0^8)'.\label{ModelE}
\end{eqnarray}
The normalization factor $\sqrt{\frac35}$, which leads to ${\rm sin}^2\theta_W=\frac{3}{8}$~\cite{SMZ12I} at the string scale, is determined by the current algebra in the heterotic string theory.

Another definition of the hypercharge ${\rm U(1)_Y}$ is  non-standard but the model does not have exotics \cite{SMZ12I},
\begin{eqnarray}
\textstyle
{\rm Model\ S}:\ Y=\sqrt{\frac3{11}}~(\frac13~\frac13~\frac13~\frac{-1}{2}~
\frac{-1}{2};0^3)(0~0~1;~0^5)'.\label{ModelS}
\end{eqnarray}
The normalization factor $\sqrt{\frac3{11}}$ leads to ${\rm sin}^2\theta_W=\frac{3}{14}$~\cite{SMZ12I} at the string scale.

For KK states, the projection condition $P\cdot V={\rm integer}$ is relaxed to $3P\cdot V={\rm integer}$, and $P\cdot W={\rm integer}$ is invalidated. As a result, the visible and hidden sectors' gauge groups are enhanced to ${\rm SU(8)\times U(1)}$ and ${\rm SO(12)'\times SU(2)'\times U(1)'}$, respectively. Their roots are
\begin{eqnarray} \label{roots}
&&{\rm SU(8)}\quad\left\{
\begin{array}{l}
(\underline{1~-1~0~0~0}~;~0~0~0)(0^8)' \quad :~{\rm SU(5)}
\\
(0~0~0~0~0~;~\underline{1~-1~0})(0^8)' \quad :~{\rm SU(3)}
\\
~\pm (\underline{1~0~0~0~0}~;\underline{1~0~0})(0^8)' \quad
:~({\bf 5},{\bf 3}),~({\bf\overline{5}},{\bf\overline{3}})
\end{array}\right\} , \nonumber
\\  [0.3em]
&&{\rm SO(12)'}~\left\{
\begin{array}{l}
(0^8)(0~0~;~0~;\underline{\pm 1~\pm 1~0^3})' \quad :~{\rm SO(10)'}
\\
(0^8)(0~0~;~\pm 1~;\underline{\pm 1~0^4})' \quad :~{\bf
10}_1,~{\bf 10}_{-1}
\end{array}\right\} ,
\\ [0.5em]
&&{\rm SU(2)'} \quad~ \bigg\{\pm (0^8)(1~ 1~;~0^6)'\bigg\} .
\nonumber
\end{eqnarray}
Note that for Model E, Eq. (\ref{ModelE}), the standard model gauge group of the model is embedded in a simple group SU(5) and further in SU(8) by including the KK modes. Accordingly, the value of ${\rm sin}^2\theta_W$ ($=\frac{3}{8}$) can be protected down to the compactification scale, which is identified with the GUT scale.

\begin{table}
\begin{tabular}{c||c|c|c|c}
\hline\noalign{\smallskip}
$P\cdot V$ & States ($P$) & ~$\chi$~ & ~SM~ & $\Gamma$   \\
\noalign{\smallskip}\hline\noalign{\smallskip} $\frac{-5}{12}$ &
$(\underline{++-};\underline{+-};+++)(0^8)'$ & L
& $Q_3$ & +1 \\
($U_1$) & $(---;\underline{+-};+--)(0^8)'$ & L & $L_3$ & $-3$
\\
\noalign{\smallskip} \hline \noalign{\smallskip}
$$ & $(\underline{+--};--;+++)(0^8)'$ & L & $d^c_3$ & $+1$
\\
$\frac{1}{12}$ & $(+++;++;+++)(0^8)'$ & L &$\nu^c_3$ & $+1$
\\
($U_3$) & $(\underline{+--};++;+--)(0^8)'$ & L & $u^c_3$ & $-3$
\\
$$  & $(+++;--;-+-)(0^8)'$ & L & $e^c_3$ & $+5$ \\
\noalign{\smallskip}\hline \noalign{\smallskip}
$$ & $(0,0,0;\underline{1,0};0,0,1)(0^8)'$ &
L & $H_d$ & $-2$
\\
$\frac{4}{12}$ ($U_2$) & $(0,0,0;\underline{-1,0};-1,0,0)(0^8)'$ &
L & $H_u$ & $+2$
\\
$$ & $(0,0,0;0,0;1,0,-1)(0^8)'$ & L & ${\bf
1}_0$ & $0$
\\
\noalign{\smallskip}\hline
\end{tabular}
\caption{Massless matter states from the $U$ sector} \label{tab:1}
\vspace*{0.5cm}
\end{table}

\begin{table}
\begin{tabular}{c||c|c|c|c}
\hline\noalign{\smallskip}
States ($P+4V$) & ~$\chi$~ & ~${\cal P}_4$~ & ~SM~ & $\Gamma$   \\
\noalign{\smallskip}\hline\noalign{\smallskip}
$\left(\underline{+--};--;\frac{1}{6},\frac{1}{6},\frac{-1}{6}\right)
(0^8)'$ & L & $2$ & ~$2\cdot d^c$~ & $+1$
\\
$\left(---;\underline{+-};\frac{1}{6},\frac{1}{6},\frac{-1}{6}\right)
(0^8)'$ & L & $2$ & $2\cdot L$ & $-3$
\\
$\left(\underline{+--}++;\frac{1}{6},\frac{1}{6},\frac{-1}{6}\right)
(0^8)'$ & L & $2$ & $2\cdot u^c$ & $-3$
\\
$\left(\underline{++-};\underline{+-};
\frac{1}{6},\frac{1}{6},\frac{-1}{6}\right) (0^8)'$ & L & $2$& $2\cdot Q$ & $+1$
\\
$\left(+++;--;\frac{1}{6},\frac{1}{6},\frac{-1}{6}\right) (0^8)'$
& L & $2$ & $2\cdot e^c$ & $+5$
\\
$\left(+++;++;\frac{1}{6},\frac{1}{6},\frac{-1}{6}\right) (0^8)'$
& L & $2$ & $2\cdot \nu^c$  & $+1$
\\
\noalign{\smallskip}\hline
\end{tabular}
\caption{Some massless matter states from the $T_4^0$ sector}
\label{tab:2}
\vspace*{0.5cm}
\end{table}
In the model of Eq.~(\ref{Z12ISM}), one family of MSSM matter and electroweak Higgs doublets come from the untwisted sector, and the other two families of MSSM matter from the $T_4^0$ sector. As shown in Tables~\ref{tab:1} and \ref{tab:2}, they have exactly the structure of the SO(10) spinor~\cite{SMZ12I}. All other matter states in
this model turn out to be vector-like under the SM gauge symmetry, and have been explicitly shown to achieve string scale masses by vacuum expectation values (VEVs) of SM singlets. In this model, such singlets' VEVs can avoid to break the R-parity. As mentioned before, the two MSSM Higgs doublets in this model become a part of the ${\cal N}=2$ gauge multiplet, being unified with the MSSM gauge fields above the compactification scale.

All the states of an SO(16) spinor ${\bf 128}$ among the root vectors of the first ${\rm E_8}$ revive as KK massive states satisfying Eq.~(\ref{KKUmatter}). Under SU(8), they split as ${\bf 8}$, ${\bf \overline{8}}$, ${\bf 56}$, and ${\bf \overline{56}}$:
\begin{equation}
\begin{array}{l}
P\cdot W={\rm integer}\\
P\cdot V-s\cdot\phi=\\
\quad\quad\frac13\cdot(\rm integer)
\end{array}
\left[
\begin{array}{cl}
{\bf 8}&=\left\{ (\underline{+----};+++)~,~~
(-----;\underline{-++})\right\}
\\
{\bf \overline{8}}&=\left\{ (\underline{-++++};---)~ ,~~
(+++++;\underline{+--}) \right\}
\\
{\bf 56}&=\bigg\{
\begin{array}{l}
(-----;---) ~, \quad (\underline{+----};\underline{+--})
\\
(\underline{++---};\underline{++-}) ,\quad
(\underline{+++--};\underline{+++})
\end{array}\bigg\}
\\
{\bf \overline{56}}&=\bigg\{
\begin{array}{l}
(+++++;+++) ~,\quad (\underline{-++++};\underline{-++})
\\
(\underline{--+++};\underline{--+}) ~,\quad
(\underline{---++};---)
\end{array} \bigg\}
\end{array}
\right. \label{SM6DbH}
\end{equation}
They form vector-like representations and compose the \N=2
hypermultiplets. From the hidden sector, we have the following SO(12)$'\times$SU(2)$'$ representations contributing to the KK states,
\begin{eqnarray}
2\times ({\bf 12},{\bf 2})'&=&\bigg\{(0^8)(\pm 1~0~;\underline{\pm
1,0^5})'~,~~(0^8)(0~\pm 1~;\underline{\pm 1,0^5})'\bigg\}
\label{Hidden6DbH}
\\
({\bf 32},{\bf 1})',~(\overline{\bf 32},{\bf 1})'&=&\left\{
\begin{array}{l}
\pm (0^8)(-+;\underline{+-----})',~\pm
(0^8)(-+;\underline{+++---})',
\\
\quad\quad\quad\quad\quad\quad \pm (0^8)(-+;\underline{+++++-})'
\end{array}\right\}\nonumber
\end{eqnarray}
which satisfy $P\cdot W={\rm integer},\ P\cdot V-s\cdot\phi=
\frac13\cdot(\rm integer)$. Here we see that the matter content in the untwisted sector coincides with the 6D results in Table \ref{tab:6Duntwst}.  One can check that from $T_3$ there are $4\times \{({\bf 8};{\bf 2'},{\bf 1}) +{\rm c.c}\}$, $4\times \{({\bf 1};{\bf 12'},{\bf 1})+{\rm c.c}\}$, and $8\times \{({\bf 1};{\bf 1},{\bf 2'})+{\rm c.c}\}$ under ${\rm SU(8)\times SO(12)'\times SU(2)'}$. From $T_6$, there are $(6+10)\times \{({\bf 8};{\bf 1},{\bf 1})+ {\rm c.c}\}$. They are coincident with the 6D results in Table \ref{tab:6Dtwisted}.

%%%%%%%%%%%%%%%%%%%%%%%%%%%%%%%%%%%%%%%%%%%%%%%%%%%%%%%%%%%%%%
%%%%%%%%%%%%%%%%%%%%%%%%%%%%%%%%%%%%%%%%%%%%%%%%%%%%%%%%%%%%%%
\section{threshold correction}\label{sec:couplings}
%%%%%%%%%%%%%%%%%%%%%%%%%%%%%%%%%%%%%%%%%%%%%%%%%%%%%%%%%%%%%

In the ${\bf Z}_{12-I}$ ($={\bf Z}_4\times {\bf Z}_3\equiv G$) orbifold, the six dimensional torus is factorized as $T_4\times T_2$. The ${\cal N}=2$ SUSY sector relevant to the threshold corrections to the four dimensional gauge couplings are associated only with the $T_2$ sub-lattice, which is fixed under the ${\bf Z}_3$ ($\equiv G'$) rotation.
The moduli in the $T_2$ sub-lattice, $g_{ab}$ and $B_{ab}$
($\equiv B\epsilon_{ab}$) can be reexpressed as
\begin{equation}
T=2\left(B+i\frac{R^2}{\alpha'}\sqrt{{\rm det}g_{ab}}\right)\equiv T_R+iT_I ,\quad U=\frac{1}{g_{11}}\left(g_{12}+i\sqrt{{\rm
det}g_{ab}}\right)\equiv U_R+iU_I
\end{equation}
where $g_{ab}$ in the $T_2$ torus is given by Eq.~(\ref{metric}) and $\sqrt{{\rm
det}g_{ab}}=\sqrt3$. The magnitudes of $T$ and $U$ are representing the size and the shape of the compactified manifold. Note that the modulus $R$ appears only in the imaginary part of $T$ in ${\bf Z}_{12-I}$, i.e. $T_I\propto R^2$. If the compactified manifolds have a fixed torus represented by $T$ and $U$, that torus can be considered
large as the second torus of Fig. \ref{fig:Z12I}, i.e. the
(34)-torus\footnote{The simplest nontrivial moduli dependence is for the the case of only one moduli dependence, e.g. by $T_I$, which is possible for $\Z_{6-I}$, $\Z_{8-I}$, and $\Z_{12-I}$ models ~\cite{Dixon:1990pc}.}.

Now, let us discuss the one-loop renormalized gauge couplings in the effective 4D field theory:
\begin{eqnarray}
\frac{4\pi}{\alpha_i(\mu)}=k_a\frac{4\pi}{\alpha_*}+b_i^0~{\rm
log}\frac{M_*^2}{\mu^2}+\Delta_i ~,
\end{eqnarray}
where $k_a=1$ for the case of level one as in our case, $\alpha_*$ is the unified gauge coupling $\frac{g_*^2}{4\pi}$ at the string scale $M_*$, and $b_i^0$ denotes the beta function coefficient by massless modes, satisfying $P\cdot W={\rm integer}$. The general expression for the moduli dependent threshold correction to the gauge couplings $\Delta_i$ can be obtained as in Refs.
\cite{Kaplunovsky,Dixon:1990pc,Mayr}
\begin{eqnarray}
\Delta_i=\frac{|G'|}{|G|}\cdot b_i^{N=2}
\int_\Gamma\frac{d^2\tau}{\tau_2}\left(\hat{Z}_{\rm
torus}(\tau,\bar{\tau})-1\right) ,\label{threshold6D}
\end{eqnarray}
where $b_i^{N=2}$ in Eqs.~(\ref{threshold6D}) denotes the beta function coefficient of ${\cal N}=2$ SUSY sector by the KK modes. Since $G={\bf Z}_4\times {\bf Z}_3$ and $G'={\bf Z}_3$ in our case, $\frac{|G'|}{|G|}=\frac{1}{4}$. The $\hat{Z}_{\rm torus}(\tau,\bar{\tau})$ is defined as
\begin{eqnarray} \label{Ztorus0}
\hat{Z}_{\rm torus}(\tau,\bar{\tau})\equiv
\sum_{\vec{P}_L,\vec{P}_R}q^{P_L^2/2}\bar{q}^{P_R^2/2}
=\sum_{\vec{\mu},\vec{\zeta}}\frac{1}{\lambda^2}~
q^{(\vec{L}+\vec{\zeta})^2/2} \bar{q}^{(\vec{L}-\vec{\zeta})^2/2},
\end{eqnarray}
where $\vec{L}$ is given in Eq.~(\ref{L^i}). Note that regardless of $(\vec{\delta}_4;\vec{\delta}_2)=(0;0)$ or not, $\hat{Z}_{\rm torus}(\tau,\bar{\tau})$ is always given by Eq.~(\ref{Ztorus0}). Using the Poisson resummation formula, $\hat{Z}_{\rm torus}(\tau,\bar{\tau})$ is shown to be proportional to Eq.~(\ref{2dpart}).  It is just the process to trace back Eq.~(\ref{full}) in Sec. VI. In terms of $T$ and $U$, $\hat{Z}_{\rm torus}(\tau,\bar{\tau})$ can be eventually written as
\begin{eqnarray} \label{Ztorus}
\tau_2\hat{Z}_{\rm torus}= \sum_{A\in M(2\times 2, Z)} ~T_I ~{\rm exp}\left[\scriptsize \frac{-\pi
T_I}{\tau_2U_I}\left|\left(1,U\right)A \left(
\begin{array}{c}
\tau \\
1
\end{array}\right)
 \right|^2\right]~e^{-2\pi i T{\rm det}A}~
 e^{2\pi i(\zeta^{\prime 3}+\zeta^{\prime 4})\theta_\zeta}
\end{eqnarray}
where $\theta_\zeta\equiv
[P^I+\frac{k}{2}V^I+\frac{(\zeta^3+\zeta^4)}{2}W^I] W^I$.  Note that  $\zeta^{\prime
b}\frac{k}{2}V^IW_b^I=\lambda n^{\prime b}\frac{k}{2}V^IW_b^I$ in the phase of Eq.~(\ref{Ztorus}) or (\ref{2dpart}) becomes integer due to Eq.~(\ref{vw}), where the values of $k$ and $\lambda$ are found in Table V. Thus, we will drop the $\frac{k}{2}V^I$ from $\theta_\zeta$. $A$ is a matrix representing winding numbers:
\begin{eqnarray} \label{A}
A=\left(
\begin{array}{cc}
\zeta^3 & \zeta^{\prime 3} \\
\zeta^4 & \zeta^{\prime 4} \end{array} \right) . ~
\end{eqnarray}
Thus, the modular transformations are equivalent to the
transformations of $A$: $A\rightarrow A'=A\cdot V_{2\times 2}$ with $V_{2\times 2}\in {\rm SL}(2,Z)$. Instead of integrating the contribution of $A'$ over the fundamental region $\Gamma$, one can integrate the contribution $A$ over $V\Gamma$, the image of $\Gamma$ under the ${\rm PSL}(2,Z)$ modular transformation~\cite{Dixon:1990pc}. The matrices $A$s are classified to three types:\\
\indent (1) $A=0$ (``zero orbit''),\\
\indent  (2) ${\rm det}A\neq 0$ (``non-degenerate orbit''),\\
\indent  (3) ${\rm det}A=0$ (``degenerate orbit'').\\
For $A=0,\vec\zeta=\vec\zeta'=0$, and for Det.$A=0$ one can set $\vec\zeta=0$ \cite{Dixon:1990pc}.

%%%%%%%%%%%%%%%%%%%%%%%%%%%%%%%%%%%%%%%%%%%%%%%%%%%%%%%%%%%%%%%%%
\paragraph{$\theta_\zeta={\rm integer}$:}

For the case that the six dimensional torus is factorized as
$T_4\times T_2$ and the KK massive states arise only from the $T_2$ torus as in our case, and $\theta_\zeta={\rm integer}$, then the threshold corrections $I_1$, $I_2$, and $I_3$ for (1) $A=0$,
(2) ${\rm det}A\neq 0$, and (3) ${\rm det}A= 0$, respectively, turn out to be proportional to~\cite{Dixon:1990pc}
\begin{align}
& I_1 = \frac{\pi}{3}T_I ~ , \quad\quad
I_2 = -2\sum_{n>0}{\rm log}\left|1-q_T^n\right|^2 ~,
\nonumber\\
& I_3 = \frac{\pi}{3}U_I -2\sum_{n=1}^{\infty}{\rm
log}\left|1-q_U^n\right|^2 -{\rm log}(T_IU_I)+\left(\gamma_E-1-{\rm
log}\frac{8\pi}{3\sqrt{3}}\right) , \label{Ishere}
\end{align}
where $q_T\equiv e^{2\pi iT}$, $q_U\equiv e^{2\pi iU}$, and
$\gamma_E$ ($\approx 0.58$) is the Euler-Mascheroni constant. The detailed calculation of $I_3$ is presented in Appendix.
Thus, $\Delta_i$ is given by
\begin{align}
\Delta_i^{(1)} &=b_i^{N=2}\frac{|G'|}{|G|}(I_1+I_2+I_3)
\nonumber\\
&=-b_i^{N=2}\frac{|G'|}{|G|}~{\rm log}\bigg[ \frac{8\pi
e^{1-\gamma_E}}{3\sqrt{3}}\left(\left|\eta(T)\right|^4{\rm
Im}T\left|\eta(U)\right|^4{\rm Im}U\right)\bigg] .
\label{threshold}
\end{align}
For a large $R$ ($\gtrsim\sqrt{\alpha'}$), we obtain
$|\eta(T)|^4\approx e^{-\pi 2\sqrt{3}R^2/3\alpha'}$ and
$|\eta(U)|^4\approx  e^{-\pi \sqrt{3}/6}$, respectively. Thus, for
$R\gtrsim\sqrt{\alpha'}$, Eq.~(\ref{threshold}) is approximated as
\begin{eqnarray} \label{thresholdapprx}
\Delta_i^{(1)} \approx \frac{b_i^{N=2}}{4}~\left[\frac{2\pi
R^2}{\sqrt{3}\alpha'}-{\rm log}\frac{R^2}{\alpha'}-2.19\right] .
\end{eqnarray}
Note that $b_i$ in Eqs.~(\ref{threshold}) and
(\ref{thresholdapprx}) denotes the beta function coefficient of the ${\cal N}=2$ SUSY sector by the KK modes satisfying
$\theta_\zeta={\rm integer}$. The terms proportional to $R^2$ and ${\rm log}R^2$ in Eq.~(\ref{thresholdapprx}) are coming from $I_1$ and $I_3$ in Eq.~(\ref{Ishere}), respectively, i.e. from the $A=0$ and ${\rm det}A=0$ cases, respectively. The contribution from $I_2$ is negligible and is not included in Eq.~(\ref{thresholdapprx}).

%%%%%%%%%%%%%%%%%%%%%%%%%%%%%%%%%%%%%%%%%%%%%%%%%%%%%%%%%%%%%
\paragraph{$\theta_\zeta=\pm\frac13~({\rm mod~Z})$:}

For $A=0$, i.e. $\vec{\zeta}=\vec{\zeta}^\prime=0$, the factor $e^{2\pi i(\zeta^{\prime 3}+\zeta^{\prime 4})\theta_\zeta}$ in Eq.~(\ref{Ztorus}) is still unity. $I_1$ for $\theta_\zeta\neq 0$ is, hence, the same as for the $\theta_\zeta = 0$ case of the preceding paragraph, and $\theta_\zeta\neq 0$ does not affect the $R^2$ term of Eq.~(\ref{thresholdapprx}). Since the sub-leading term proportional to ${\rm log}R$ (and also the constant term) in
Eq.~(\ref{thresholdapprx}) is coming from $I_3$, it is enough to calculate only $I_3$ for the case $\theta_\zeta\neq {\rm
integer}$. We present the $I_3$  calculation in Appendix. The result for the case of $P^I W^I=\pm\frac{1}{3}$ (mod Z) is
\begin{eqnarray}
\Delta_i^{(2)} \approx \frac{c_i}{4}~\left[\frac{2\pi
R^2}{\sqrt{3}\alpha'} -0.30\right]\label{Delthe}
\end{eqnarray}
for $R\gg\sqrt{\alpha'}$, where $c_i$ denotes the beta function coefficient of ${\cal N}=2$ SUSY sector by the KK states with $\theta_\zeta=\pm\frac13$ (mod Z).   Note the ${\rm log}R^2$ term of Eq.~(\ref{thresholdapprx}) is absent in Eq. (\ref{Delthe}) for the states with $P^I
W^I=\pm\frac{1}{3}$ (mod Z).  As discussed in Subsec. \ref{subsec:massive}, the states with $[P^I+\frac{(\zeta^3+\zeta^4)}{2}W^I]W^I\neq {\rm integer}$ are always KK massive states.

Thus, if a gauge group ${\cal G}$ is broken to  ${\cal H}$ by the Wilson lines and further broken to ${\cal H}_0$ by orbifolding, the renormalized gauge couplings of ${\cal H}$ at low energies is
\begin{eqnarray}
\frac{4\pi}{\alpha_{\cal H}(\mu)}
=\frac{4\pi}{\alpha_*}+b_{{\cal H}_0}^0~{\rm
log}\frac{M_*^2}{\mu^2}-\frac{b_{\cal H}}{4}\left[{\rm
log}\frac{R^2}{\alpha'}+1.89\right]+\frac{b_{\cal H+G/H}}{4} \left[\frac{2\pi R^2}{\sqrt{3}\alpha'}-0.30\right] .
\label{gaugerun}
\end{eqnarray}
$b_{\cal H}^0$ is the beta function coefficient contributed by ${\cal N}=1$ SUSY sector states projected by $P^I W^I={\rm integer}$ and $1\times \Theta_k$.  $b_{\cal H}$ is by the states projected by $P^IW^I={\rm integer}$ and $3\times \Theta_k$, and $b_{\cal G/H}$ by the states projected by $P^IW^I\neq {\rm integer}$ and $3\times\Theta_k$.

Hence, the $R^2$ coefficient $b_{\cal H+G/H}(\equiv b_{\cal G}+b_{\cal G/H})$ is
contributed by all the states charged under ${\cal G}$. Therefore, the difference ${\alpha_{{\cal H}_0}(\mu)}_i^{-1} -{\alpha_{{\cal H}_0}(\mu)}_j^{-1}$ does not get an $R^2$ dependent piece if the corresponding groups are unified to ${\cal G}$ above the scale $R^{-1}$. Even for this case, the logarithmic contribution is present from the third term in the right-hand side (RHS) of Eq. (\ref{gaugerun}). The absolute values of the constant and the $R^2$ term are more or less reliable since the 10D \EE\ heterotic string gives an exact spectrum at the compactification scale, say at $M_s$ and below $M_s$, and we can apply the result to the absolute values of gauge
couplings such as in the strength of the hidden sector scale
\cite{GMSBun}.

Before discussing some phenomenological applications, let us
comment on the efforts on threshold correction in field theoretic orbifolds. In field theory calculations, the constant and the $R^2$ dependent term cannot be pinpointed exactly \cite{Dienes} because of the unknown dynamics near the cutoff scale. For example, from Ref. \cite{HMLee} one can obtain the $R_1$ and $R_2$ dependent terms in field theoretic $T_2/\Z_2$ orbifold (with radius $R_1$ and $R_2$) in terms of 4D and 6D beta function coefficients $b_1$ and $b_2$. The equation should read, at the 4D low momentum scale $k^2\ll 1/R^2_{1,2}$ \cite{HMLeeprivate},
\begin{align}
\frac{1}{g^2_{a,{\rm eff}}(k^2)}&=\frac{1}{g^2_{a,*}}+\frac{1}{16\pi^2}
b^a_2\bigg\{ e^{-\gamma-2}\pi\frac{M^2_* R^2_1 U_2}{N}+2\gamma+2-\ln(|\eta(U)|^4U_2)
-\ln\frac{4\pi^2 M^2_* R_1^2 U_2}{N}\bigg\}
\nonumber \\
&\quad+\frac{1}{16\pi^2}b_1^a \ln\frac{M^2_*}{k^2}\nonumber
\end{align}
where $N=2$ and $U=e^{i\theta}R_2/R_1=U_1+iU_2$ in terms of the angle between $R_1$ and $R_2$, i.e. $\theta=\frac{\pi}{2}$ so that $U_1=0$ and $U_2=R_2/R_1$. This expression has some corresponding terms in Eq. (\ref{gaugerun}) but the constant and the $R^2$ term
are not reliable because of unknown physics at the cutoff scale $M_*$, e.g. in the uncertainty of $g^2_{a,*}$.
Actually the quadratically divergent part in field theory is regularization scheme dependent. Thus, only when a regularization scheme is properly chosen, the quadratically divergent part in field theory can reproduce the result from the fundamental theory such as string theory \cite{Ghilencea}.

The $T_2/\Z_2$ orbifold compactification breaks only ${\cal N}=2$ SUSY (not a gauge symmetry). A gauge symmetry ${\cal G}$ could be also broken down to ${\cal H}$ by employing $T_2/(\Z_2\times \Z_2')$ or another $T_2/\Z_N$ orbifold instead of $T_2/\Z_2$~\cite{Asaka}. Then the beta function coefficient of the ${\rm log} R_1^2$ term would be replaced by a coefficient calculated with the KK massive modes originating from the bulk fields providing massless modes~\cite{HMLeeprivate}. However, the $R_1^2$ term coefficient would be still calculated by all the KK
modes preserving ${\cal G}$.  It is a common feature observed also in the threshold corrections calculated in 5D orbifold field theory of $S^1/(\Z_2\times \Z_2')$~\cite{Hall-Nomura}.

In string calculation, however, such a deviation in the ${\rm log}R^2$ coefficient from the value of the $R^2$ coefficient
arises via Wilson line breaking as seen in Eq.~(\ref{gaugerun}). It is obvious from the observation for the KK mass-squared in Eq.~(\ref{kkmass}). Namely, in orbifolded string theory, only the KK masses of the states with $P^IW^I\neq {\rm integer}$ are shifted.  Since gauge symmetry breaking effects by orbifolding appear just through the GSO projection (or the phase), e.g. in the untwisted sector, $3\times (P^IV^I-s\cdot\phi)={\rm integer}\rightarrow 1\times (P^IV^I-s\cdot\phi)={\rm integer}$, the mass-squared is not affected just by orbifolding, and so eventually the coefficient of $R^2$ remains the same with that
of ${\rm log}R^2$ as seen in Eq.~(\ref{threshold}) or
(\ref{thresholdapprx}).

%%%%%%%%%%%%%%%%%%%%%%%%%%%%%%%%%%%%%%%%%%%%%%%%%%%%%%%%%%%%%%
%%%%%%%%%%%%%%%%%%%%%%%%%%%%%%%%%%%%%%%%%%%%%%%%%%%%%%%%%%%%%%
\subsection{Coefficients of $\beta$ functions } \label{subsec:6Dcouplings}

In this subsection, we present an application of the $R$ dependence of the evolution of gauge couplings in 6D. For a 6D \N=1 supersymmetric model (\N=2 in 4D), there is no bulk mass.
The compactification scale is $R^{-1}$ which will be fitted below such that at the electroweak scale $\sin^2\theta_W$ is phenomenologically viable.

\begin{figure}[t]
\resizebox{0.9\columnwidth}{!}
{\includegraphics{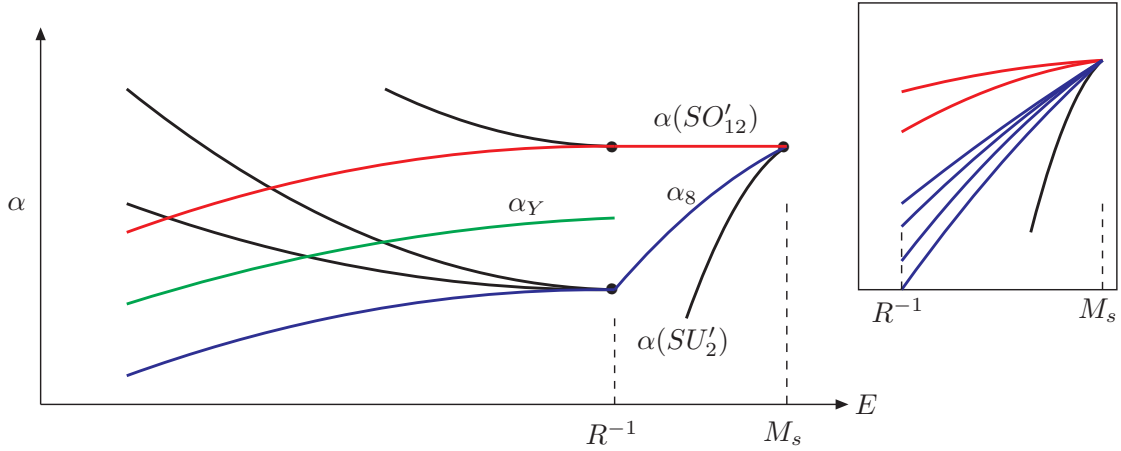}}
\caption{A schematic view of gauge coupling evolution without KK modes correction. $\alpha_8$ is the 6D SU(8) coupling and the green line for $\alpha_Y$ is the hypercharge coupling in Model S. The KK modes split couplings above $R^{-1}$ as depicted within the square.}\label{fig:evolution}
\end{figure}
%%%%%%%%%%%%%%%%%%%%%%%%%%%%%%%%%%%%%%%%%%%%%%%%%%%%%%%%%%%%%%%%%%%%%%%%%%%%%%
%%%%%%%%%%%%%%%%%%%%%%%%%%%%%%%%%%%%%%%%%%%%%%%%%%%%%%%%%%%%%%%%%%

For SU$(N)$, the quadratic invariants for a given representation are
\begin{align}
&{\rm tr}_{\rm Adj}F^2=2N {\rm tr}_N F^2, \\
&{\rm tr}_{a_2}F^2= (N-2) {\rm tr}_N F^2, \\
&{\rm tr}_{a_3}F^2= \frac{1}{2}(N-2)(N-3) {\rm tr}_N F^2
\end{align}
where $\rm Adj$, $a_2$ and $a_3$ are adjoint, second and third rank totally anti-symmetric tensor representations. In the normalization for a fundamental $N$ such that ${\rm tr}_N T^aT^b=l(N)\delta^{ab}$ with $l(N)=\frac{1}{2}$, we obtain $l_2({\rm Adj})=N$ and the indices for other representations, $l(a_2)=\frac{1}{2}(N-2)$ and $l(a_3)=\frac{1}{4}(N-2)(N-3)$.

For SO$(2n)$, the quadratic invariants for a given representation are
\begin{align}
&{\rm tr}_{\rm Adj}F^2= 2(n-1) {\rm tr}_{2n}F^2, \\
&{\rm tr}_{2^{n-1}}F^2= 2^{n-4} {\rm tr}_{2n} F^2.
\end{align}
In the normalization for a vector representation $2n$ such that ${\rm tr}_{2n}T^a T^b=l(2n)\delta^{ab}$ with $l(2n)=1$, we obtain $l_2({\rm Adj})=2n-2$ and the index for a spinor representation, $l(2^{n-1})=2^{n-4}$.

Thus, for the model discussed in Sec. \ref{sec:6DGUT} we obtain
\begin{equation}
\begin{array}{l}
b^{N=2}_{\rm SU(8)}=-2\times 8+\frac{1}{2}\times 50
+\frac{15}{2}\times 2 =24 ,\\
b^{N=2}_{\rm SO(12)'}=-2\times 10 + 1\times 12+ 4\times 2 = 0 ,\\
b^{N=2}_{\rm SU(2)'}=-2\times 2+ \frac{1}{2}\times 104 =48 .
\end{array}\label{b6DfullG}
\end{equation}
These contribute to $b_{\cal H+G/H}$ in Eq. (\ref{gaugerun}).

Let us calculate $b_{\cal H}$ of Eq.~(\ref{gaugerun}) for the model discussed in Sec.~III and Subsec. \ref{subsec:Themodel}. The condition $P^IW^I={\rm
integer}$ by Wilson line Eq.~(\ref{ShiftSM}) breaks the 6D gauge symmetry ${\cal G}={\rm SU(8)\times U(1) \times [SU(2)\times SO(12)\times U(1)]'}$ to ${\cal H}={\rm SU(3)_c\times SU(4)\times U(1)^3\times [SO(10)\times U(1)^{3}]'}$ whose root vectors are
\begin{eqnarray}
 {\rm SU(3)_c} &:&~~ (\underline{1~-1~0}~;~0~0~;0^3)(0^8)',
\\
 {\rm SU(4)} ~&:&~ \left\{
\begin{array}{l}
(0~0~0~;\underline{1~-1}~;~0~0~0)(0^8)' ,
\\
\pm (0~0~0~;\underline{1~0}; 1~0~0)(0^8)',
\\
\pm (0~0~0~;\underline{1,0};~0~0~1)(0^8)'
\\
\pm (0~0~0~;~0~0~;1~0~-1)(0^8)'
\end{array}\right\} , \nonumber
\\  [0.3em]
{\rm SO(10)'} &:&~~(0^8)(0~0~0~;\underline{\pm 1~\pm1~0~0~0})'
\end{eqnarray}

From Tables \ref{tab:6Duntwst} and \ref{tab:6Dtwisted}, or from Eqs. (\ref{roots}), (\ref{SM6DbH}), (\ref{Hidden6DbH}), which display the states of $P^IW^I=\frac13\cdot({\rm integer})$, we pick up the states fulfilling $P\cdot W={\rm integer}$. The results are shown in Tables \ref{tab:HU} and \ref{tab:HT}.
%%%%%%%%%%%%%%%%%%%%%%%%%%%%%%%%%%%%%%%%%%%%%%%%%%%%%%%%%%%%%%%
\begin{table}
\begin{tabular}{c|c|c|c|c}
\hline %\noalign{\smallskip}
States ($P$) & $({\rm SM})_Y$ &$\sqrt{2}Q_c$ &~4D $\chi$~
& ~$(SU(3), SU(4); SO(10)')$~ \\
\hline   ~$\left(---;\underline{+-};+--\right) (0^8)'$ ~&
~$L_{-1/2}$~& $0$ & L, R &
\\
 $\left(---;\underline{+-};--+\right) (0^8)'$  & $L_{-1/2}$
& $0$ & L, R &
\\
$\left(---;--;---\right) (0^8)'$ & $\one_0$
&$0$ & L, R & $ (\one,{\bf 6};\one)$
\\
$\left(---;++;+-+\right) (0^8)'$ & $e_{-1}$
& $0$ & L, R &
\\
 \hline
 $\left(\underline{++-};--;++-\right) (0^8)'$ & $u_{2/3}$
& $0$ & L, R &
\\
$\left(\underline{++-};--;-++\right) (0^8)'$ & $u_{2/3}$
& $0$ & L, R & $ (\overline{\bf 3},{\bf 4};\one)$
\\
$\left(\underline{++-};\underline{+-};+++\right) (0^8)'$ &
$Q_{1/6}$
& $0$ & L, R &
\\
 \hline $\left(\underline{+--};--;+++\right) (0^8)'$ & $d^c_{1/3}$
& $0$ & L, R & $ (\bf{3},\one;\one)$
\\
\hline  $(0^8)(1~0~0;\underline{\pm 1,0^4})'$ & $\one_0$
& $0$ & L, R & $ (\one,\one;{\bf 10'})$
\\ \hline
 $(0^8)(0~-1~1;0^5)'$ & $\one_0$
&$1$ & L, R& $ (\one,\one;\one)$
\\
\hline  $(0^8)(-+-;\underline{+----})'$ & $\one_0$
& $-1/2$ & L, R &
\\
 $(0^8)(-+-;\underline{+++--})'$ & $\one_0$
& $-1/2$ & L, R & $ (\one,\one;{\bf 16'})$
\\
 $(0^8)(-+-;\underline{+++++})'$ & $\one_0$
& $-1/2$ & L, R &
\\ \hline
\end{tabular}
\caption{6D untwisted matter states satisfying $P\cdot W={\rm integer}, 3P\cdot V=\pm\frac14$ (mod Z). $\sqrt{2}Q_c$ is the unnormalized U(1)$_c$ charge of the L states, i.e. composed of some multiples of integers from Eq. (\ref{U1charges}). We do not display the CTP conjugates here.} \label{tab:HU}
\end{table}

%%%%%%%%%%%%%%%%%%%%%%%%%%%%%%%%%%%%%%%%%%%%%%%%%%%%%%%%%%%%%%%
\begin{table}
\begin{tabular}{c|c|c|c|c|c|c}
\hline States ($P+kV$) & $({\rm SM})_{Y}$
&$\sqrt{2}Q_c$ & ~Sector~ & ~${\cal P}_k$~ & ~4D $\chi$~
& ~Representations~\\
\hline
$\left(\frac14~\frac14~\frac14~;\underline{\frac{-3}{4}
~\frac{1}{4}}~;
~\frac{-1}{4}~\frac{-1}{4}~\frac{-1}{4}\right)
(\frac{1}{4}~\frac{3}{4}~;0^6)'$ & $H_{u1/2}$
& $0$ & $T_3$ & 4 & L, R &
\\
$\left(\frac{1}{4}~\frac{1}{4}~\frac{1}{4}~;\frac{1}{4}
~\frac{1}{4}~;\frac{3}{4}~ \frac{-1}{4}~\frac{-1}{4}\right)
(\frac{1}{4}~\frac{3}{4}~;0^6)'$ & $\one_0$
& $0$ & $T_3$ & 4 & L, R & $4\times (\one,\overline{\bf 4};\one)$
\\
$\left(\frac{1}{4}~\frac{1}{4}~\frac{1}{4}~;\frac{1}{4}
~\frac{1}{4}~;\frac{-1}{4}~ \frac{-1}{4}~\frac{3}{4}\right)
(\frac{1}{4}~\frac{3}{4}~;0^6)'$ & $\one_0$
& $0$ & $T_3$ & 4 & L, R &
\\  \hline
$\left(\underline{\frac{-3}{4}~\frac14~\frac14}~;
~\frac14~\frac{1}{4}~;
~\frac{-1}{4}~\frac{-1}{4}~\frac{-1}{4}\right)
(\frac{-3}{4}~\frac{-1}{4}~;0^6)'$ & $D_{-1/3}$
& $0$ & $T_3$ & 4 & L, R & $4\times (\overline{\bf 3},\one;\one)$
\\ [0.3em]
\hline $\left(\frac{-1}{4}~\frac{-1}{4}~\frac{-1}{4}~;\frac{-1}{4}
~\frac{-1}{4}~;\frac{1}{4}~ \frac{1}{4}~\frac{1}{4}\right)
(\frac{-3}{4}~\frac{-1}{4}~;0^6)'$ & $\one_0$
& $0$ & $T_3(1_1,1_3)$ & 4 & L, R & $8\times (\one,\one;\one)$
\\ \hline
$\left((\frac{-1}{4})^3~; \frac{-1}{4}~\frac{-1}{4}~;~
(\frac{1}{4})^3\right) (\frac{1}{4}~\frac{-1}{4}~0~;\underline{\pm
1~0~0~0~0})'$ & $\one_0$
& $0$ & $T_3$ & 4 & L, R & $4\times (\one,\one;{\bf 10}')$
\\
\hline
$\left(0~0~0~;\underline{1~0}~;~0~0~0\right)(+-;0~0^6)'$ &
$H_{d-1/2}$
& $0$ & $T_6$ & 6 & L, R &
\\
$\left(0~0~0~;0~0~;-1~0~0\right)(+-;0~0^6)'$ & $\one_0$
& $0$ & $T_6$ & 6 & L, R & $6\times (\one,{\bf 4};\one)$
\\
$\left(0~0~0~;0~0~;~0~0~-1\right)(+-;0~0^6)'$ & $\one_0$
& $0$ & $T_6$ & 6 & L, R &
\\ \hline
$\left(\underline{1~0~0}~;~0~0~;~0~0~0\right)(-+;0^6)'$ &
$\overline{D}_{1/3}$
& $0$ & $T_6$ & 10& L, R & $10\times ({\bf 3},\one ;\one)$
\\ \hline
\end{tabular}
\caption{6D twisted sector matter states under
SU(3)$\times$SU(4)$\times$SO(10)$'$ satisfying $P\cdot W={\rm integer}$. $\sqrt{2}Q_c$ is the unnormalized U(1)$_c$ charge, i.e. composed of integers from Eq. (\ref{U1charges}). We do not display the CTP conjugates here.} \label{tab:HT}
\end{table}

In the model in Ref.~\cite{SMZ12I}, a lot of the standard model singlets develop VEVs through non-renormalizable Yukawa couplings. By such singlets' VEVs, extra fields unobserved in the MSSM become superheavy, and extra U(1)s are broken, leaving just the SM gauge symmetry. For simplicity and reality, let us assume that SU(4) is broken to SU(2)$_L$ through the couplings of the SM singlets, which achieve VEVs of the string scale at the fixed points. Such localized masses would shift up the KK masses of the coset ${\rm SU(4)/SU(2)_L}$ gauge sector by the amount of order of the compactification scale $1/R$ \cite{Numuraetal}.  For lower lying KK masses, this mechanism works more effectively. It is because in the limit $R\rightarrow\infty$, the effect by the fixed points should disappear and the 6D result be restored. Hence such
localized masses do not affect $b^{N=2}_{\rm SU(8)}$ in
Eq.~(\ref{b6DfullG}) or $b_{\cal H+ G/H }$ in
Eq.~(\ref{gaugerun}). As seen in Eqs.~(\ref{gaugerun}) and
(\ref{kkmass}), the KK mass-shifting (by Wilson line or a localized mass or whatever) leads to removing the contribution by states of the shifted KK masses from $b_{\cal H}$ in Eq.~(\ref{gaugerun}). Hence, in our case $b_{\cal H}$ becomes $b^{N=2}_{\rm SU(2)_L}$ with keeping intact $b_{\cal H+ G/H }=b^{N=2}_{\rm
SU(8)}$.

In ``Model E'' in Ref.~\cite{SMZ12I}, the hypercharge is
defined using only with the E$_8$ part:
$Y_e=\sqrt{\frac{3}{5}}(\frac13,\frac13,\frac13;-\frac12,
-\frac12;0^3)(0^8)'$. We obtain the E$_8$ part beta function coefficients by KK states as
\begin{equation}
\begin{array}{l}
b^{N=2}_{\rm SU(3)_c}=-2\times 3+\frac12\times 2\times (3+2+4+10)=13 ,\\
b^{N=2}_{\rm SU(2)_L}=-2\times 2+\frac12\times 2\times (2+3+4+6)=11 ,\\
b^{N=2}_{\rm U(1)_{Y_e}}=\frac35\times 2\times (\frac14\times
2\times 12+\frac{1}{36}\times 6+\frac49\times 3\times
2+\frac19\times 3\times 15 +1)=\frac{89}{5} .
\end{array}\label{b6DKKSM}
\end{equation}

Although we kept here all the states shown in Tables~\ref{tab:HU} and \ref{tab:HT}, one could consider the possibility that they get localized masses at the fixed points. Then their contribution would be removed from the $b_{\cal H}$ with leaving intact $b_{\cal H+ G/H}$ as in the gauge sector.

In ``Model S'' of Ref.~\cite{SMZ12I}, the SM hypercharge is
defined as
$Y_s=\sqrt{\frac{3}{11}}(\frac13,\frac13,\frac13;-\frac12,
-\frac12;0^3)(0,0,1;0^5)'$. That is to say, U(1)$_{Y_s}$ is a linear combinations of U(1)$_{Y_e}$ and U(1)$_c$, which is a part of the hidden sector E$_8'$ and its charge is defined by
\begin{equation}\label{U1charges}
Q_c=\frac{1}{\sqrt{2}}(0^8)(0~0~1~;~0^5)' .
\end{equation}
In terms of U(1)$_{Y_e}$ and U(1)$_c$ gauge couplings $g_{Y_e}$ and $g_c$, the gauge coupling of U(1)$_{Y_s}$ is given by
\begin{eqnarray} \label{g_s}
\frac{1}{g_{Y_s}^2}=\frac{1}{11}\left(\frac{5}{g_{Y_e}^2}
+\frac{6}{g_c^2}\right) .
\end{eqnarray}
For the hidden sector E$_8'$ part, we have ten vectors and two spinors of SO(10)$'$. The beta function coefficients for SO(10)$'$ and U(1)$_c$ are
\begin{equation}
\begin{array}{l}
b^{N=2}_{\rm SO(10)'}=-2\times 8+ 1\times 2\times 5+2\times
2\times 1 =-2 ,
\\
b^{N=2}_{c}=\frac12\times 2\times [1\times 1+\frac14\times 16]=5 .
\end{array}\label{b6DKKhid}
\end{equation}
Note that SO(10)$'$ and U(1)$_c$ are the subgroups of SO(12)$'$. As seen in Eq.~(\ref{b6DfullG}), the $R^2$ term's coefficient for SO(12)$'$ in the threshold correction vanishes. Therefore, these coefficients of the ${\rm log} R^2$ term in Eq.~(\ref{gaugerun}) are leading terms.  Particularly, from Eq.~(\ref{gaugerun}), the gauge coupling of  U(1)$_c$ (SO(10)$'$) decreases (increases) with energy above the compactification scale $M_R$. Hence, around the
compactification scale, the U(1)$_{Y_s}$ gauge coupling becomes $g_{Y_s}\approx\sqrt{\frac{11}{5}}\times g_{Y_e}$ from Eq.~(\ref{g_s}).

Now we can use the result (\ref{b6DfullG}) for $b_{\cal H+G/H}$. And for $b_{\cal H}$ we use Eqs. (\ref{b6DKKSM}) and (\ref{b6DKKhid}). For $b^0_{\cal H}$ we use the 4D spectrum of Ref. \cite{SMZ12I} or Tables \ref{tab:1} and \ref{tab:2},
\begin{equation}
b_3^0= -3,\quad b_2^0=1,\quad b_1^0= \frac{33}{5}.
\end{equation}
As mentioned before, however, we have a lot of superheavy extra states, which form vector-like representations under the SM gauge symmetry. If their masses are lighter slightly than the string scale, they would contribute to the threshold corrections of the gauge couplings. We can parametrize the contributions by heavy vector-like states as
\begin{eqnarray}
b_i=h_i\times \left[{\rm log}\frac{M_s^2}{M_R^2}+1.89\right] ,
\end{eqnarray}
where $i=1,2,3$, and $M_s$ and $M_R$ denote the string and
compactification scale. $h_i$, which should be positive definite, are parameters depending on the number of states lighter than string scale and their masses (and also charges for $h_1$).

%%%%%%%%%%%%%%%%%%%%%%%%%%%%%%%%%%%%%%%%%%%%%%%%%%%%%%%%%%%%%%%
\subsection{Fitting $\sin^2\theta_W$ by gauge coupling running in 6D on $T^2/Z_4$}

The electroweak hypercharge $Y$ has different combinations in Model E and Model S. Model E has $\sin^2\theta_W^0=\frac38$ at the string scale, for which our emphasis is to find the allowed range for the masses and number of the superheavy extra vector-like states. Model S has $\sin^2\theta_W^0=\frac3{14}$ at the string scale,
for which we try to fit $R$ such that the the electroweak scale $\sin^2\theta_W$ is increased to an acceptable one. We use the following observed values \cite{PData},
\begin{equation}
\sin^2\theta_W(M_Z)=0.22306\pm 0.00033,\quad
\alpha_3(M_Z)=0.1216\pm 0.0017.\label{dataswac}
\end{equation}

%%%%%%%%%%%%%%%%%%%%%%%%%%%%%%%%%%%%%%%%%%%%%%%%%%%%%%%%%%%%%
\subsubsection{Model E}

In this model, the hypercharge coupling $g_Y$ evolves below
$R^{-1}$ as the blue line of Fig. \ref{fig:evolution} up to the U(1)$_Y$ normalization, $g_Y=\sqrt{{3}/{5}}g_{1,\rm SU(5)}$ and the SU(3) and SU(2) couplings as the black lines of Fig. \ref{fig:evolution}. Thus, at $R^{-1}\equiv M_R$ the weak mixing angle is $\frac38$. The three MSSM couplings at $M_Z$ are given by
\begin{eqnarray}
&&\frac{4\pi}{\alpha_{3}(M_Z)}
=\frac{4\pi}{\alpha_*}-3\ln\frac{M_s^2}{M_Z^2}
+\left(h_3-\frac{13}{4}\right)\left[\ln\frac{M_s^2}{M_R^2}
+1.89\right]+6
\left[\frac{2\pi}{\sqrt{3}}\frac{M_s^2}{M_R^2}-0.30\right] ,
\nonumber \\
&&\frac{4\pi}{\alpha_{2}(M_Z)}
=\frac{4\pi}{\alpha_*}+\ln\frac{M_s^2}{M_Z^2}
+\left(h_2-\frac{11}4\right)
\left[\ln\frac{M_s^2}{M_R^2}+1.89\right]+6
\left[\frac{2\pi}{\sqrt{3}}\frac{M_s^2}{M_R^2}-0.30\right] ,
\\
&&\frac{4\pi}{\alpha_{Y_e}(M_Z)}=\frac{4\pi}{\alpha_*}
+\frac{33}{5}\ln\frac{M_s^2}{M_Z^2}+\left(h_1-\frac{89}{4\cdot
5}\right)\left[ \ln\frac{M_s^2}{M_R^2}+1.89\right]+6
\left[\frac{2\pi}{\sqrt{3}}\frac{M_s^2}{M_R^2}-0.30\right] ,
\nonumber
\label{runE}
\end{eqnarray}
Since the $M_s^2/M_R^2$ term is a common term, the differences between two gauge couplings, e.g.
$\alpha^{-1}_3-\alpha^{-1}_{Y_e}$ and
$\alpha^{-1}_2-\alpha^{-1}_{Y_e}$ shows the logarithmic behavior. Let us take their linear combination of form,
\begin{align}
(2\alpha^{-1}_3+3\alpha^{-1}_2-5\alpha^{-1}_{Y_e})\vert_{M_Z}
=-\frac{36}{4\pi}\ln\frac{M_s^2}{M_Z^2}
+\frac{1}{4\pi}\left(2h_3+3h_2-5h_1+\frac{15}{2}\right)\left[
\ln\frac{M_s^2}{M_R^2}+1.89\right]
\end{align}
which can be decomposed as $2(\alpha^{-1}_3-\alpha^{-1}_{Y_e})
+3(\alpha^{-1}_2-\alpha^{-1}_{Y_e})$. The comparison with the
minimal SU(5) model unification, in which
$(2\alpha^{-1}_3+3\alpha^{-1}_2-5\alpha^{-1}_{Y_e})\vert_{M_Z}
=-\frac{36}{4\pi}\ln\frac{M_G^2}{M_Z^2}$, yields
\begin{eqnarray}
\ln\frac{M_s^2}{M_G^2}=\frac{1}{36}\left(2h_3+3h_2-5h_1
+\frac{15}{2}\right)\left[\ln\frac{M_s^2}{M_R^2}+1.89\right] ,
\end{eqnarray}
where the grand unification scale $M_G$ is restricted
experimentally in the range $1\times 10^{16} ~{\rm GeV}\lesssim M_G\lesssim 3\times 10^{16}$ GeV. To avoid rapid proton decay~\cite{Shiozawa}, the compactification scale needs to fulfill the bound $M_R\gtrsim 5\times 10^{15}$ GeV~\cite{Hebecker}. With $M_R<M_s\approx 5\times 10^{17}$ GeV, we obtain the quite broad restriction:
\begin{eqnarray}
10.75\lesssim (2h_3+3h_2-5h_1)\lesssim 141.53 .
\end{eqnarray}
Another combination $7(\alpha^{-1}_3-\alpha^{-1}_{Y_e})
-12(\alpha^{-1}_2-\alpha^{-1}_{Y_e})$ provides
\begin{eqnarray} \label{restrict1}
(7\alpha^{-1}_3-12\alpha^{-1}_2+5\alpha^{-1}_{Y_e})\vert_{M_Z}
=\frac{1}{4\pi}(7h_3-12h_2+5h_1-12)\left[
\ln\frac{M_s^2}{M_R^2}+1.89\right] .
\end{eqnarray}
From the experimental uncertainty of $\alpha_3(M_Z)$ and $5\times 10^{15}~{\rm GeV} \lesssim M_R\lesssim M_s$, we get one more restriction,
\begin{eqnarray} \label{restrict2}
22.78\lesssim (7h_3-12h_2+5h_1)\lesssim 86.02.
\end{eqnarray}
The restrictions Eqs.~(\ref{restrict1}) and (\ref{restrict2}) are easily satisfied, for example, if $6\lesssim h_3\lesssim 12$ and $h_2\approx h_1\approx 0$. It means that some colored states heavier than compactification scale could lead to gauge coupling unification at the string scale.  As seen in Ref.~\cite{SMZ12I}, there are sufficient vector-like superheavy colored states in the model. Particularly, four pairs among them are electrically neutral SU(2) singlets.

%%%%%%%%%%%%%%%%%%%%%%%%%%%%%%%%%%%%%%%%%%%%%%%%%%%%%%%%%%%%%%
\subsubsection{Model S}

At the compactificatin scale, the U(1)$_c$ gauge coupling is
estimated as
\begin{eqnarray}
\frac{4\pi}{\alpha_{c}(M_R)}=\frac{4\pi}{\alpha_*}+
\left(h_c-\frac{5}{4}\right)\left[\ln\frac{M_s^2}{M_R^2}
+1.89\right]
,
\end{eqnarray}
where there is no $R^2$ (i.e. $\frac{M_*^2}{M_R^2}$) term, and $h_c$ parametrizes the states charged under U(1)$_c$, which are from the twist sectors preserving only ${\cal N}=1$ SUSY. In Model S, the gauge coupling of the hypercharge is given by that of  Model E and the U(1)$_c$ gauge coupling as seen in Eq.~(\ref{g_s}). Accordingly, at the $M_Z$ scale we have
\begin{eqnarray} \label{eq0}
\frac{4\pi}{\alpha_{Y_s}(M_Z)} \approx \frac{4\pi}{\alpha_*}+
3\ln\frac{M_*^2}{M_Z^2} +\frac{30}{11}
\left[\frac{2\pi}{\sqrt{3}}\frac{M_s^2}{M_R^2}-0.30\right] .
\end{eqnarray}
Here we dropped the $\ln\frac{M_s^2}{M_R^2}$ terms, because the $\frac{M_s^2}{M_R^2}$ term is dominant over it, and moreover such $\frac{M_s^2}{M_R^2}$ term is not cancelled even if we take the differences of the $\alpha_i^{-1}$s unlike in Model E. Note that in Eq.~(\ref{eq0}) the hypercharges of the SM model states should be normalized with $\sqrt{\frac{3}{11}}$ rather than $\sqrt{\frac35}$.  Let us take the linear combinations discussed in Model E:
\begin{eqnarray} \label{combi1}
&&(2\alpha^{-1}_3+3\alpha^{-1}_2-5\alpha^{-1}_{Y_s})\vert_{M_Z}
\approx
-\frac{18}{4\pi}\ln\frac{M_*^2}{M_Z^2}+\frac{180}{4\pi\cdot 11}
\left[\frac{2\pi}{\sqrt{3}}\frac{M_s^2}{M_R^2}-0.30\right] ,
\\ \label{combi2}
&&(7\alpha^{-1}_3-12\alpha^{-1}_2+5\alpha^{-1}_{Y_s})\vert_{M_Z}
\approx
-\frac{18}{4\pi}\ln\frac{M_*^2}{M_Z^2}-\frac{180}{4\pi\cdot 11}
\left[\frac{2\pi}{\sqrt{3}}\frac{M_s^2}{M_R^2}-0.30\right] ,
\end{eqnarray}
where we neglect again the $\ln\frac{M_s^2}{M_R^2}$ terms. With $1/\alpha_3\approx (0.1216)^{-1}$, $1/\alpha_2=30.55$, and $1/\alpha_{Y_s}=106.45\times \frac{3}{11}=29.03$, the left-hand side (LHS) of Eqs. ~(\ref{combi1}) and (\ref{combi2}) are
$-37.05$ and $-163.88$, respectively. Therefore, for Model S we obtain $\frac{M_*}{M_Z}\approx 1.70\times 10^{15}$ and
$\frac{M_s}{M_R}\approx 3.68$.

%%%%%%%%%%%%%%%%%%%%%%%%%%%%%%%%%%%%%%%%%%%%%%%%%%%%%%%%%%%%
%%%%%%%%%%%%%%%%%%%%%%%%%%%%%%%%%%%%%%%%%%%%%%%%%%%%%%%%%%%%
\section{Conclusion}\label{sec:Conclusion}

In this paper, we discussed the KK states in the orbifold
compactification of the heterotic string theory by analyzing the one-loop partition function. The ${\cal N}=2$ SUSY KK massive states associated with (relatively) large extra dimensions can arise only in non-prime orbifolds. In the ${\bf Z}_{12-I}$ orbifold we consider as an example, the partition function relaxes the GSO projection condition defined with $\Theta_k=(P^I+\frac{k}{2}V^I)V^I
-(s+\frac{k}{2}\phi)\phi$ to that defined with $3\times \Theta_k$ above the compactification scale ($1/R$). Accordingly, a 4D gauge symmetry is enhanced above $1/R$
energy scale.  By the analysis of the partition function, it is shown that the other condition on the Wilson line, $P\cdot W={\rm integer}$, turns out to be just a 4D massless condition. Thus, it is invalidated for KK massive states. The masses for the states with $P\cdot W\neq {\rm integer}$ are shifted up by the Wilson line without leaving massless modes. Because of the presence of such KK states of $P\cdot W\neq {\rm integer}$, a 4D gauge symmetry is more enhanced above $1/R$.

One can consider a 6D theory as the limit of $R\to\infty$ from a 4D theory in nonprime orbifolds, and calculate 6D massless spectrum in such a 6D theory. For a consistency check, we compared the KK spectrum obtained from the partition function approach with such a 6D massless spectrum. We have explicitly showed that in a ${\bf Z}_{12-I}$ model the KK spectrum obtained with the relaxed orbifold condition is in general coincident with the massless spectrum in the 6D theory. Hence, one can confirm that more gauge
and matter fields appearing above the compactification scale
indeed become 6D massless fields in the limit of
$R\rightarrow\infty$, and so gauge symmetry and SUSY are enhanced to those of the 6D theory.

We considered a phenomenologically viable model \cite{SMZ12I}, in which the gauge group is ${\rm SU(3)_c\times SU(2)_L\times U(1)_Y}$ with three ${\bf 16}$ chiral matter states. The gauge group in this model turns out to be enhanced to a simple group SU(8). Thus, it is a realization of the 6D SUSY GUT in the context of the heterotic string theory.

In the model we consider, the two MSSM Higgs doublets are from the untwisted sector (i.e. from the 10D bulk).  They become a part of the ${\cal N}=2$ gauge multiplet in 6D, while the MSSM matter fields in the untwisted sector form ${\cal N}=2$ hypermultiplets. Thus, this model realizes the idea of ``gauge-Higgs unification.''

The knowledge on the string partition function enables us to obtain the threshold corrections to the gauge couplings. We explicitly calculated the effect by  Wilson lines on the threshold correction in ${\bf Z}_{12-I}$ compactification. Since the orbifold symmetry breaking mechanism in string theory works as the GSO projection rather than as the massless condition unlike in orbifold {\it field} theory, breaking of a 6D gauge symmetry just by orbifolding does not affect the logarithmic threshold correction at all. It is a feature different from the orbifold field theory. However, 6D gauge symmetry breaking by the Wilson line ($P\cdot W={\rm integer}$) in string theory works as a masslessness condition.  Hence, such a breaking by the Wilson line affects the logarithmic threshold correction.

We also showed that the threshold corrections by KK massive states can lead to gauge coupling unification at the string scale even for models with $\sin^2\theta_W^0\ne \frac38$ at the string scale.

%%%%%%%%%%%%%%%%%%%%%%%%%%%%%%%%%%%%%%%%%%%%%%%%%%%%%%%%%%%%%%%%%%%%%%%%%%

\acknowledgments{}

This work is supported in part by the KRF Grants, No. R14-2003-012-01001-0 and No. KRF-2005-084-C00001.

%%%%%%%%%%%%%%%%%%%%%%%%%%%%%%%%%%%%%%%%%%%%%%%%%%%%%%%%%%%%%%%%%%%%%%%%%
%%%%%%%%%%%%%%%%%%%%%%%%%%%%%%%%%%%%%%%%%%%%%%%%%%%%%%%%%%%%%%%%%%%%%%%%%

\vskip 1cm \centerline{\bf Appendix} \vskip 0.5cm

In this appendix, we present the effect of Wilson lines on the threshold correction, basically following the calculation of Appendix B of Ref.~\cite{Dixon:1990pc}. For the case of ${\rm det}A=0$ (``degenerate orbit'') to consider the threshold correction, one can choose in general the following ``representative matrix''~\cite{Dixon:1990pc}:
\begin{eqnarray} \label{A_D}
A_D=\left(
\begin{array}{cc}
0 & j \\
0 & p \end{array} \right) ~.
\end{eqnarray}
That is to say $\vec{\zeta}=0$ and $\zeta^{\prime 3}\equiv j$, $\zeta^{\prime 4}\equiv p$. Then the contribution by $A_D$ to the threshold correction $\Delta_i$ is
\begin{eqnarray}
I_3=\int_{-1/2}^{1/2}d\tau_1\int_0^\infty
\frac{d\tau_2}{\tau_2^2}
\left\{T_I\sum_{j,p}{\rm exp}\left[-\frac{\pi
T_I}{\tau_2U_I}\left|j+Up\right|^2+2\pi i(j+p)\theta_{0}\right]
-\tau_2\cdot\theta(\tau\in\Gamma)\right\} , \nonumber
\end{eqnarray}
where $\theta(\tau\in\Gamma)$ is defined as 1 for $\tau\in \Gamma$ but 0 otherwise. Note that the summation here is over all $(j,p)\neq 0$. Since $|j+Up|^2$ is invariant under $j\rightarrow -j$ and $p\rightarrow -p$, $e^{2\pi i(j+p)\theta_0}$ can be replaced by ${\rm cos}[2\pi (j+p)\theta_0]$. Since $\vec{\zeta}=0$, $\theta_0$ is given by $P^IW^I$. In $\Z_{12-I}$, $\theta_0=P^IW^I=0,\pm\frac13$ (mod Z).

As argued in VII, $\hat{Z}_{\rm torus}$ in Eq.~(\ref{Ztorus}) is valid regardless whether $(\vec{\delta}_4;\vec{\delta}_2)=(0;0)$ or not. For all the sectors $[k,l]$ in Fig.~3, the above integrations give indeed the same values. For instance, the $[k=3,9;l]$ sector, $j$ and $p$ are $4\times {\rm integer}$s. Instead, there are 15 more sectors giving the same integrations for each sector of $[k=3,9;l]$ by including also $\delta_4^a= 1,2,3$ ($a=3,4$) sectors, which cancel the effect by $j$, $p=4\times {\rm integers}$. Similarly, in the $[k=6,l]$ sectors, $j$ and $p$ are $2\times {\rm integer}$s, and there are 3 more sectors for each $[k=6;l]$ from the sectors $(\delta_2^3,\delta_2^4)=(1,0)$ $(0,1)$, and $(1,1)$. They compensate the effects by $j$, $p=2\times {\rm integers}$ in the integration. On the other hand, in $[k=0;l]$ sectors, $j$ and $p$ are just ordinary integers. However, the sectors from $(\vec{\delta}_4;\vec{\delta}_2)\neq (0;0)$ are not consistent with $A_D$: Only the sectors of
$(\vec{\delta}_4;\vec{\delta}_2)=(0;0)$ should be counted in the above integration.

After integrations, $I_3$ becomes
\begin{eqnarray} \label{I3}
I_3=\lim_{N\to\infty}\left[\frac{U_I}{\pi}
\sum_{j,p}\left\{\frac{c_jc_p-s_js_p}{\left|j+Up\right|^2}
-\frac{c_jc_p-s_js_p}{\left|j+Up\right|^2+NU_I/\pi T_I
}\right\}-\int_\Gamma d^2\tau\frac{1-e^{-N/\tau_2}}{\tau_2}\right]
,
\end{eqnarray}
where $c_jc_p-s_js_p$ indicates $\cos(2\pi j\theta_0)\cos(2\pi
p\theta_0)-\sin(2\pi j\theta_0)\sin(2\pi p\theta_0)$ with
$\theta_0=P\cdot W=0,\pm\frac{1}{3}$. The regulator
$(1-e^{-N/\tau_2})$ will be eventually removed by taking
$N\rightarrow\infty$.

For calculations of Eq.~(\ref{I3}), let us list some useful
formulae:
\begin{eqnarray} \label{formula}
\sum_{n=-\infty}^{+\infty}\frac{1}{(n+B)^2+C^2}
&=&\frac{\pi}{2C}\left[i{\rm cot}(\pi {\cal Z})-i{\rm
cot}(\pi\bar{\cal Z})\right] =\frac{\pi}{2C}\left[2+\frac{2q_{z
}}{1-q_{ z}}+\frac{2\bar{q}_{z}}{1-\bar{q}_{z}}\right] ,
\\ \label{cos}
 \sum_{n=-\infty}^{+\infty}\frac{{\rm cos}\frac{2n
}{3}\pi}{(n+B)^2+C^2}&=&\sum_{n=-\infty}^{+\infty}
\left[\frac{1}{(3n+B)^2+C^2}-\frac{1/2}{(3n+1+B)^2+C^2}
-\frac{1/2}{(3n-1+B)^2+C^2}\right]
\nonumber \\
&=&\frac{\pi}{6C}\left\{i{\rm cot}(\pi{\cal Z}/3)-\frac{i}{2}{\rm
cot}[\pi({\cal Z}+1)/3]-\frac{i}{2}{\rm cot}[\pi({\cal
Z}-1)/3]+{\rm c.c. }\right\}
\nonumber \\
&=&\frac{\pi}{6C}\left[\frac{2q_{z }^{1/3}}{1-q_{
z}^{1/3}}-\frac{q_{z}^{1/3}e^{2\pi i/3}}{1-q_{z}^{1/3}e^{2\pi
i/3}}-\frac{q_{z }^{1/3}e^{-2\pi i/3}}{1-q_{z}^{1/3}e^{-2\pi
i/3}}+{\rm c.c.}\right] ,
\\ \label{sin}
\sum_{n=-\infty}^{+\infty}\frac{{\rm sin}\frac{2n
}{3}\pi}{(n+B)^2+C^2}
&=&\frac{\pi}{2\sqrt{3}C}\left\{\frac{i}{2}{\rm cot}[\pi({\cal
Z}+1)/3]-\frac{i}{2}{\rm cot}[\pi({\cal Z}-1)/3]+{\rm c.c.
}\right\}
\nonumber \\
&=&\frac{\pi}{2\sqrt{3}C}\left[\frac{q_{z}^{1/3}e^{2\pi
i/3}}{1-q_{z}^{1/3}e^{2\pi i/3}}-\frac{q_{z }^{1/3}e^{-2\pi
i/3}}{1-q_{z}^{1/3}e^{-2\pi i/3}}+{\rm c.c.}\right] ,
\end{eqnarray}
where ${\cal Z}\equiv B+iC$ and $q_z\equiv e^{2\pi i {\cal Z}}$. For $C\rightarrow +\infty$, $q_z\rightarrow 0$, and so the asymptotic behaviors of Eqs.~(\ref{formula}), (\ref{cos}), and (\ref{sin}) are
\begin{eqnarray}
&&\quad\quad\quad\quad\quad\quad
\sum_{n=-\infty}^{+\infty}\frac{1}{(n+B)^2+C^2}\longrightarrow
\frac{\pi}{C} ,
\\
&&\sum_{n=-\infty}^{+\infty}\frac{{\rm cos}\frac{2n
}{3}\pi}{(n+B)^2+C^2} \longrightarrow 0 ~, ~~
\sum_{n=-\infty}^{+\infty}\frac{{\rm sin}\frac{2n
}{3}\pi}{(n+B)^2+C^2} \longrightarrow 0 .
\end{eqnarray}
For $B=C=0$, Eqs.~(\ref{formula}), (\ref{cos}), and (\ref{sin}) should be replaced by
\begin{eqnarray}
\sum_{n\neq 0}\frac{1}{n^2}=\frac{\pi^2}{3} , \quad\quad
\sum_{n\neq 0}\frac{{\rm
cos}\frac{2n}{3}\pi}{n^2}=-\frac{\pi^2}{9} , \quad\quad
\sum_{n\neq 0}\frac{{\rm sin}\frac{2n}{3}\pi}{n^2}= 0 .
\end{eqnarray}

With the above formulae, in the case $\theta_0=P\cdot W={\rm
integer}$, one can calculate
\begin{eqnarray}
&&\quad\quad\quad\quad
\frac{U_I}{\pi}\sum_{j,p}\left(\frac{1}{\left|j+Up\right|^2}
-\frac{1}{\left|j+Up\right|^2
+NU_I/\pi T_I }\right) \nonumber \\ \label{I3interm}
&&=\frac{\pi}{3}U_I+\sum_{p>0}\frac{2}{p}
\left(\frac{q_U^p}{1-q_U^p}+\frac{\bar{q}_U^p}{1-\bar{q}_U^p}
\right)
+\sum_{p>0}\left(\frac{2}{p}-\frac{2}{\sqrt{p^2+(N/\pi
T_IU_I)}}\right) ,
\end{eqnarray}
where $q_U\equiv e^{2\pi i U}$. The first term on the right-hand side is the result coming from $p=0$. Thus, in the limit $N\rightarrow \infty$, Eq.~(\ref{I3}) becomes \cite{Dixon:1990pc}
\begin{eqnarray}
I_3 = \frac{\pi}{3}U_I -2\sum_{n=1}^{\infty}{\rm
log}\left|1-q_U^n\right|^2 -{\rm
log}(T_IU_I)+\left(\gamma_E-1-{\rm
log}\frac{8\pi}{3\sqrt{3}}\right) \label{Is}
\end{eqnarray}
for $\theta_\zeta=P\cdot W={\rm integer}$. Here the formula
$\sum_{p>0}\frac{1}{p}\frac{q^p}{1-q^p}=-\sum_{n>0}{\rm
log}(1-q^n)$ was utilized. Note that the ${\rm log}(T_IU_I)$ term and the last three constant terms are originated from the last two terms in Eq.~(\ref{I3interm}).  The second term in Eq.~(\ref{I3interm}) is much smaller than the other terms and so does not contribute to Eq.~(\ref{thresholdapprx}).

For $\theta_0=P\cdot W=\pm\frac13$ (mod Z), we have
\begin{eqnarray} \label{I3interm2}
I_3&=&-\frac{\pi}{9}U_I +\sum_{p>0}\frac{{\rm
cos}\frac{2p}{3}\pi}{3p}\left[\frac{2q_U^{p/3}}{1-q_U^{p/3}}
-\frac{q_U^{p/3}e^{2\pi i/3}}{1-q_U^{p/3}e^{2\pi i/3}}
-\frac{q_U^{p/3}e^{-2\pi i/3}}{1-q_U^{p/3}e^{-2\pi i/3}}+{\rm
c.c.}\right]
\nonumber \\
&&\quad\quad-\sum_{p>0}\frac{{\rm
sin}\frac{2p}{3}\pi}{\sqrt{3}p}\left[\frac{q_U^{p/3}e^{2\pi
i/3}}{1-q_U^{p/3}e^{2\pi i/3}} -\frac{q_U^{p/3}e^{-2\pi
i/3}}{1-q_U^{p/3}e^{-2\pi i/3}}+{\rm c.c. }\right] .
\end{eqnarray}
Note that in Eq.~(\ref{I3interm2}) there are no terms
corresponding to the terms in the last big bracket of Eq.
(\ref{I3interm}). Hence, for the states with $\theta_0=P^I
W^I=\pm\frac13$ (mod Z), there does not appear ${\rm log}R^2$ corrections in $\Delta_i$, and only $-\frac{\pi}{9}U_I$, which is just a number ($\approx -0.3$) in the $\Z_{12-I}$ orbifiold compactification, contributes to Eq.~(\ref{thresholdapprx}).

%%%%%%%%%%%%%%%%%%%%%%%%%%%%%%%%%%%%%%%%%%%%%%%%%%%%%%%%%%%%%%%%%%%%%%%%%
%%%%%%%%%%%%%%%%%%%%%%%%%%%%%%%%%%%%%%%%%%%%%%%%%%%%%%%%%%%%%%%%%%%%%%%%%

%%%%%%%%%%%%%%%%%%%%%%%%%%%%%%%%%%%%%%%%%%%%%%%%%%%%%%%%%%%%%%%%%%%%%%%%%%%
%%%%%%%%%%%%%%%%%%%%%%%%%%%%%%%%%%%%%%%%%%%%%%%%%%%%%%%%%%%%%%%%%%%%%%%%%%%

\end{document}